\documentclass[fleqn,usenatbib]{mnras}

\usepackage{orcidlink}
\usepackage{newtxtext,newtxmath}
\usepackage{xspace}
\usepackage{bm}
\usepackage{xcolor}

\usepackage[T1]{fontenc}
\usepackage{booktabs}
\usepackage{tabularx}
\usepackage{siunitx}
\usepackage{hyperref}
\usepackage{textcomp}
\usepackage[capitalise]{cleveref}
\usepackage{adjustbox}  
\usepackage{wasysym}    
\crefname{figure}{Fig.}{Figs.}
\crefname{table}{Table}{Tables}
\DeclareRobustCommand{\VAN}[3]{#2}
\let\VANthebibliography\thebibliography
\def\thebibliography{\DeclareRobustCommand{\VAN}[3]{##3}\VANthebibliography}

\usepackage{graphicx} 
\usepackage{amsmath}  

\makeatletter 
  \patchcmd{\NAT@citex}
    {\@citea\NAT@hyper@{%
      \NAT@nmfmt{\NAT@nm}%
      \hyper@natlinkbreak{\NAT@aysep\NAT@spacechar}{\@citeb\@extra@b@citeb}%
      \NAT@date}}
    {\@citea\NAT@nmfmt{\NAT@nm}%
    \NAT@aysep\NAT@spacechar\NAT@hyper@{\NAT@date}}{}{}

  \patchcmd{\NAT@citex}
    {\@citea\NAT@hyper@{%
      \NAT@nmfmt{\NAT@nm}%
      \hyper@natlinkbreak{\NAT@spacechar\NAT@@open\if*#1*\else#1\NAT@spacechar\fi}%
        {\@citeb\@extra@b@citeb}%
      \NAT@date}}
    {\@citea\NAT@nmfmt{\NAT@nm}%
    \NAT@spacechar\NAT@@open\if*#1*\else#1\NAT@spacechar\fi\NAT@hyper@{\NAT@date}}
    {}{}
\makeatother

\newcommand{\lumina}{\mbox{\textsc{Lumina}}\xspace}
\newcommand{\luminafifty}{\mbox{\textsc{Lumina-50}}\xspace}
\newcommand{\luminafiftynox}{\mbox{\textsc{Lumina-50-NoX}}\xspace}
\newcommand\thesanone{\mbox{\textsc{Thesan-1}}\xspace}
\newcommand\thesantwo{\mbox{\textsc{Thesan-2}}\xspace}

\newcommand{\illustrisTNG}{\mbox{\textsc{IllustrisTNG}}\xspace}

\newcommand\Msun{\text{M}_{\astrosun}}

\newcommand\Zsun{\text{Z}_{\astrosun}}

\newcommand{\Mpc}{{\rm Mpc}}
\newcommand{\kms}{{\rm km\, s^{-1}}}

\newcommand{\HI}{\ion{H}{I}\xspace}
\newcommand{\HII}{\ion{H}{II}\xspace}
\newcommand{\HeI}{\ion{He}{I}\xspace}
\newcommand{\HeII}{\ion{He}{II}\xspace}
\newcommand{\HeIII}{\ion{He}{III}\xspace}

\newcommand{\arepo}{\textsc{arepo}\xspace}
\newcommand{\camb}{\textsc{camb}\xspace}
\newcommand{\areport}{\textsc{arepo-rt}\xspace}

\newcommand{\thesan}{\textsc{Thesan}\xspace}
\newcommand{\jwst}{\texttt{JWST}\xspace}
\newcommand{\alma}{\texttt{ALMA}\xspace}

\newcommand{\thzoom}{\mbox{\textsc{Thesan-Zoom}}\xspace}

\renewcommand{\vec}[1]{ {\bmath #1} } 
\DeclareSIUnit\erg{erg}
\DeclareSIUnit\angstrom{\text{\AA}}

\newcommand\orcid[1]{\href{https://orcid.org/#1}{\adjustbox{trim={-.15\width} 0 {-.15\width} 0,clip}{\includegraphics[height=9pt]{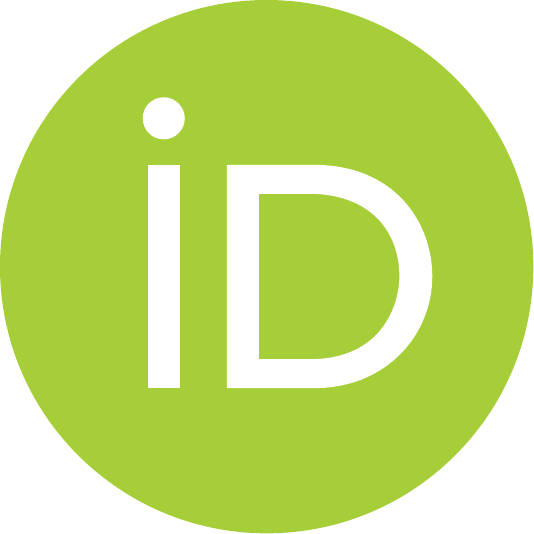}}}}
\newcommand{\appropto}{\mathrel{\vcenter{
  \offinterlineskip\halign{\hfil$##$\cr
    \propto\cr\noalign{\kern2pt}\sim\cr\noalign{\kern-2pt}}}}}

\title[\lumina]{Introducing the \lumina project: large-volume radiation-hydrodynamic simulations of the epochs of hydrogen and helium reionization}

\author[Zier et al.]{%
Oliver Zier\orcid{0000-0003-1811-8915},$^{1}$\thanks{E-mail: \href{mailto:oliver.zier@cfa.harvard.edu}{oliver.zier@cfa.harvard.edu}}
Aaron Smith\orcid{0000-0002-2838-9033},$^{2}$ Xuejian Shen\orcid{0000-0002-6196-823X},$^{3,4}$ Rongrong Liu\orcid{0000-0003-0685-3525},$^{1}$ Rahul Kannan\orcid{0000-0001-6092-2187},$^{5}$
\newauthor Sonja M. Koehler\orcid{0009-0008-0814-3328},$^{1}$ Volker Springel\orcid{0000-0001-5976-4599},$^{6}$ R\"udiger Pakmor\orcid{0000-0003-3308-2420},$^{6}$ Mark Vogelsberger\orcid{0000-0001-8593-7692},$^{3,4}$ \newauthor Teodora-Elena Bulichi\orcid{0000-0001-8174-6389}$^{3,4}$ and Lars Hernquist\orcid{0000-0001-6950-1629}$^{1}$
\vspace*{0.1cm}\\%
$^{1}$ Center for Astrophysics $|$ Harvard $\&$ Smithsonian, 60 Garden Street, Cambridge, MA 02138, USA\\%
$^{2}$ Department of Physics, The University of Texas at Dallas, Richardson, TX 75080, USA \\%
$^{3}$ Massachusetts Institute of Technology, Kavli Institute for Astrophysics and Space Research, 77 Massachusetts Avenue, Cambridge, MA 02139, USA\\%
$^{4}$ Department of Physics, Massachusetts Institute of Technology, 77 Massachusetts Avenue, Cambridge, MA 02139, USA\\%
$^{5}$ Department of Physics and Astronomy, York University, 4700 Keele Street, Toronto, ON M3J 1P3, Canada\\%
$^{6}$ Max Planck Institute for Astrophysics, Karl-Schwarzschild-Str. 1, D-85741 Garching, Germany
}

\date{Accepted XXX. Received YYY; in original form ZZZ}

\pubyear{\the\year{}}

\begin{document}
\label{firstpage}
\pagerange{\pageref{firstpage}--\pageref{lastpage}}
\maketitle

\begin{abstract}
Understanding how galaxies and active galactic nuclei (AGN) jointly drive the reionization of the intergalactic medium (IGM) across cosmic time remains a major challenge in cosmology. We present \lumina, a large-volume radiation-hydrodynamic simulation that self-consistently follows the coupled evolution of the intergalactic medium, galaxies, and AGN through \HI, \HeI, and \HeII reionization down to redshift $z=3$. \lumina evolves a cosmological volume of comoving side length $L_{\mathrm{box}}=500\,\mathrm{cMpc}$ with $2\times 6000^{3}$ resolution elements, corresponding to baryonic and dark-matter mass resolutions of $3.6\times 10^{6}\,\Msun$ and $1.9\times 10^{7}\,\Msun$, respectively. The simulation uses the moving-mesh code \arepo, combining the IllustrisTNG galaxy-formation model with a GPU-accelerated M1 radiation-transport solver in six frequency bins. The initial conditions employ separate transfer functions for baryons and dark matter and include their relative streaming velocity. \lumina predicts a late, predominantly stellar-driven hydrogen reionization, with the median sub-volume fully ionized by $z\approx 5.2$ and residual neutral \HI patches persisting until $z\approx 4.75$. \HeII reionization is driven self-consistently by AGN and is nearly complete by $z=3$. The simulation yields a Thomson-scattering optical depth in excellent agreement with Planck, an IGM thermal history and photoionization background broadly consistent with observational constraints, and a clear late-time thermal boost associated with \HeII reionization. Its galaxy population remains consistent with the original IllustrisTNG project, while the larger volume improves statistics for rare objects, large-scale environments, and cosmic variance, enabling forward modelling of observables linking \HI and \HeII topologies to the evolving galaxy and AGN populations.
\end{abstract}




\section{Introduction}
After the Big Bang, the Universe was filled with a hot, dense plasma in which baryons and radiation were tightly coupled. As it expanded and cooled, electrons recombined with primordial nuclei---mostly hydrogen and helium---allowing photons to decouple from matter $\sim 3.8\times10^{5}\,\mathrm{yr}$ after the Big Bang ($z\simeq 1100$). This relic radiation subsequently free-streamed and cooled to become the cosmic microwave background (CMB; \citealt{alpher1948relative,penzias1965measurement, Dicke1965}), observed today as a nearly perfect blackbody with temperature $T_{\rm CMB}\simeq 2.7255\,\mathrm{K}$ \citep{fixsen2009}. The CMB encodes the initial conditions for cosmic structure formation and provides stringent constraints on the cosmological model  \citep[e.g.][]{planck2020}.

The growth of structures is driven primarily by collisionless cold dark matter (CDM), whose density perturbations begin to grow efficiently once the energy density of the Universe becomes matter-dominated \citep{zeldovich1970}. Prior to recombination, baryons are supported by radiation pressure and undergo baryon acoustic oscillations \citep[BAO;][]{Peebles1970, Eisenstein2005}, while CDM already clusters gravitationally. After recombination, baryons decouple from the photon bath and fall into CDM potential wells; because their clustering is delayed, the baryonic density field is initially smoother, with reduced power relative to the CDM field across a broad range of scales (e.g.\ \citealt{eisenstein1998baryonic}). The delay also imprints a coherent relative (``streaming'') velocity between baryons and dark matter that is coherent over several comoving Mpc and decays slowly with time \citep{tseliakhovich2010relative, dalal2010large}. These streaming motions can suppress gas collapse and star formation in the smallest haloes, thereby modulating the abundance and clustering of the first luminous sources \citep[e.g.][]{greif2011delay, schauer2019influence, schauer2021influence, lake2024supersonic}.

Following recombination, the Universe entered the \emph{dark ages}, during which the primordial gas cooled---first via adiabatic expansion and later through $\mathrm{H_2}$ cooling in dense regions---while the first non-linear structures assembled. The earliest star-forming sites are expected to be dark-matter minihaloes, in which molecular-hydrogen cooling enables gas collapse once the halo mass exceeds a threshold that depends on redshift and on the local radiation background \citep{tegmark1997small,machacek2001simulations,abel2002formation,Klessen2023}. The first generation of stars, Population~III (Pop~III), formed from metal-free gas and is thought to have had a top-heavy initial mass function, producing copious Lyman-continuum (LyC; $h\nu\ge 13.6\,\mathrm{eV}$) photons. Through radiative, chemical, and mechanical feedback, Pop~III stars ionized and enriched their surroundings, enabling the transition to metal-enriched Population~II star formation and the emergence of the first protogalaxies \citep{bromm2011}.

As galaxies grew in abundance and luminosity, their ionizing radiation carved out expanding ionized regions in the intergalactic medium (IGM). Within these bubbles, hydrogen and helium were predominantly singly ionized, while outside these bubbles the IGM remained largely neutral but could be heated by long-mean-free-path photons such as X-rays \citep{madau199721,venkatesan2001heating,Pritchard2007} from high-mass X-ray binaries \citep{mirabel2011stellar,fragos2013x,jeon2014radiative} and from shock-heated gas in the interstellar medium \citep[ISM,][]{2014Pacucci,meiksin2017galactic}. The resulting Epoch of Reionization (EoR) was highly inhomogeneous: ionized regions formed around biased sources, expanded into the cosmic web, and eventually merged \citep{neyer2024thesan, jamieson2025thesan}. Consequently, reionization is best described as an extended, patchy process rather than a sudden global ``phase transition'' \citep{shapiro1986cosmological,haardt1995radiative,gnedin1997reionization,madau1999radiative,gnedin2000effect}.

Reionization also affects galaxy formation by reducing gas inflows \citep[e.g.][]{Rees1986,Shapiro2004,Okamoto2008} and suppressing star formation in low-mass systems \citep{Okamoto2008,Wu2019a,gutcke2022lyra,Zier2025, ZierPopIII}. Recent observations, particularly based on Lyman-$\alpha$ transmission in high-redshift quasar spectra, favour a late and extended reionization history, with the volume-averaged neutral fraction falling below $\sim$5 per cent by $z\simeq 5.3$--5.8 \citep[e.g.][]{becker2015evidence,Kulkarni2019,zhu2021chasing,zhu2022long,bosman2022hydrogen,Jin2023,davies2024constraints,Becker2024}.

The reionization of singly ionized helium (\HeII; requiring $h\nu\ge 54.4\,\mathrm{eV}$) constitutes a second major phase transition of the IGM. The stellar populations that dominate the hydrogen-ionizing budget during the EoR typically have spectra that are too soft to drive efficient \HeII reionization on cosmological scales \citep[e.g.][]{mcquinn2016evolution}. Instead, \HeII reionization is commonly attributed to hard ultraviolet emission from accreting supermassive black holes (quasars), whose abundance peaks at later times. In addition to changing the ionization balance, helium reionization injects substantial heat into the low-density IGM \citep{hui1997equation,schaye2000thermal,mcquinn2009he,garzilli2012intergalactic} and leaves observable imprints on the IGM temperature--density relation and on the small-scale structure of the hydrogen Lyman-$\alpha$ forest \citep[e.g.][]{becker2011detection,boera2014thermal,gaikwad2021consistent}. Despite the difficulty of observing the \HeII Lyman-$\alpha$ forest, a consensus has emerged that helium reionization occurred mainly over $2.7\lesssim z\lesssim 4.5$ \citep[e.g.][]{kriss2001resolving,shull2010hst,becker2011detection,syphers2013hst,worseck2011end,worseck2016early,worseck2019evolution}.

Over the past decade, our empirical view of galaxies in the EoR has advanced rapidly. Even the deepest \textit{Hubble Space Telescope} (\textit{HST}) imaging could identify only $\mathcal{O}(10^{3})$ galaxy candidates at $z\simeq 6$--$8$, and only a handful at higher redshift, with limited spectroscopic confirmation \citep[e.g.][]{bouwens2015uv,livermore2017directly,atek2018extreme,oesch2018dearth}. The launch of the \textit{James Webb Space Telescope} (\textit{JWST}) has dramatically expanded this landscape, yielding large samples of high-redshift galaxy candidates. Early observations reported a surprisingly high abundance of UV-bright galaxies at $z\gtrsim 10$ \citep[e.g.][]{Naidu2022,Castellano2022,Finkelstein2022,Adams2023,Atek2023,bouwens2023evolution,Donnan2023,harikane2023comprehensive,Robertson2023,Yan2023,Hainline2024}, subsequently supported by spectroscopic confirmations extending to $z\sim 14$ \citep[e.g.][]{CurtisLake2023,Carniani2024,Roberts-Borsani2024,Harikane2024-spec,harikane2025jwst}.

This apparent UV luminosity function (UVLF) tension may arise from a combination of observational systematics and genuine astrophysical effects. Possibilities that have been considered include bursty star formation boosting bright-end counts through Eddington bias \citep{Mason2023,Mirocha2023,Shen2023,Sun2023}, a top-heavy initial mass function \citep{Inayoshi2022,Yung2023,Cueto2023,Trinca2024,Lu2024}, enhanced star formation in extreme high-redshift environments \citep{Dekel2023,Li2023,BK2025,Shen2025}, and modifications to the underlying cosmology \citep[e.g.][]{Para2023,Sabti2024,Shen2024b,Shen2025_EDE}. Related tensions include reports of extremely massive galaxy candidates across $z\sim 4$--$12$ \citep[e.g.][]{Labbe2023,Akins2023,Xiao2023,deGraaff2025,Casey2024,Glazebrook2024}, which raise the question of whether such systems can assemble their inferred stellar masses within the baryon budget available in $\Lambda$CDM \citep{BK2023,Lovell2023}; see, however, \citet{Cochrane2025}. Many candidates were subsequently identified as low-redshift interlopers or were shown to receive a potentially substantial contribution from active galactic nuclei (AGN), and the stellar-mass estimates remain sensitive to assumptions about the initial mass function, the star-formation history, and nebular emission \citep[e.g.][]{Endsley2022,Larson2022,Kocevski2023,Desprez2024,Narayanan2024,Wang2024a,Wang2024c,Turner2025}. Constraints from large-area \textit{JWST} surveys have reduced the tension for individual objects but still indicate number densities of massive galaxies that exceed most pre-\textit{JWST} model predictions \citep[e.g.][]{Weibel2024,WangT2024,Harvey2025}. \textit{JWST} has also uncovered a surprisingly large population of quenched galaxy candidates as early as $z\simeq 7$, hinting at an early assembly of stellar mass \citep[e.g.][]{Carnall2023a,Carnall2023b,Valentino2023,Weaver2023,Alberts2024,deGraaff2025,Weibel2025}.

Complementary constraints on the ISM of reionization-era galaxies come from the Atacama Large Millimeter/submillimeter Array (\alma), which probes the dust continuum and rest-frame far-infrared fine-structure lines such as [C\,\textsc{ii}] $158\,\mu$m and [O\,\textsc{iii}] $88\,\mu$m. \alma observations of bright, lensed galaxies reveal dynamically cold, rotating gaseous discs out to $z\gtrsim 7$ \citep[e.g.][]{Rizzo2021,Jones2021,Roman-Oliveira2023,Fujimoto2024,Rowland2024}, characterised by high rotation-to-dispersion ratios ($V_{\rm rot}/\sigma \gtrsim 10$) and prominent clumpy sub-structure \citep[e.g.][]{Fujimoto2024}. Evidence for ordered ionized-gas kinematics in individual high-redshift galaxies has also emerged from \textit{JWST} observations \citep[e.g.][]{Nelson2024,Xu2024}. Further constraints on the sources of reionization will come from the \textit{Nancy Grace Roman Space Telescope}, whose wide field of view is expected to deliver very large samples of $z>6$ galaxies and quasars, particularly at the bright end \citep{Yung2023_roman,Tee2023,2025arXiv250510574O}.

Inferring the state of the IGM itself is more challenging. The CMB provides an integral constraint on reionization through the Thomson optical depth $\tau_{\rm CMB}$ to free electrons \citep{planck2020}, but is only weakly sensitive to the detailed timing and topology of ionized regions. It has recently been emphasised that $\tau_{\rm CMB}$ can be degenerate with $\Omega_{\rm m}$ (and related late-time parameters) in certain CMB--large-scale structure (LSS) combinations \citep{sailer2025,Jhaveri2025}. In particular, combining CMB data with BAO constraints can shift the preferred $\tau_{\rm CMB}$ relative to analyses anchored by large-scale polarization measurements \citep[e.g.][]{sailer2025,Cain2025,Elbers2025}.

At $z\lesssim 8$, the IGM can be probed through Lyman-$\alpha$ absorption along sightlines to bright background sources, producing the Lyman-$\alpha$ forest at lower redshift and extended saturated absorption (the Gunn--Peterson trough; \citealt{gunn1965density}) at higher redshift. During the late stages of reionization, additional constraints come from damping-wing absorption and from observations of Lyman-break galaxies (LBGs) and Lyman-$\alpha$ emitters (LAEs). These measurements constrain the tail end of hydrogen reionization and the post-reionization fluctuations of the UV background \citep[e.g.][]{fan2006constraining,becker2015evidence,Kulkarni2019,zhu2021chasing,zhu2022long,bosman2022hydrogen,Jin2023}, as well as IGM heating \citep[e.g.][]{1999Madau_IGM,2022Burkhart_Lya,2024Montero-Camacho_IGM}, but they remain limited by the scarcity of bright background sources and by the one-dimensional nature of individual sightlines. Probing helium reionization through the \HeII Lyman-$\alpha$ forest is even more challenging: at $z\gtrsim 2$ the relevant absorption lies in the far-UV and is strongly attenuated by intervening \HI Lyman-limit systems \citep{2005pgqa.conf..484Z}. As a result, only a small number of quasar sightlines are usable, with many studies relying on fewer than ${\sim}25$ objects \citep[e.g.][]{worseck2016early,worseck2019evolution,makan2021new,makan2022he}.

A major goal for upcoming surveys is to move beyond sparse absorption constraints towards statistical and tomographic probes of the IGM in emission, as well as towards three-dimensional reconstructions of its ionization structure \citep[e.g.][]{Eilers2025}. In particular, redshifted 21\,cm radiation from the hyperfine transition of neutral hydrogen offers a direct, three-dimensional view of the neutral IGM during cosmic dawn and reionization \citep[e.g.][]{scott199021,madau199721,shaver1999can,furlanetto2004growth,loeb2004measuring,furlanetto2006cosmology,morales2010reionization,pritchard201221,Shimabukuro2023}. Interferometric experiments targeting the 21\,cm power spectrum (and, ultimately, imaging) are designed to measure the evolving characteristic scale and morphology of ionized regions, providing strong complementarity to galaxy surveys. Although no definitive detection has yet been reported, current experiments such as the Hydrogen Epoch of Reionization Array (HERA; \citealt{HERA_2017}) and the Low-Frequency Array (LOFAR; \citealt{LOFAR_2013}) already place stringent upper limits \citep[e.g.][]{mertens2020improved,abdurashidova2022hera,adams2023improved,ceccotti2025first,acharya2024revised}. In parallel, line-intensity mapping \citep[LIM;][]{Kovetz2019,Bernal2022} aims to measure aggregate emission from unresolved galaxies and from the IGM in lines such as [C\,\textsc{ii}], the CO rotational lines, Ly$\alpha$, and H$\alpha$, enabling efficient large-volume measurements and cross-correlations with 21\,cm and galaxy samples \citep[e.g.][]{roy2025investigating}. In the longer term, it may also be possible to probe helium reionization through the $^3$He$^{+}$ 3.5\,cm hyperfine transition \citep[e.g.][]{McQuinn2009_3he,khullar2020probing,basu2025probing}.

Interpreting current and future observations of the high-redshift Universe requires comparably realistic theoretical models. However, self-consistently modelling the EoR in cosmological simulations is extremely challenging because it requires: (\textit{i}) sufficient resolution to identify and model the ionizing sources and the escape of ionizing radiation from their host galaxies; (\textit{ii}) radiative transfer that captures the patchy propagation of ionization fronts, the associated photoheating, and the hydrodynamic response of the IGM; and (\textit{iii}) a simulation volume large enough to sample rare objects and reduce cosmic variance. The third requirement is particularly acute for helium reionization, which is driven by rare, highly biased quasars and therefore benefits from boxes of several hundred comoving Mpc (e.g.\ $L=186$ and $429\,\mathrm{cMpc}$ in \citealt{mcquinn2009he}). Similarly, although a robust hydrogen-reionization history emerges already at $L\gtrsim 100\,h^{-1}\,\mathrm{cMpc}$ \citep{iliev2014simulating}, convergence of large-scale 21\,cm statistics can require box lengths of at least $L\gtrsim 250\,\mathrm{cMpc}$ \citep{kaur2020minimum}.

On-the-fly radiation hydrodynamics (RHD) substantially increases the computational cost, owing to an additional time-step constraint and to the need to evolve non-equilibrium ionization and heating. An alternative is to use \emph{semi-numerical} or \emph{semi-analytic} models, which generate reionization fields from approximate source prescriptions and excursion-set-based ionization criteria \citep[e.g.][]{furlanetto2004growth,mesinger2011,Fialkov2020,Munoz2020}. More accurate---but still computationally efficient---approaches populate haloes in pure $N$-body simulations with idealised galaxy models and post-process them with radiative transfer \citep[e.g.][]{Ciardi2003,Iliev2007,McQuinn2007,mcquinn2009he}, or couple semi-analytic galaxy-formation models to an evolving, spatially inhomogeneous reionization field \citep[e.g.][]{Mutch2016,Seiler2019,Hutter2021}. These methods make it feasible to model large, representative volumes ($\sim 500\,\mathrm{cMpc}$) and to explore parameter space efficiently, but they typically introduce multiple tunable ingredients (e.g.\ source emissivities, escape fractions, and sub-grid prescriptions for sinks and clumping). A complementary strategy is to post-process hydrodynamical simulations with radiative transfer, which resolves the gas density field but neglects the dynamical back-reaction of reionization on galaxy formation; this remains a common state-of-the-art approach for \HeII reionization \citep[e.g.][]{puchwein2023sherwood,basu2024helium}.

To study galaxy formation and the role of galaxies as sources of hydrogen reionization, several notable simulation programmes focus on resolving the ISM either in small cosmological volumes or in zoom-in regions. Such simulations are valuable for constraining feedback and the escape fraction of ionizing photons, but their limited volumes make it difficult to obtain a converged global reionization history and robust statistics for rare, bright galaxies. Examples of small-box simulations include \textsc{SPHINX} \citep{SPHINX2018,SPHINX2022}, \textsc{Obelisk} \citep{Obelisk2021}, and \textsc{SPICE} \citep{bhagwat2023spice,Bhagwat2024}, while examples of zoom-ins include \textsc{Renaissance} \citep{xu2016galaxy}, \textsc{FIRE} \citep{Ma2020_FIREesc}, \thzoom\ \citep{kannan2025introducing}, and \textsc{Megatron} \citep{katz2025megatron}.

Despite the high computational cost, several large-volume ($L\sim 100\,\mathrm{cMpc}$) RHD simulations have begun to model the EoR. One prominent example is the Cosmic Reionization On Computers project \citep[CROC;][]{Gnedin2014design}, which uses the adaptive-mesh-refinement code \textsc{ART} \citep{Kravtsov1997ART} together with the optically thin variable Eddington tensor (OTVET) moment method for radiative transfer \citep{Gnedin2001}. While the initial calibration suite used smaller boxes, the largest production runs reach side lengths of $L=80\,h^{-1}\,\mathrm{cMpc}\approx 120\,\mathrm{cMpc}$ \citep[e.g.][]{croc120cMpc}. Another project is \textsc{CoDa} \citep{ocvirk2016cosmic,ocvirk2020cosmic,lewis2022short}, which uses the hybrid CPU--GPU code \textsc{RAMSES-CUDATON}, coupling \textsc{RAMSES} for gravity and hydrodynamics \citep{teyssier2002cosmological} to the \textsc{ATON} radiative-transfer module \citep{aubert2008radiative}. The latest iteration, \textsc{CoDa III}, evolves a periodic box of side length $L=94.4\,\mathrm{cMpc}$ on a uniform grid with $8192^{3}$ cells, corresponding to a spatial resolution of $\Delta x\simeq 11.5\,\mathrm{ckpc}$ \citep{lewis2022short}. This provides excellent uniform resolution in the IGM (and very high dark-matter mass resolution), but the fixed ${\sim}12\,\mathrm{ckpc}$ cell size is insufficient to resolve the internal properties of galaxies.

A major step towards fully self-consistent reionization modelling was the \thesan project \citep{Thesan1,ThesanAaron,ThesanEnrico,ThesanDR}, which coupled the \textsc{IllustrisTNG} galaxy-formation model \citep{Weinberger2017,Pillepich2018Model} to on-the-fly radiation hydrodynamics using the moving-mesh code \arepo\ \citep{springel2010pur,pakmor2016improving,weinberger2020arepo} and its moment-based radiative-transfer module \areport\ \citep{Kannan2019}. The flagship run, \thesanone, evolved $2\times 2100^{3}$ resolution elements in a box of side length $L_{\mathrm{box}}=95.5\,\mathrm{cMpc}$, corresponding to dark-matter and baryonic mass resolutions of $m_{\rm DM}\simeq 3.1\times 10^{6}\,\Msun$ and $m_{\rm b}\simeq 5.8\times 10^{5}\,\Msun$. Combined with the quasi-Lagrangian adaptivity of the moving-mesh approach, this enabled \thesan to resolve the galaxy population down to the atomic-cooling limit while simultaneously capturing a representative hydrogen-reionization topology on ${\sim}100\,\mathrm{cMpc}$ scales. The simulations were evolved to $z=5.5$, focusing on the EoR and the late stages of hydrogen reionization, but did not extend into the subsequent epoch of \HeII reionization and the post-reionization IGM. Moreover, while a ${\sim}100\,\mathrm{cMpc}$ volume captures many aspects of patchy hydrogen reionization, the largest-scale bubble statistics and long-wavelength fluctuations in the radiation field are inevitably affected by finite-volume effects \citep[e.g.][]{iliev2014simulating}; this becomes even more acute for \HeII reionization given the rarity and strong bias of its dominant sources.

\begin{figure*}
    \centering
    \includegraphics[width=0.92\linewidth]{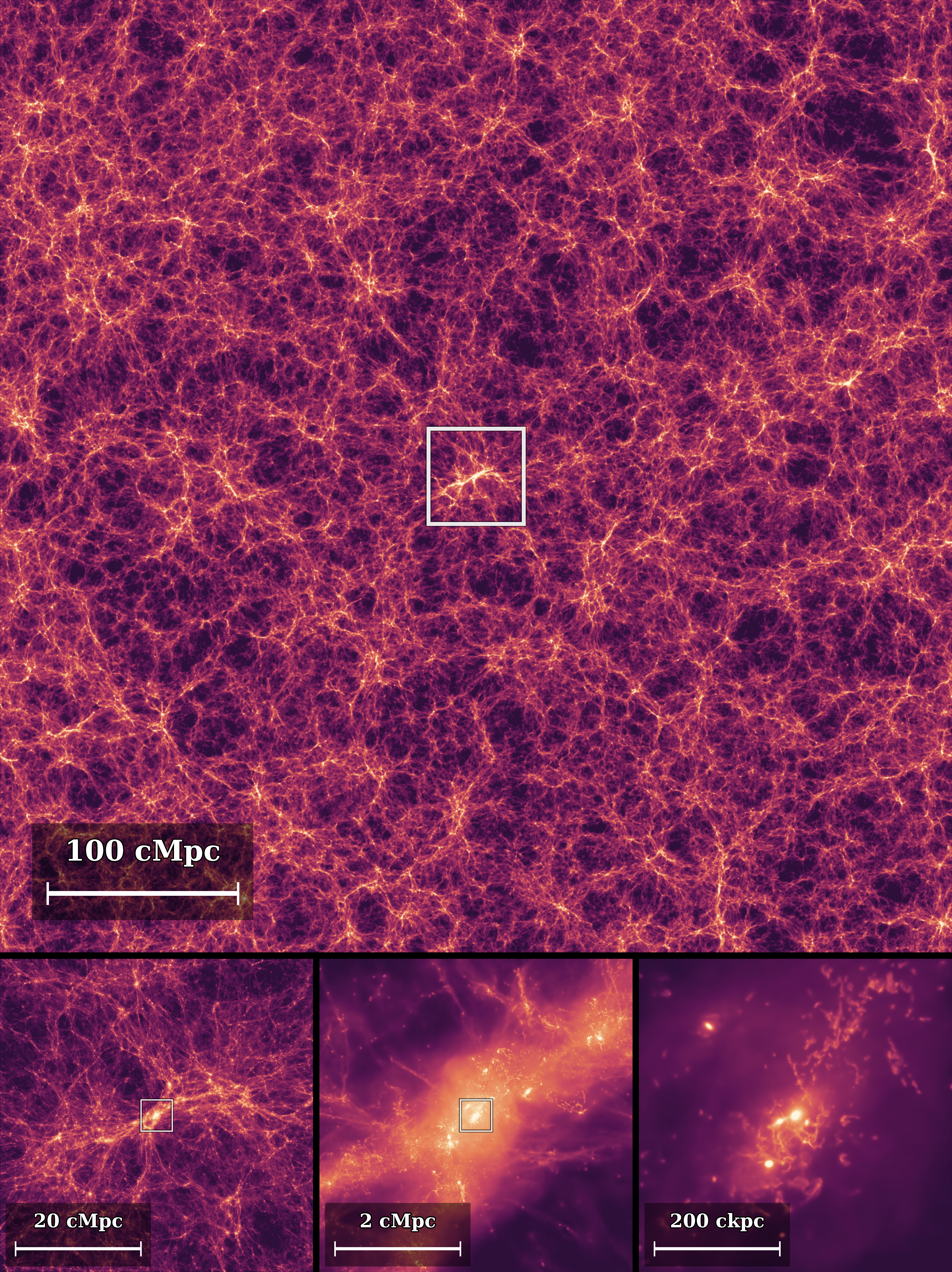}
\caption{Projected baryonic density (gas + stars) in a multi-scale zoom-in centred on the most massive halo in \lumina at $z=3$ ($M_{200{\rm c}} = 5.6 \times 10^{13}\,\mathrm{M_\odot}$). The four panels show regions of side length $500\,\mathrm{cMpc}$, $50\,\mathrm{cMpc}$, $5\,\mathrm{cMpc}$, and $500\,\mathrm{ckpc}$, respectively. The white square in each panel marks the region shown in the next zoom level. The figure illustrates the embedding of the halo in the cosmic web, its filamentary gas supply, and the compact stellar structure at its centre.}
    \label{fig:lumina_zoom_ladder}
\end{figure*}

\begin{figure*}
    \centering
    \includegraphics[width=\linewidth]{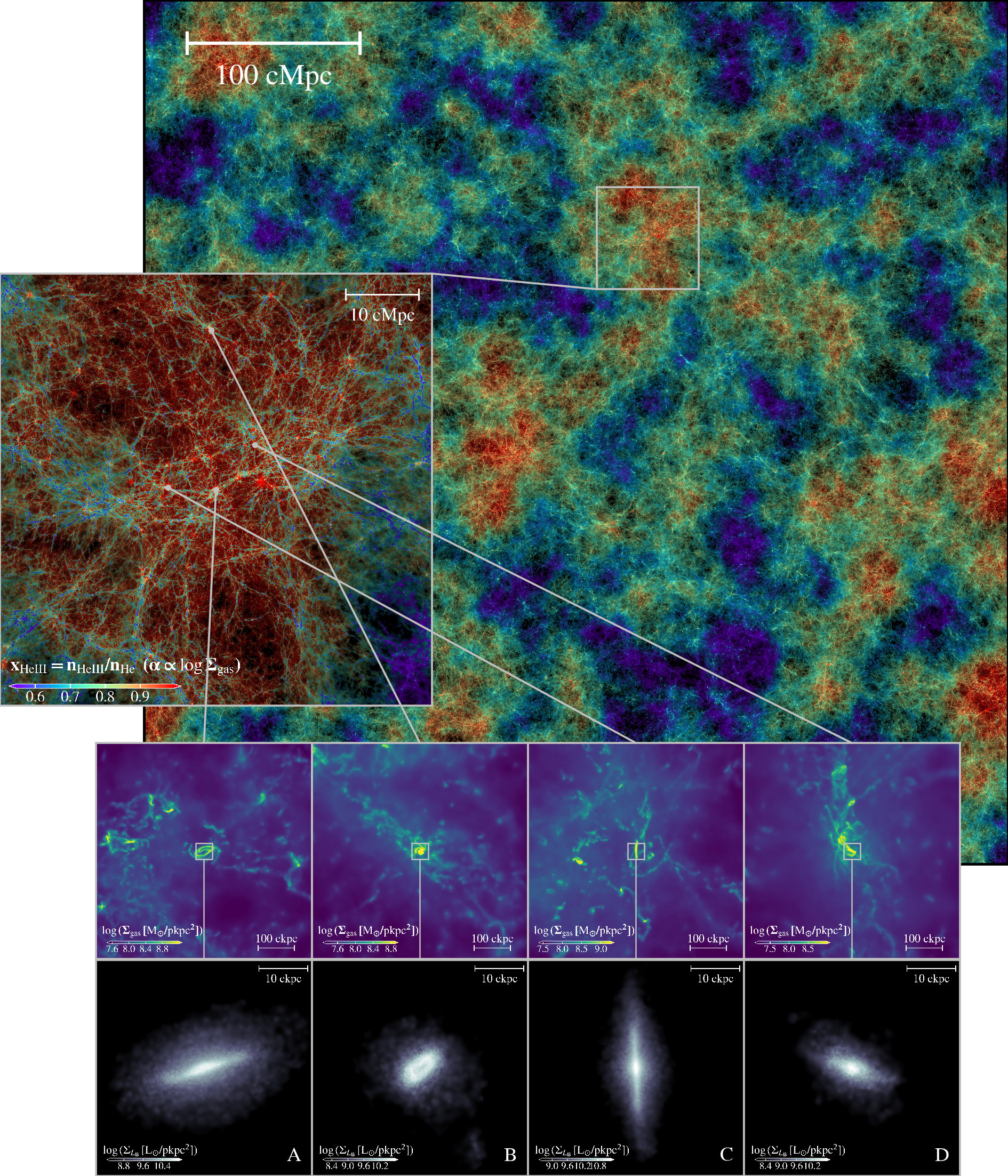}
    \caption{A multi-scale view of \lumina at $z\simeq 4$. The main panel shows the projected gas distribution in a slice of the simulation volume with thickness $150\,\mathrm{cMpc}$. It is color-coded by the mass-weighted \HeIII fraction, while transparency encodes the gas surface density. We first zoom in on a sub-field of $60\times 60\,\mathrm{cMpc}^2$, showing a large \HeIII bubble around a cluster of massive galaxies. Then we further zoom in on four selected galaxies, displaying their optical B-band light distribution along with the gas surface density field in the circumgalactic medium. Galaxies A and C exhibit extended disky structures, while galaxies B and D are more spheroidal and compact.}
    \label{fig:lumina_zoom_ladder2}
\end{figure*}

In this paper we introduce \lumina, a simulation that extends fully coupled radiation hydrodynamics into a new regime of volume and resolution, and that enables a unified treatment of galaxies, quasars, and the IGM from cosmic dawn to the end of \HeII reionization. \lumina evolves $2\times 6000^{3}$ resolution elements in a box of side length $L_{\mathrm{box}}=500\,\mathrm{cMpc}$ down to $z=3$, providing (\textit{i}) the dynamic range required to model the galaxy population that dominates hydrogen reionization and (\textit{ii}) the cosmological volume needed to sample the rare, bright quasars that drive \HeII reionization and produce very large ionized regions. The volume also captures the long-wavelength modes that can bias large-scale 21\,cm statistics in smaller boxes \citep{kaur2020minimum}. We additionally include coherent baryon--CDM streaming velocities in the initial conditions, enabling a self-consistent assessment of their impact on early star formation and on the ionizing emissivity, which we will present in a future paper together with complementary smaller-volume, higher-resolution simulations. This combination allows \lumina to connect reionization topology, the IGM thermal history, and source demographics within a single framework, and to forward-model observables spanning the EoR and the post-reionization Universe. To illustrate the dynamic range of \lumina, \cref{fig:lumina_zoom_ladder} shows a zoom-in on the most massive halo at $z=3$, while \cref{fig:lumina_zoom_ladder2} shows a zoom-in on four galaxies within a \HeIII bubble at $z=4$.

The paper is organised as follows.
\Cref{sec:methods} summarizes the numerical framework and the galaxy-formation model.
\Cref{sec:radTreatment} describes the radiative-transfer solver and its coupling to the galaxy-formation model.
\Cref{sec:changed_at_z_4p75} describes the changes to the model adopted at $z=4.75$ for simulating \HeII reionization.
\Cref{sec:GeneratingICs} presents the initial conditions and cosmological parameters.
\Cref{sec:TechnicalDetails} discusses technical aspects of the simulation and basic global diagnostics.
In \cref{sec:galaxies,sec:IGM} we present the evolution of the galaxy population and of the IGM across hydrogen and helium reionization.
\Cref{sec:dataProducts} describes the data products, and \cref{sec:summary} presents our conclusions.

\section{Simulation methods}
\label{sec:methods}
\lumina is run with the massively parallel \arepo\ code \citep{springel2010pur,pakmor2016improving,weinberger2020arepo}, which solves the Euler equations on an unstructured, moving Voronoi mesh with a finite-volume Godunov scheme, coupled to an efficient adaptive gravity solver. We additionally use the GPU-accelerated \areport\ module \citep{Kannan2019,zier2024adapting} to solve the radiative-transfer moment equations on the same mesh using the M1 closure.

Our codebase incorporates several key improvements from the MillenniumTNG (MTNG) project \citep{pakmor2023millenniumtng} that increase scalability and reduce the memory footprint. Specifically, the code uses MPI-3 shared memory so that data identical across MPI ranks are stored only once per node (e.g.\ domain-decomposition metadata, the top-level tree, and tables of stellar yields). Global MPI operations are split into a node-local step followed by an inter-node step, which improves efficiency and reduces network congestion. The original domain decomposition has been replaced by a hierarchical scheme that first subdivides the simulation volume among compute nodes and then further partitions each subdomain within a node. The gravity-tree algorithm has been restructured to front-load the import of external data, reducing wait times during force evaluations. We also adopt the {\small SUBFIND-HBT} algorithm \citep{Gadget4} in place of the original {\small SUBFIND} algorithm \citep{springel2001populating}, and we compute merger trees and matter power spectra on the fly.

\subsection{Hydrodynamics and self-gravity}
\arepo\ solves the Euler equations with a second-order, finite-volume Godunov scheme \citep{godunov1959finite} on a moving, unstructured Voronoi mesh. The mesh is defined by a set of mesh-generating points that move approximately with the local fluid velocity \citep{vogelsberger2012moving}. This yields quasi-Lagrangian behaviour, reduces advection errors relative to static grids, and preserves Galilean invariance---a key advantage for problems with large bulk flows. 
To maintain an approximately constant gas-mass resolution, cells are adaptively refined or derefined: cells more massive than $2\times$ the target mass $m_{\rm target}$ are split, whereas cells below $0.5\times$ the target mass are merged. In our setup, the quasi-Lagrangian character is primarily provided by mesh motion, while refinement and derefinement serve to keep cell masses within the adopted range around $m_{\rm target}$. The spatial resolution is therefore a direct function of the local gas density and varies by several orders of magnitude between cosmic voids and dense, star-forming gas (see \cref{sec:TechnicalDetails}).
For efficiency, cells are integrated with a second-order accurate Runge--Kutta scheme using local time steps organised in a power-of-two hierarchy \citep{pakmor2016improving}.
Gradients required for linear interpolation are calculated using local least-square fits \citep{pakmor2016improving}.

Following the MTNG setup \citep{pakmor2023millenniumtng}, we disable magnetic fields to reduce the memory footprint; at our target resolution we expect saturated fields only in the most massive clusters by the end of the run \citep{Pakmor2024,Tevlin2025}. Hydrodynamic fluxes are computed by default with the HLLC Riemann solver \citep{toro1994restoration}; in practice, we first attempt the HLLC solver and evaluate the interface pressure $p^\star$, falling back to the HLL solver \citep{harten1983upstream} if $p^\star < 0$ and, if the pressure remains negative, to the Rusanov solver \citep{rusanov1961calculation}, which guarantees a physically valid solution, following \citet{pakmor2011magnetohydrodynamics}.

Gravitational forces between gas cells, collisionless dark-matter, stellar, and black-hole particles are computed with an adaptive TreePM scheme \citep{Bagla2002,Bagla2003}: short-range forces are evaluated with an octree \citep{Barnes1986}, while long-range forces are evaluated with a Fourier-space particle--mesh method \citep{Aarseth2003}. We employ hierarchical time integration for gravity \citep{Gadget4}, which yields substantial speed-ups for deep time-step hierarchies. As discussed in \citet{Gadget4} for dark-matter-only simulations with hierarchical gravity, it is advantageous to move all particles from a given time bin into the next-lower one whenever that lower bin already contains at least one-third of the particles of the higher bin. This criterion, however, does not account for the additional computational cost introduced by baryonic physics. We therefore replace the factor of one-third with two-thirds. In practice this condition is never met during our simulation, which also helps to avoid gaps in the time-step hierarchy.

Collisionless particles use a fixed gravitational softening length, while gas cells use an adaptive softening of $\simeq 2.5$ times the effective cell radius, limited from below by a minimum softening length. For gas cells we associate each cell's mass with its mesh-generating point when calculating gravitational forces; this choice helps to preserve small-scale power in the initial conditions (see \cref{sec:GeneratingICs}).

\subsection{The IllustrisTNG subgrid model}
Our galaxy-formation model follows the well-tested \illustrisTNG model described by \citet{Weinberger2017} and \citet{Pillepich2018Model}, itself an updated version of the original Illustris framework \citep{Vogelsberger2013,IllustrisIntro,IllustrisNature}. A further discussion of the \illustrisTNG model can be found in \citet{burger2025applying}. Below we briefly summarize the relevant components and highlight the modifications adopted for \lumina.

\subsubsection{Gas treatment}
Most large-scale cosmological simulations do not resolve the multi-phase structure of the interstellar medium. In \illustrisTNG, cold gas cells with densities $n_{\rm H} \ge 0.106\,\mathrm{cm}^{-3} \equiv n_{\rm H,SF}$ are assumed to host a two-phase ISM consisting of cold clouds embedded in a hot ambient medium \citep{springel2003cosmological}. Within each such cell the two phases are maintained in thermodynamic equilibrium, with supernova heating balancing radiative cooling; this yields an effective equation of state (EoS). To soften this EoS, \illustrisTNG combines two pressures: $0.3$ of the \citet{springel2003cosmological} model and $0.7$ of an isothermal EoS at $T=10^{4}\,\mathrm{K}$ \citep{Springel2005_EOS}. The EoS acts only as a temperature floor for gas above the density threshold, so cells that are heated above the EoS are modelled as a single hot phase.

Gas below the star-formation density threshold and gas above the EoS are treated with primordial \citep{katz1995cosmological} and metal-line cooling in ionization equilibrium \citep{Wiersma2009,Vogelsberger2013}, assuming a spatially uniform, time-dependent metagalactic UV background \citep{FG09}. As discussed in \cref{subsec:ConnectingTNGModel}, in \lumina we replace the equilibrium primordial-cooling routines for low-density gas with a non-equilibrium chemical network, while retaining the equilibrium treatment for metal-line cooling.

\subsubsection{Star formation and stellar feedback}
In the \illustrisTNG model, all gas on the effective EoS is eligible to form stars with a density-dependent gas depletion time,
\begin{equation}
    t_\star(n_{\rm H}) = t_0^\star \left( \frac{n_{\rm H}}{n_{\rm H, SF}} \right)^{-1/2},
    \label{eq:depletionTime}
\end{equation}
where the normalisation $t_0^\star = 2.2\,\mathrm{Gyr}$ is chosen to reproduce the observed Kennicutt--Schmidt relation \citep{Kennicutt1998}. The scaling $\propto n_{\rm H}^{-1/2}$ corresponds to a constant star-formation efficiency per free-fall time.

At densities above $n_{\rm H,crit}\simeq 24.35\,\mathrm{cm}^{-3}$, the effective EoS in \illustrisTNG can become softer than the critical slope of $4/3$, implying a Jeans mass that decreases with increasing density. This promotes runaway collapse in dense gas and forces extremely small hydrodynamic time steps. The problem is more severe in simulations that include magnetic fields and was first encountered in TNG50 \citep{Pillepich2019,nelson2019first}. To mitigate it, TNG50 adopted a shortened depletion time above the threshold density $n_{\rm H,crit}$,
\begin{equation}
    t_\star(n_{\rm H}) = t_0^\star \left( \frac{n_{\rm H, crit}}{n_{\rm H, SF}} \right)^{-1/2}
    \left( \frac{n_{\rm H}}{n_{\rm H, crit}} \right)^{-1}.
    \label{eq:fasterDepletionTime}
\end{equation}
\citet{nelson2019first} found no obvious impact of this modification on the star-formation rates or morphologies of lower-mass galaxies, while \citet{nelson2024introducing} recently showed in the TNG-Cluster simulations that it can suppress black-hole growth in massive clusters, leading to weaker AGN feedback and higher stellar masses.

In \lumina we adopt the rapid star-formation prescription of \cref{eq:fasterDepletionTime} for $z>4.75$. This reduces the maximum density reached in the simulation and therefore increases the minimum time step. For $z\leq 4.75$, once the most massive black holes (with masses of a few $\times 10^{8}\,\Msun$) have exited their near-Eddington growth phase, we revert to the fiducial depletion time of \cref{eq:depletionTime}, so that black-hole growth during \HeII reionization remains consistent with the calibrated \textsc{IllustrisTNG} model. \HeII reionization is dominated by the most luminous quasars and is therefore sensitive to the upper end of the black-hole mass function.

Star formation is implemented stochastically, spawning stellar particles with masses up to $2\,m_{\rm target}$. Because the cold and hot ISM phases are tightly coupled within each cell, the model cannot self-consistently launch galactic winds. Instead, it uses an \emph{ad hoc} kinetic wind scheme in which randomly selected star-forming cells are converted into wind particles that briefly decouple from the hydrodynamics, leave the dense ISM, and re-couple to the Voronoi mesh at low density. The mass and energy loading factors of the wind are set directly by the model and are described in detail by \citet{Pillepich2018Model}.

Star particles in \illustrisTNG evolve with time and inject metals into the surrounding gas according to the yield tables of \citet{Pillepich2018Model}. Unlike the original \illustrisTNG implementation, we track only the total metallicity rather than individual elemental abundances (as in MTNG), which substantially reduces memory usage, disk space, and the number of gradient calculations required. This choice has no impact at run time because the cooling and yield tables are scaled by the total metallicity.

For $z>4.75$ we distribute newly produced metals from each star particle into the 64 nearest gas cells using an SPH-kernel weighting. For $z\leq 4.75$ we instead deposit all metals into the single nearest gas cell, thereby avoiding the expensive iterative neighbour search \citep{van2020neutron}. We have verified that this change does not affect the metallicity gradients of resolved galaxies.

\subsubsection{Black hole treatment}
The black-hole (BH) model follows \citet{Weinberger2017}. BHs are seeded in haloes identified by a friends-of-friends (FoF) algorithm whenever the halo mass exceeds $5\times10^{10}\,h^{-1}\,\Msun$ and no BH is yet present. The gas cell with the lowest gravitational potential is then converted into a BH particle with an initial mass of $M_{\rm BH}=8\times10^{5}\,h^{-1}\,\Msun$.

BHs grow by accreting mass from nearby gas cells according to the Bondi--Hoyle prescription \citep{Bondi1944,Bondi1952}:
\begin{equation}
\dot{M}_{\rm Bondi} = \frac{4\pi G^2 M_{\rm BH}^2 \rho}{c_\text{s}^3}\,,
\label{eq:bondiRate}
\end{equation}
where $G$ is the gravitational constant, $M_{\rm BH}$ is the BH mass, $\rho$ is the ambient gas density, and $c_\text{s}$ is the local sound speed. Ambient quantities are estimated with an SPH-like kernel over the 128 nearest gas cells. To prevent unphysically high Bondi accretion rates in low-density environments---which can arise when the smoothing length grows excessively large---we reduce the accretion rate in the high-accretion (``quasar'') mode introduced below by a factor of
$
\left( P_{\rm ext} / P_{\rm ref} \right)^{2}
$
whenever the kernel-weighted external pressure $P_{\rm ext}$ falls below the reference pressure $P_{\rm ref}$ defined in equation~(23) of \citet{Vogelsberger2013}. We note that $P_{\rm ext}$ in the original \illustrisTNG simulations also includes the magnetic pressure, which can lead to higher accretion rates in low-density, magnetically dominated regions.

Accretion is capped at the Eddington limit,
\begin{equation}
\dot{M}_{\rm Edd} = \frac{4\pi G M_{\rm BH} m_p}{\epsilon_{\rm rad}\,\sigma_{\rm T}\,c}\,,
\label{eq:eddingtonRate}
\end{equation}
where $m_p$ is the proton mass, $c$ is the speed of light, $\sigma_{\rm T}$ is the Thomson cross-section, and $\epsilon_{\rm rad}=0.2$ is the radiative efficiency. 
The value for $\epsilon_{\rm rad}$ is adopted as the fiducial one in the calibrated TNG model, which was chosen to reproduce a realistic black-hole population at $z=0$ \citep{Weinberger2017}.
The mass accretion rate is therefore
\begin{equation}
\dot{M}_{\rm BH} = \min\!\left(\dot{M}_{\rm Bondi},\,\dot{M}_{\rm Edd}\right)\,.
\end{equation}
BHs also grow through mergers with nearby BHs. To mimic unresolved dynamical friction, BHs are repositioned at each globally synchronised time step to the local minimum of the gravitational potential. Frequent repositioning is essential to ensure that BHs remain near the centres of their host galaxies, where they can efficiently accrete gas and avoid premature growth suppression. To this end we impose a maximum global time step of $\Delta\ln a = 0.005$ for $z>4.75$, $\Delta\ln a = 0.0025$ for $3.72<z<4.75$, and $\Delta\ln a = 0.00125$ for $z<3.72$. In future simulations we plan to extend BH repositioning to non-global time steps as well.

The model has two feedback modes. At high accretion rates (the ``quasar'' mode), the feedback is purely thermal,
\begin{equation}
\dot{E}_{\rm therm} = \epsilon_{f,{\rm high}}\;\epsilon_{\rm rad}\,\dot{M}_{\rm BH}\,c^2\,,
\label{eq:etherm}
\end{equation}
with efficiency $\epsilon_{f,{\rm high}}=0.1$. At low accretion rates (the ``radio'' mode), the feedback energy is injected in pulsed, kinetic form,
\begin{equation}
\dot{E}_{\rm kin} = \epsilon_{f,{\rm kin}}\;\dot{M}_{\rm BH}\,c^2\,,
\end{equation}
where the efficiency is
\begin{equation}
\epsilon_{f,{\rm kin}} = \min\!\left(\frac{n_{\rm H}}{0.05\,n_{\rm H, SF}},\,0.2\right)\,.
\end{equation}
The transition between modes depends on the Eddington ratio, with the switch happening when
\begin{equation}
\frac{\dot{M}_{\rm BH}}{\dot{M}_{\rm Edd}} = \chi = \min\!\left[\,0.002\left(\frac{M_{\rm BH}}{10^8\,\Msun}\right)^{\!2},\,0.1\,\right].
\end{equation}
The mass dependence allows higher-mass BHs to enter the more efficient radio mode at larger Eddington ratios---and therefore earlier.

\subsection{Structure finding}
\label{subsec:structureFinding}
We identify dark-matter haloes and their substructures in two stages. First, we run a friends-of-friends (FoF) algorithm \citep{Davis1985} on the dark-matter particles with a linking length of $0.2$ times the mean inter-particle separation. Gas, stellar, and black-hole particles are then associated with the FoF group of their nearest dark-matter particle, provided they lie within one linking length; otherwise they remain unassigned. FoF groups with fewer than 32 particles are discarded.

Second, each FoF group is processed with the \textsc{SUBFIND-HBT} algorithm \citep{springel2001populating,Gadget4}, which iteratively identifies gravitationally bound substructures (subhaloes). \textsc{SUBFIND-HBT} extends the original \textsc{SUBFIND} by carrying subhalo membership information across snapshots, which improves the tracking of infalling systems through mergers \citep{hbt2012}. Subhaloes with fewer than 20 bound particles are removed. In addition, we post-process the full snapshots with the original \textsc{SUBFIND} algorithm (see \cref{subsec:haloCatalogue}).

Following common practice, we associate galaxies with subhaloes, identifying the most massive subhalo as the central galaxy and classifying the remaining subhaloes as satellites. At high redshift this classification is not ideal, because FoF groups often contain several physically distinct haloes. Since FoF masses can also be inflated by percolation through ``particle bridges'', we additionally compute spherical-overdensity halo properties. Specifically, we define $R_{200{\rm c}}$ as the radius enclosing a mean density of $200$ times the critical density $\bar{\rho}_{\rm c}(z) = 3H^{2}(z)/(8\pi G)$, and $M_{200{\rm c}}$ as the mass enclosed within this radius. Here $H(z)$ is the Hubble parameter; unless stated otherwise, we adopt $M_{200{\rm c}}$ as the halo mass throughout the paper.

\section{Radiation treatment}
\label{sec:radTreatment}
In the following we describe, in more detail, the GPU-accelerated radiative-transfer solver (\cref{subsec:RTTransport}); the radiation sources in \lumina (\cref{subsec:RTSources}); the coupling to the \illustrisTNG galaxy-formation model (\cref{subsec:ConnectingTNGModel}); the discretization of the radiation field into frequency bins (\cref{subsec:discretizationRadiation}); and the differences relative to the \thesan project (\cref{subsec:diffThesan}).

\subsection{GPU-accelerated transport with the M1 closure}
\label{subsec:RTTransport}
The radiation field is fully specified by the specific intensity
\(I_\nu(\bm{x},t,\bm{n},\nu)\), defined such that the energy \( \mathrm{d}E_\nu \)
crossing an oriented area element \(\mathrm{d}\bm{A}\) at position \(\bm{x}\) and time \(t\),
within frequency interval \([\nu,\nu+\mathrm{d}\nu]\) and solid angle \(\mathrm{d}\Omega\) about
direction \(\bm{n}\), in time \(\mathrm{d}t\), is
\[
\mathrm{d}E_\nu = I_\nu(\bm{x},t,\bm{n},\nu)\, (\bm{n} \cdot \mathrm{d}\bm{A})\, \mathrm{d}t\, \mathrm{d}\nu\, \mathrm{d}\Omega \, .
\]

In an expanding universe the comoving specific intensity obeys the radiative-transfer equation \citep{mihalas1999foundations}
\begin{equation}
\frac{1}{c}\frac{\partial I_\nu}{\partial t}
+\frac{1}{a}\,\bm{n}\!\cdot\!\bm{\nabla} I_\nu
-\frac{H}{c}\!\left(\nu\frac{\partial I_\nu}{\partial \nu}-3 I_\nu\right)
= j_\nu - \kappa_\nu \rho\, I_\nu \, ,
\label{eq:RT}
\end{equation}
where \(a(t)\) is the scale factor, \(H=\dot a/a\), \(j_\nu\) is the emissivity, and
\(\kappa_\nu\rho\) is the extinction (absorption plus out-scattering) coefficient.
The third term on the left accounts for cosmological redshifting and dilution.
We do not include cosmological redshifting and instead assume that photons are absorbed close to their source before being significantly redshifted. Only the hardest photons in our X-ray band ($h\nu\gtrsim 1.5\,\mathrm{keV}$) can traverse a sizeable fraction of a Hubble length before being absorbed \citep{furlanetto2006cosmology}, so neglecting their redshifting may slightly overestimate the associated heating. The dilution term disappears when we switch to the comoving intensity $\tilde{I}_\nu = a^{3} I_\nu$. Solving \cref{eq:RT} directly is prohibitively expensive because of its high dimensionality, and ray-tracing methods scale poorly with the number of sources, which is large in cosmological simulations. We therefore adopt a moment-based scheme.
Defining frequency-integrated comoving moments,
\begin{equation}
\{\tilde c\,E_r,\ \bm{F}_r,\ \tilde c\,\mathbb{P}_r\}
= \int_{\nu_1}^{\nu_2}\!\int_{4\pi}\{1,\ \bm{n},\ \bm{n}\otimes\bm{n}\}\,
\tilde{I}_\nu\, \mathrm{d}\Omega\, \mathrm{d}\nu \, ,
\end{equation}
the equations for each radiation bin read
\begin{equation}
\frac{\partial E_r}{\partial t}
+\frac{1}{a}\,\bm{\nabla}\!\cdot\!\bm{F}_r
= S - \kappa_{\rm E}\,\rho\,\tilde c\,E_r \, ,
\label{eq:E}
\end{equation}
\begin{equation}
\frac{\partial \bm{F}_r}{\partial t}
+\frac{\tilde c^{\,2}}{a}\,\bm{\nabla}\!\cdot\!\mathbb{P}_r
= - \kappa_{\rm F}\,\rho\,\tilde c\,\bm{F}_r \, .
\label{eq:F}
\end{equation}
Here $S$ is the radiative energy source term, $\kappa_{\rm E}$ and $\kappa_{\rm F}$ are the energy- and flux-weighted mean opacities over $[\nu_1,\nu_2]$, and $\tilde{c}$ is the signal speed of the radiation. The signal speed equals $c$ physically, but as discussed below it is sometimes approximated by a smaller value to reduce the numerical cost. We close the system with the Eddington relation
\begin{equation}
\mathbb{P}_r = E_r\,\mathbb{D} \, ,
\label{eq:edd}
\end{equation}
where the Eddington tensor $\mathbb{D}$ is approximated with the M1 closure \citep{Levermore1984,Dubroca1999} as a function of the radiation energy $E_r$ and the radiation flux $\bm{F}_r$. The transport step (for the four conserved variables $E_r$ and the components of $\bm{F}_r$) is hyperbolic. For time integration we operator-split the transport step from the local source/sink terms.

We solve \cref{eq:E,eq:F} with a finite-volume scheme on the moving Voronoi mesh, following \citet{Kannan2019} and \citet{zier2024adapting}. Second-order spatial accuracy is obtained by linear reconstruction with gradients from local least-squares fits \citep{pakmor2016improving}. We do not apply a time extrapolation in the second Runge--Kutta stage; this reduces the computational cost by lowering the number of gradient evaluations required, while having only a minor impact on accuracy. As discussed in \cref{subsec:discretizationRadiation}, the radiation field is discretised into multiple frequency bins, yielding four independent conservation laws per bin. Unlike \citet{Kannan2019}, we store the radiation energy directly rather than the photon number. The two formulations are mathematically equivalent, but storing the energy reduces the magnitude of the numerical values involved and thus the risk of floating-point overflow; in \thesan the same problem was mitigated by normalising the photon number by $10^{63}$.

Absorption is coupled to a non-equilibrium thermo-chemical network for primordial gas, in which the photoionization and photoheating rates are modified by the local radiation field. The full chemical network, including all cooling processes, is described in \citet{Kannan2019}. We use a semi-implicit integrator with adaptive subcycling and enforce a 10\% per-step limit on the change of the specific thermal energy and of each ionization fraction, as described in \citet{zier2024adapting}. The resulting ion abundances are advected with the local flow. Metal-line cooling is not coupled to the radiation field; we use the tables of \citet{Vogelsberger2013}, which scale linearly with the gas metallicity and are computed with \textsc{CLOUDY} \citep{Gunasekera2025}.

Because the explicit time-step constraint scales inversely with the speed of light, we adopt the reduced-speed-of-light approximation of \citet{Gnedin2001} and set $\tilde c = 0.2\,c$ in the radiation solver. This can bias the late-time propagation of ionization fronts and the post-overlap neutral fraction, particularly in underdense regions \citep{Deparis2019,Ocvirk2019}. We nevertheless adopt this value because a full-speed-of-light treatment would be computationally prohibitive given the large dynamic range in spatial and temporal scales of the simulation, and because previous tests indicate that $\tilde c\sim 0.2\,c$ reproduces the global reionization history in similar setups \citep{Thesan1}. Radiative transfer and chemistry are subcycled with factors of 64 during hydrogen reionization and 256 at $z<4.75$. Since $\tilde c$ and the number of subcycles are fixed globally, we enforce a maximum radiation Courant number of $C_{\rm RT}=0.3$ by reducing the hydrodynamical time step of a cell whenever this limit would otherwise be exceeded.

During the RT subcycles the Voronoi mesh is held fixed, which enables efficient GPU acceleration. We use a further-optimised variant of the GPU solver of \citet{zier2024adapting}, rewritten in AMD HIP for this project. On Frontier, the machine used for this work, the new version achieves a ${\sim}20\times$ speed-up relative to the original CPU-based solver of \citet{Kannan2019}.

In \citet{zier2024adapting} we also introduced an all-to-all communication scheme that aggregates all task-to-task messages destined for the same remote node into a single node-to-node message. This substantially reduces the number of messages (and hence network congestion) and improves scalability, at the cost of increased shared-memory usage. In the version of \arepo\ used in \lumina, the memory available to each task must be partitioned into private and shared segments at the start of a run, so it is advantageous to keep the shared-memory footprint roughly constant across runs. For this reason we do not enable the node-aggregated scheme here and instead use a traditional task-to-task scheme that maximises the available private memory per task. In an upcoming version of \arepo\ the distinction between private and shared memory will be removed, which will make the node-to-node scheme significantly more efficient for some of the physics modules.

\subsection{Radiation sources}
\label{subsec:RTSources}
\begin{figure}
    \centering
    \includegraphics[width=1\linewidth]{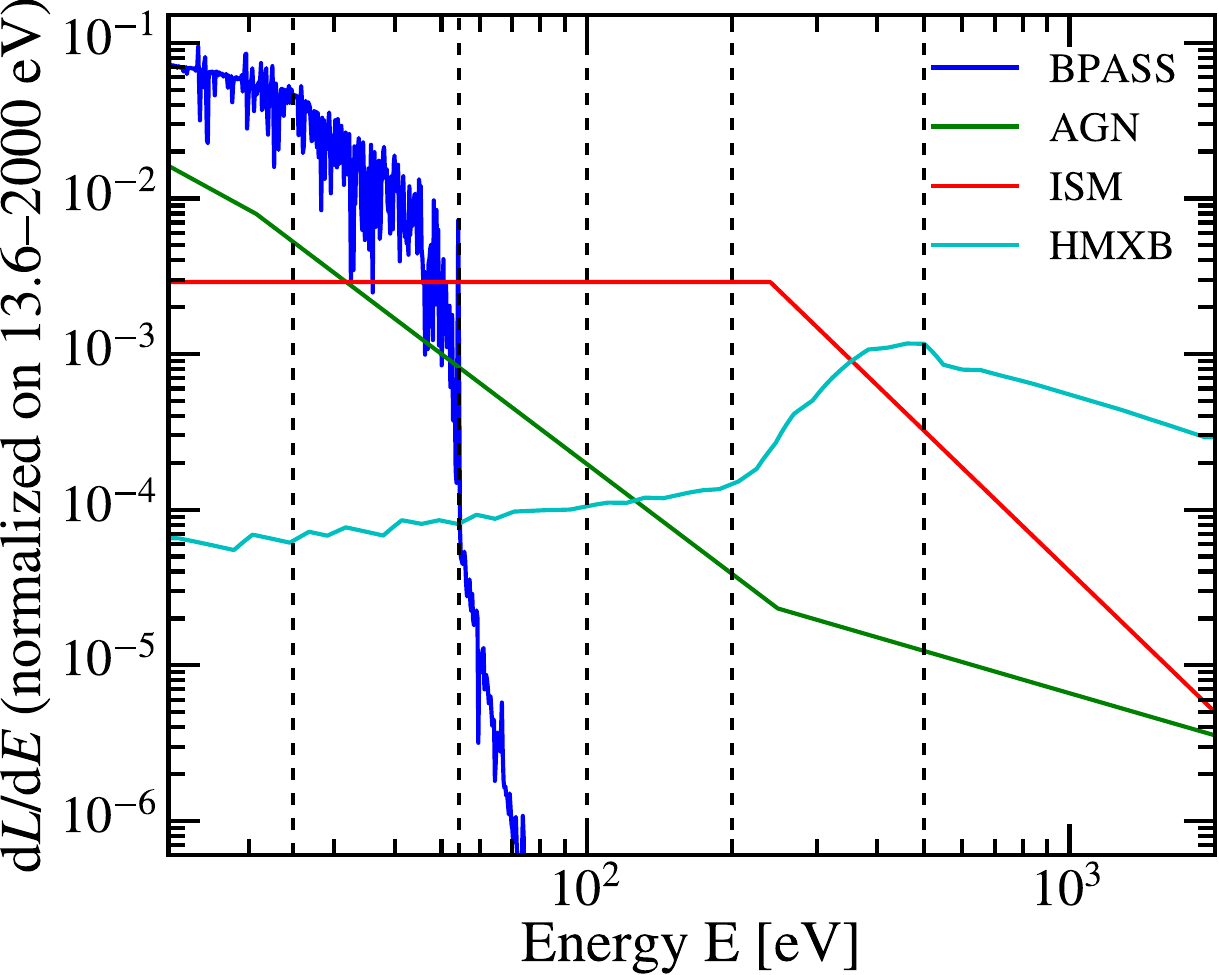}
    \caption{Spectral energy distributions (SEDs) for the four radiation sources used in this work: a $1\,\mathrm{Myr}$-old, $0.25\,\Zsun$ stellar population from BPASS v2.2.1 (blue); an AGN from \protect\citet{Shen2020} (green); a shock-heated ISM component described by Eq.~\ref{eq:shockHeated} with effective temperature $k_{\rm B} T = 240\,\mathrm{eV}$ (red); and an HMXB at $z=10$ from \protect\citet{fragos2013energy} (cyan). All SEDs are normalised over the simulated energy band $13.6\,\mathrm{eV}$--$2\,\mathrm{keV}$. The vertical dashed lines mark the six energy bins used for $z>4.75$. For $z<4.75$, the four highest-energy bins are merged into a single bin, and radiative transfer in the two lowest-energy bins is replaced by a uniform UV background.}
    \label{fig:sed}
\end{figure}

\lumina includes three classes of radiation sources: (\textit{i}) stellar populations, which are the primary drivers of hydrogen reionization; (\textit{ii}) accreting black holes, which power \HeII reionization \citep{mcquinn2009he}; and (\textit{iii}) high-mass X-ray binaries \citep{Madau2017} and the shock-heated interstellar medium \citep{2014Pacucci}, which pre-heat the IGM before hydrogen reionization \citep{Pritchard2007,ma2018,2018Eide,2020Eide}. Representative spectra of all four source classes are shown in \cref{fig:sed}: stars have the softest spectrum, followed by AGN, the shock-heated ISM, and high-mass X-ray binaries.

\subsubsection{Stellar radiation}
The ionizing radiation output of a stellar population depends sensitively on its age and metallicity. We adopt the \textsc{BPASS} v2.2.1 binary population synthesis models \citep{BPASS2017,Stanway2018}, assuming a Chabrier IMF \citep{Chabrier2003} with an upper-mass cutoff of $100\,\Msun$ and including binary evolution. Whenever a star particle is active, we compute the number of photons emitted since its previous activation from pre-computed cumulative yields tabulated as a function of stellar age and metallicity, and scaled by the particle mass. The resulting photons are deposited into the 64 nearest gas cells using an SPH kernel.

In contrast to \thesan, which approximated the emission during a step as the instantaneous luminosity times the time-step length, we guarantee exact photon-number injections even for coarse time steps by tabulating the cumulative photon yields and injecting their differences. As discussed in the next section, photons are not absorbed inside cells lying on the star-forming equation of state. To account for unresolved ISM absorption we therefore apply a constant stellar escape fraction $f_{\mathrm{esc},\star}=0.18$, i.e.\ we inject $f_{\mathrm{esc},\star}$ times the raw photon number. $f_{\mathrm{esc},\star}$ has been calibrated in boxes of side length $100\,\mathrm{cMpc}$ to yield a realistic end to hydrogen reionization. We further assume that the escape fraction is independent of photon frequency.

Because stellar spectra are substantially softer than those of accreting black holes and contribute negligibly to \HeII reionization, we disable stellar emission for $z<4.75$. In a $100\,\mathrm{cMpc}$ test box this reduces the computational cost down to $z=3.5$ by about $10\%$, with even larger savings expected at lower redshift owing to the growing number of star particles. Stellar radiation below $54.42\,\mathrm{eV}$ is still accounted for indirectly through the UVB adopted for these energies at $z<4.75$.

\subsubsection{Black holes}
To model AGN spectra, we replace the \citet{Lusso2015} SED used in \thesan with the composite quasar SED of \citet{Shen2020}, which explicitly covers the X‑ray regime required here. The far-UV ($\lambda > 912$\,\AA), optical, and IR portions of the SED are based on the broad-line AGN template of \citet{Krawczyk2013}. The extreme-UV ($\lambda < 912$\,\AA) SED follows the power law of \citet{Lusso2015} and is extended to $\lambda = 600$\,\AA. The X-ray SED is modelled as a cut-off power law, as described in \citet{Shen2020}, with a normalisation fixed by the optical-to-X-ray spectral slope $\alpha_{\rm ox}$. For simplicity we assume a constant value $\alpha_{\rm ox}\simeq -1.5$, which is typical of low-redshift AGN \citep{Steffen2006,Lusso2010}. We connect the extreme-UV and X-ray SEDs through a simple power law (between $\lambda = 600$\,\AA\ and $\lambda = 50$\,\AA) owing to the lack of direct observational constraints in this range. In practice, the combined uncertainty associated with the intrinsic SED, the unresolved nuclear region, and the local environment in setting the emergent \HeII-ionizing output is effectively absorbed into a luminosity-dependent obscuration correction. We therefore do not apply any additional subgrid, source-level escape fraction.

For each active black hole we compute the intrinsic bolometric luminosity $L_{\rm bol}^{\rm AGN} = \epsilon_{\rm rad}\,(1-\epsilon_{f,{\rm high}})\,\dot{M}_{\rm BH}\,c^{2}$, in which we have already subtracted the energy allocated to AGN feedback in the quasar mode (Eq.~\ref{eq:etherm}). To account for obscuration by unresolved gas and dust we apply a luminosity-dependent correction, motivated by the functional form of \citet{hopkins2007observational},
\begin{equation}
\frac{L_{\rm bol}^{\rm AGN, obs}}{L_{\rm bol}^{\rm AGN}} = \omega_{1}\left(\frac{L_{\rm bol}^{\rm AGN}}{10^{46}\,\rm erg~s^{-1}}\right)^{\omega_{2}} \, ,
\end{equation}
where the parameters $\omega_{1}=0.3$ and $\omega_{2}=0.07$ are taken from \citet{Vogelsberger2013}. We integrate the SED described above over our radiation bins to obtain the fractional luminosity injected into each bin, and we deposit the radiation into the 64 nearest gas cells using a standard SPH-kernel weighting.

Following the Illustris/IllustrisTNG prescription, we suppress radiation from BHs accreting at Eddington ratios $\lambda_{\rm Edd}<2\times 10^{-3}$ (i.e.\ less than $0.2\%$ of the Eddington rate), where we assume an advection-dominated accretion flow.

\subsubsection{Additional X-ray sources}
\label{subsubsec:xraySource}
We include two additional X-ray sources: thermal bremsstrahlung emission from the shock-heated ISM and radiation from high-mass X-ray binaries (HMXBs).  
For the ISM component, we adopt a thermal bremsstrahlung spectral energy distribution (SED) given by
\begin{equation}
\frac{\mathrm{d}L_{\mathrm{ISM}}(e)}{\mathrm{d}e} =
\begin{cases}
1 & \quad e < k_\text{B} T_{\mathrm{ISM}}\\[4pt]
\left(\dfrac{e}{k_\text{B} T_{\mathrm{ISM}}}\right)^{-3} & \quad e \ge k_\text{B} T_{\mathrm{ISM}}
\end{cases}
\label{eq:shockHeated}
\end{equation}
with a characteristic temperature of \( k_\text{B} T_{\mathrm{ISM}} = 240\,\mathrm{eV} \) \citep{2012bMineo-HotISM, 2014Pacucci, 2018Eide}.  
The normalization is
\begin{equation}
L_{\mathrm{ISM}}^{13.6\text{--}2000\,\mathrm{eV}}\,[\mathrm{erg}\,\mathrm{s}^{-1}]
= 3.3 \times 10^{40}\,\mathrm{SFR}\,[\Msun\,\mathrm{yr}^{-1}] \, ,
\end{equation}
which depends on the local star formation rate and is obtained by rescaling the estimates of \citet{2012bMineo-HotISM} for \( L_{\mathrm{ISM}}^{0.3\text{--}10\,\mathrm{keV}} \) to our photon energy range using our adopted SED (Eq.~\ref{eq:shockHeated}).  
We use the instantaneous star formation rate of individual gas cells to inject the corresponding photon emission.

Another important X-ray source population is high-mass X-ray binaries.  
We adopt the intrinsic SEDs from \citet{fragos2013energy, fragos2016erratum} at \( z \approx 10 \), which show only a weak dependence on redshift.  
These SEDs were originally obtained from large-scale population synthesis simulations \citep{fragos2013x} based on the \textsc{StarTrack} code \citep{belczynski2008compact}.  
The total luminosity depends on both metallicity and the local star formation rate, and we employ the parametrization from \citet{Madau2017}.  
We convert their reported values for \( L_{\mathrm{HMXB}}^{2\text{--}10\,\mathrm{keV}} \) (their Table~1) to \( L_{\mathrm{HMXB}}^{13.6\text{--}2000\,\mathrm{eV}} \) using our assumed SED.  
As for the shock-heated ISM, we use the local gas metallicity and star formation rate of the Voronoi cells to inject the radiation from HMXBs.

The injection rates of these X-ray sources are proportional to the star-formation rate density, which is not converged at high redshift. We also do not model Pop~III star formation, which further reduces the predictive power of our model at the highest redshifts. In addition, our implementation neglects secondary ionizations and instead deposits any remaining electron energy as heat. This approximation is most relevant in partially neutral gas and may overestimate the X-ray heating while underestimating the additional ionizations.

In \cref{app:impactPreheating} we compare two small-box test simulations performed with and without these additional X-ray sources, and show that the resulting pre-heating leaves the reionization history and the galaxy population essentially unchanged, while substantially raising the temperature of the still-neutral IGM.

\subsection{Connecting with the IllustrisTNG model}
\label{subsec:ConnectingTNGModel}
Coupling radiative transfer to the IllustrisTNG galaxy-formation model is non-trivial because gas that exceeds the star-formation threshold $n_{\rm H,SF}=0.106\,\mathrm{cm}^{-3}$ is placed on the effective equation of state. The ``temperature'' assigned to such cells is an effective pressurising quantity for the unresolved two-phase ISM, rather than a true thermodynamic temperature. A fully consistent coupling would therefore require modelling the cold and hot phases within each EoS cell in the \citet{springel2003cosmological} framework. The subgrid model does not, however, specify a unique internal configuration, which precludes a robust microphysical treatment; even if it did, the resolution required to model this phase structure explicitly in a cosmological volume would be prohibitive.

\begin{table*}
\centering
\caption{Frequency discretization of the radiation field used in our simulations.
Columns list the frequency range, photon-number-weighted mean cross sections
$\sigma$ (for \HI, \HeI, \HeII), mean excess energies per absorption $\mathcal{E}$,
the mean photon energy $e$, and the source SED. We switch at $z=4.75$ from six
bins to one.}
\label{tab:freqbins}

\setlength{\tabcolsep}{6pt}
\renewcommand{\arraystretch}{1.2}

\begin{tabular}{@{}lcccccccc@{}}
\toprule
Frequency bin [eV] &
$\sigma_{\mathrm{H\,I}}$ [cm$^{2}$] &
$\sigma_{\mathrm{He\,I}}$ [cm$^{2}$] &
$\sigma_{\mathrm{He\,II}}$ [cm$^{2}$] &
$\mathcal{E}_{\mathrm{H\,I}}$ [eV] &
$\mathcal{E}_{\mathrm{He\,I}}$ [eV] &
$\mathcal{E}_{\mathrm{He\,II}}$ [eV] &
$e$ [eV] &
Spectra \\
\midrule
\multicolumn{9}{c}{Hydrogen reionization ($z>4.75$)}\\
\hline
\addlinespace[2pt]
13.6--24.59    & $3.36\times10^{-18}$ & $0$                & $0$                & $3.19$  & $0$     & $0$    & $18.09$ & Stellar \\
24.59--54.42   & $7.13\times10^{-19}$ & $5.21\times10^{-18}$ & $0$              & $15.46$ & $5.32$  & $0$    & $31.78$ & Stellar \\
54.42--100     & $5.64\times10^{-20}$ & $9.12\times10^{-19}$ & $7.74\times10^{-19}$ & $54.84$ & $45.13$ & $14.55$ & $74.93$ & ISM \\
100--200       & $7.94\times10^{-21}$ & $1.76\times10^{-19}$ & $1.28\times10^{-19}$ & $114.4$ & $104.8$ & $74.65$ & $144.3$ & ISM \\
200--500       & $1.05\times10^{-21}$ & $2.70\times10^{-20}$ & $1.95\times10^{-20}$ & $229.4$ & $219.3$ & $189.9$ & $276.5$ & ISM \\
500--2000      & $5.49\times10^{-23}$ & $1.56\times10^{-21}$ & $1.21\times10^{-21}$ & $580.2$ & $570.4$ & $542.0$ & $714.3$ & ISM \\
\midrule
\multicolumn{9}{c}{Helium reionization ($z<4.75$)}\\
\hline
\addlinespace[2pt]
54.42--2000     & $5.38\times10^{-20}$ & $8.55\times10^{-19}$ & $7.34\times10^{-19}$ & -- & -- & $13.50$ & $96.8$ & BH \citep{Shen2020} \\
\bottomrule
\end{tabular}
\end{table*}

We therefore adopt a \emph{transparent-EoS} prescription: cells above the density threshold $n_{\rm H,SF}$ do not absorb ionizing radiation. Cooling and heating rates for gas above the EoS are computed with the equilibrium network used in IllustrisTNG, while gas below the EoS is instantaneously placed on it. This strategy, already adopted by \citet{bulichi2025high}, provides a clean definition of the stellar escape fraction as the fraction of photons that emerge from the unresolved ISM, and makes the EoS treatment fully equivalent to that of the original IllustrisTNG simulations.

This choice differs from \thesan, which applied non-equilibrium cooling to EoS gas using the (unphysical) effective temperature as input. The escape fractions in \thesan are correspondingly larger ($37\%$ in \thesan vs.\ $18\%$ in \lumina) and can be interpreted instead as the escape from the natal cloud. Looking ahead, more sophisticated subgrid prescriptions for matter--radiation coupling within the ISM---in particular models that explicitly evolve the cold-gas fraction in time \citep[e.g.][]{weinberger2023modelling,bollati2025modeling}---could enable a more physical coupling.

Below the density threshold for star formation, we replace the equilibrium cooling network by the non-equilibrium model including absorption of radiation.

\subsection{Discretization of radiation}
\label{subsec:discretizationRadiation}

We discretise the radiation field into frequency bins, listed in detail in \cref{tab:freqbins}. For hydrogen reionization we employ six bins: three photoionizing bands targeting \HI, \HeI, and \HeII, plus three additional X-ray bins that were not present in \thesan. The X-ray bins have substantially smaller photoionization cross-sections, and therefore much larger mean free paths, allowing them to penetrate the neutral IGM prior to reionization. At $z=4.75$ we switch focus to \HeII reionization and evolve only a single frequency bin. In this phase we replace the \HI and \HeI photoheating and photoionization rates by a spatially uniform UV background \citep{FG09}, complemented by the self-shielding prescription of \citet{Rahmati2013} for $z<3.35$.

We adopt frequency-dependent photoionization cross-sections from \citet{Verner1996}.
For each energy bin \(i\) with bounds \([e_i,e_{i+1}]\) we define the
\emph{effective} cross section for species \(j\) as the photon-number-weighted mean
\begin{equation}
\label{eq:meanCrossSection}
  \bar{\sigma}_{i,j}
  = \frac{\displaystyle \int_{e_i}^{e_{i+1}}
      \sigma_j(e)\,\frac{I(e)}{e}\,\mathrm{d}e}
    {\displaystyle \int_{e_i}^{e_{i+1}}
      \frac{I(e)}{e}\,\mathrm{d}e}\,,
\end{equation}
the mean excess energy per absorbed photon
\begin{equation}
\label{eq:meanExcessEnergy}
  \bar{\mathcal{E}}_{i,j}
  = \frac{\displaystyle \int_{e_i}^{e_{i+1}}
      \sigma_j(e)\,\frac{I(e)}{e}\,
      \bigl(e-e_{\mathrm{th},j}\bigr)\,\mathrm{d}e}
    {\displaystyle \int_{e_i}^{e_{i+1}}
      \sigma_j(e)\,\frac{I(e)}{e}\,\mathrm{d}e}\,,
\end{equation}
and the mean photon energy
\begin{equation}
\label{eq:meanEnergy}
  \bar{e}_i
  = \frac{\displaystyle \int_{e_i}^{e_{i+1}} I(e)\,\mathrm{d}e}
    {\displaystyle \int_{e_i}^{e_{i+1}} \frac{I(e)}{e}\,\mathrm{d}e}\,,
\end{equation}
which we use to convert between radiation energy and photon number.
Here \(e_{\mathrm{th},j}\) denotes the ionization threshold of species \(j\).

By construction, these definitions exactly preserve the photoionization rates, photoheating rates, and energy flux of a continuous spectrum (for the adopted spectral shape $I(e)$ within the bin) when that spectrum is represented by a single bin. The accuracy degrades if the spectrum varies strongly within a bin with time, or if photons from distinct source populations are mixed within the same bin. Ideally one would evolve the spectral shape itself, which would naturally accommodate, e.g.\ cosmological redshifting and wavelength-dependent attenuation, but this would substantially increase memory usage and computational cost. In practice we therefore tabulate $\bar{\sigma}_{i,j}$, $\bar{\mathcal{E}}_{i,j}$, and $\bar{e}_i$ using the SED of the source class that dominates each bin.

During hydrogen reionization, the two lowest-energy bins use a stellar SED for a $1\,\mathrm{Myr}$-old population with metallicity $0.25\,\Zsun$ (consistent with \thesan), while the higher-energy bins (including the X-ray bands) use the shock-heated ISM SED described in \cref{subsubsec:xraySource}. During \HeII reionization, when accreting black holes dominate the budget in the remaining bin, we instead adopt the composite quasar SED of \citet{Shen2020}.

\subsection{Differences relative to the \thesan project}
\label{subsec:diffThesan}
\lumina and \thesan share the goal of combining the IllustrisTNG galaxy-formation model, validated at $z=0$, with the \areport\ radiative-transfer solver to study reionization. They differ, however, in several important ways that particularly affect galaxy properties at lower redshift. We summarize these differences below.

\thesan applied the non-equilibrium chemical network to all gas cells, including those on the effective equation of state. The cells' effective temperatures were used as initial values; because of their high densities these cells cooled efficiently and, as they cooled, became increasingly neutral and absorbed more radiation. At the end of each time step, however, the cells were instantaneously returned to the EoS, so the minimum temperature reached---and the corresponding photon absorption---ended up depending on the local integration time step. As a result, \thesan adopted a larger escape fraction ($37\%$) to compensate for these losses. The cooled intermediate state was also used in the Euler equations, which reduced the pressure support provided by the EoS and required a larger gravitational softening length for the gas. The same cooled state was used to compute the Bondi accretion rate onto BHs (see \cref{eq:bondiRate}), boosting the global Bondi rate.

\begin{figure*}
    \centering
    \includegraphics[width=1\linewidth]{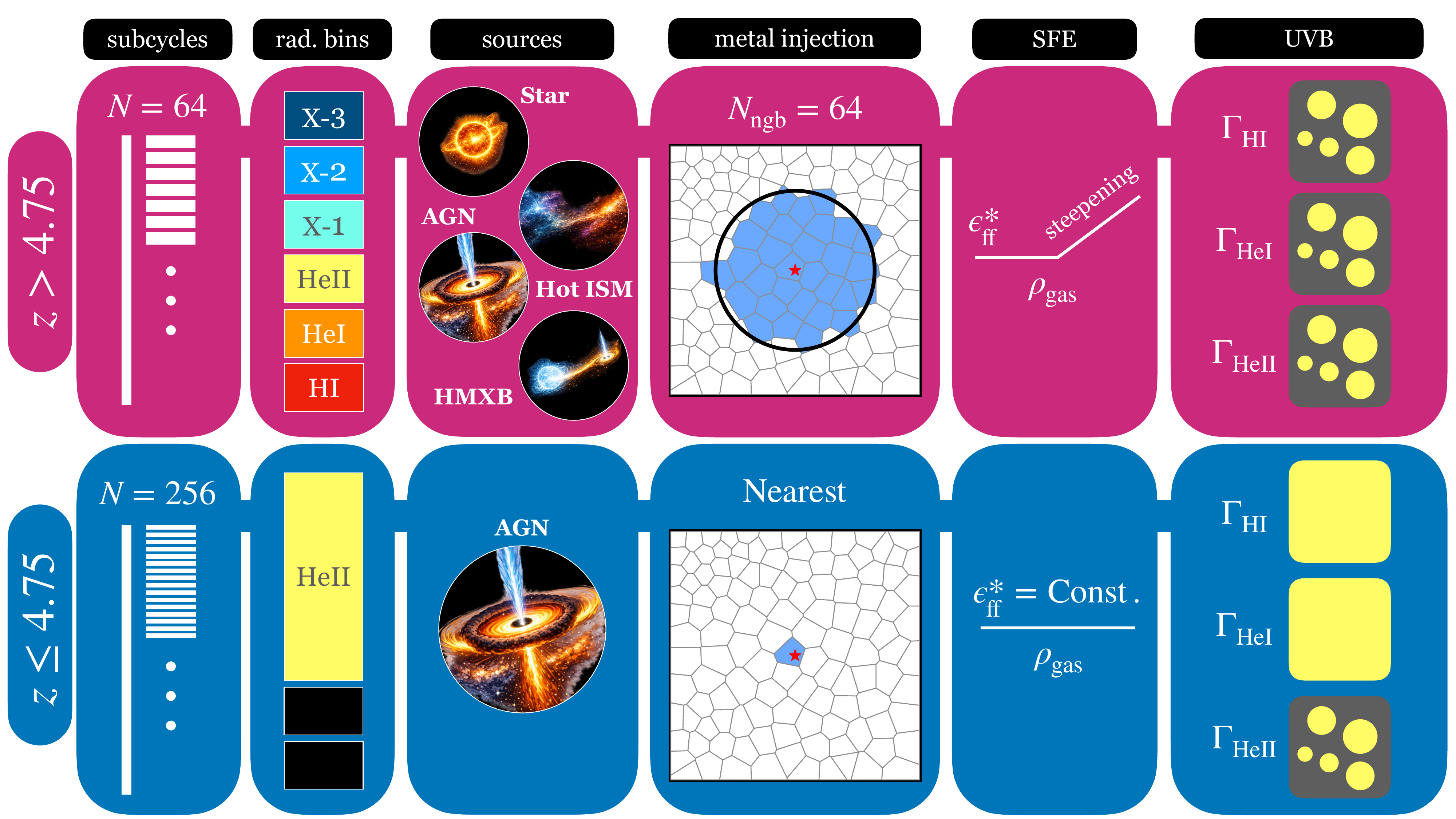}
  \caption{Schematic view of the modifications applied at the transition from hydrogen to helium reionization at $z=4.75$. We increase the number of radiative-transfer subcycles from 64 to 256, switch from six frequency bins to a single bin tracking the ionization of \HeII, and restrict the radiation sources to AGN only. Metals from star particles are deposited into the single nearest Voronoi cell rather than spread over an SPH kernel, which significantly reduces the cost of the iterative neighbour search. We also revert to the fiducial star-formation efficiency, which is required for efficient BH growth. For the ionization rates of \HI and \HeI we adopt a uniform UV background for $z\leq4.75$ instead of solving the full radiative transfer.}
  
    \label{fig:schematic_changes}
\end{figure*}

In addition, an undetected coding error in \thesan replaced the speed of light by the reduced value $0.2\,c$ in both the feedback routines and the Eddington-rate calculation \citep{garaldi2026correction}. This further accelerated BH growth, so that BHs entered the radio mode earlier and quenched galaxies sooner than in the IllustrisTNG model \citep{Chittenden2025}.

\section{Modelling the transition from hydrogen to \texorpdfstring{\HeII}{He II} reionization}
\label{sec:changed_at_z_4p75}
Modelling \HeII reionization introduces significant computational challenges. To reach $z=3$ the simulation must evolve for almost an additional gigayear after the end of \HI reionization, during which the matter distribution becomes increasingly clustered. The densities in the most massive haloes are then higher and force smaller time steps. To address these issues, we apply several changes to the model at $z=4.75$, summarized in \cref{fig:schematic_changes}. The transition redshift is set by the simulation itself: $z=4.75$ is the redshift at which the last neutral \HI island in the volume disappears (\cref{subsec:HIreionization}), which we verified by computing reionization histories in sub-boxes of side length $25\,\mathrm{cMpc}$ and requiring the evolution of the \HI fraction to have become flat in all of them. 

We increase the number of radiative-transfer subcycles from 64 to 256, which allows larger time steps for the remaining physics. Once \HI and \HeI are fully ionized in our box at $z=4.75$, we adopt a uniform UV background for photons with energies $E<54.42\,\mathrm{eV}$ in place of self-consistent radiative transfer. In practice, within the non-equilibrium chemical network we replace the photoionization and photoheating rates by those of \citet{FG09}, and for $z<3.35$ we additionally apply the density-based self-shielding prescription of \citet{Rahmati2013}. We also merge the remaining four radiation bins into a single bin spanning $54.42\,\mathrm{eV}$ to $2\,\mathrm{keV}$ to track the ionization of \HeII. Because \HeII reionization is driven primarily by AGN, below $z=4.75$ we restrict the ionizing sources to the black-hole particles only.

To reduce the computational cost we also modify the metal-injection scheme. In the default IllustrisTNG model, metals from each star particle are distributed to the 64 nearest gas cells using an SPH-kernel weighting, which requires a relatively expensive iterative neighbour search per active star particle. Below $z=4.75$ we instead deposit all metals into the single nearest Voronoi cell, which can typically be identified in one iteration by using $1.5$ times the radius of the cell enriched at the previous time step as the initial search radius. We have verified that this change does not affect the metallicity profiles of well-resolved galaxies, consistent with \citet{van2020neutron}, who compared the two metal-injection schemes in zoom-in simulations from the Auriga project \citep{grand2017auriga}.

\citet{nelson2024introducing} recently demonstrated, in the TNG-Cluster suite, that the steeper star-formation efficiency shown in \cref{fig:schematic_changes} can reduce the masses of the most massive black holes by lowering the gas density within the accretion region. In our own tests below $z=4.75$ we likewise observed an earlier transition into radio mode when adopting shorter gas-consumption times. To remain consistent with the calibrated IllustrisTNG model---and to avoid underproducing the bright end of the quasar luminosity function---we therefore disable this modification for $z\le 4.75$. The reversion produces only a small, transient dip in the star-formation rate density (see \cref{fig:SFRD} and \cref{sec:sfrd}). At earlier times the most massive black holes accreted at nearly the Eddington rate, so we do not expect our choice to use the shorter consumption time beforehand to significantly affect the results.

\section{Generation of initial conditions}
\label{sec:GeneratingICs}
On the large scales relevant to cosmological initial conditions, the Universe is nearly uniform, with small, Gaussian perturbations that grow linearly at early times. Linear perturbation theory connects the primordial power spectrum to low-redshift matter perturbations through transfer functions computed by Boltzmann solvers such as \camb\ \citep{camb2000,camb2012}. Linear theory begins to break down at shell crossing, when the density contrast $\delta \equiv \delta\rho/\bar\rho$ approaches unity.

\begin{figure}
    \centering
    \includegraphics[width=1\linewidth]{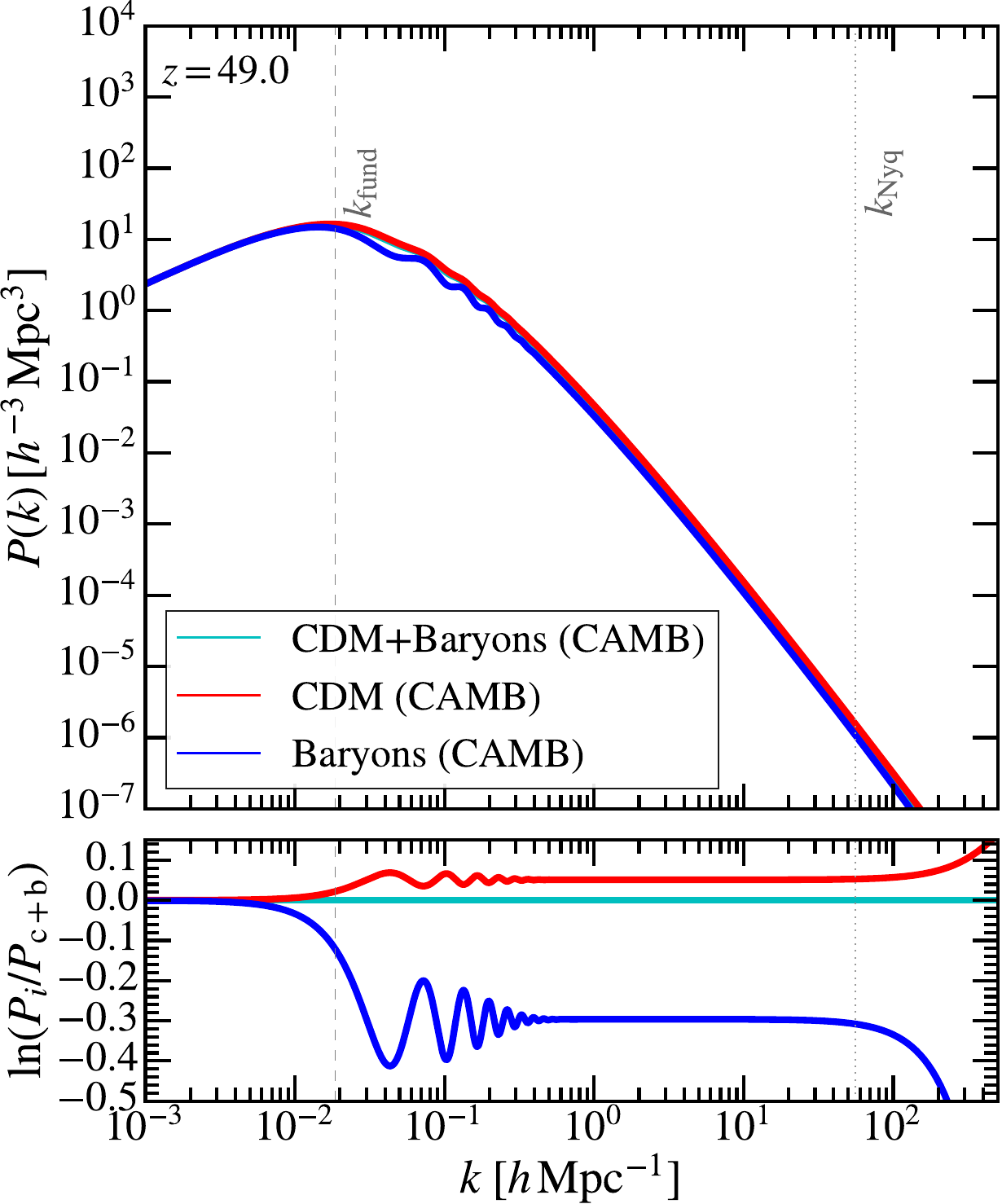}
 \caption{Linear matter power spectra from \camb \protect\citep{camb2000,camb2012} at the initial redshift $z=49$ used in our simulation. The upper panel shows the absolute spectra for cold dark matter (CDM), baryons, and their total; the lower panel shows the corresponding ratios with respect to the total spectrum. The baryon spectrum is clearly suppressed on all scales. In \lumina we generate initial conditions with separate baryon and CDM transfer functions, whereas many previous simulations used the total-matter transfer function for both species. Vertical dashed and dotted lines indicate the fundamental mode of the box, $k_{\rm fund}$, and the Nyquist frequency, $k_{\rm Nyq}$, respectively.}
    \label{fig:powerSpectraLinear}
\end{figure}

The transfer functions can be evaluated at the chosen starting redshift $z_{\rm init}$ to construct the initial linear power spectra. Because cold dark matter is not coupled to the radiation field, it can begin to form structure already before recombination. The baryons, in contrast, are tightly coupled to photons before recombination and experience pressure support and acoustic oscillations. Consequently, the dark-matter and baryon transfer functions differ \citep{eisenstein1998baryonic}, with enhanced power in the dark matter relative to the baryons, as shown in \cref{fig:powerSpectraLinear}. These relative differences diminish over time, as baryons fall into the inhomogeneous dark-matter potential, but remain at the few-per-cent level at $z\simeq 5$. The mismatch also gives rise to a large-scale modulation of the \emph{local} baryon fraction,
\begin{equation}
 {f_b}(\vec{x}) = \rho_{\rm b}(\vec{x}) / \rho_{\rm tot}(\vec{x}) = [1+\delta_{\rm b}(\vec{x})]/[1+\delta_{\rm tot}(\vec{x})]\, \Omega_{\rm b} / \Omega_{\rm tot} \, ,
\end{equation}
where $\delta_{\rm b}(\vec{x})$ and $\delta_{\rm tot}(\vec{x})$ are the local baryon and total density contrasts, respectively. This modulation directly affects observations of the \mbox{Lyman-$\alpha$} forest \citep{Fernandez2021}, the 21\,cm signal \citep{naoz2005growth}, and the baryon fraction of the earliest galaxies \citep{jessop2025ripples}.

The baryon--CDM mismatch also induces a coherent, large-scale relative (``streaming'') velocity between the two fluids \citep{tseliakhovich2010relative}, which can delay the onset of star formation in the first minihaloes \citep{Fialkov2012}. The resulting change in the clustering of the first star-forming haloes also directly affects the 21\,cm power spectrum \citep{visbal2012signature,mcquinn2012impact,fialkov201321,cohen201621,cain2020model}. 

Most large-scale cosmological simulations to date---including Illustris \citep{IllustrisNature}, IllustrisTNG \citep{pillepich2018first}, EAGLE \citep{schaye2015eagle,crain2015eagle}, and CoDa \citep{ocvirk2016cosmic,ocvirk2020cosmic,lewis2022short}---have neglected both the distinct baryon and dark-matter transfer functions and the streaming velocity. Notable exceptions are the more recent {\small ASTRID} \citep{bird2022}, {\small FLAMINGO} \citep{flamingo}, and {\small COLIBRE} \citep{schaye2025colibre} simulations, which include separate transfer functions but not the streaming velocity. In this work we introduce a new initial-conditions generation scheme that consistently incorporates both effects. To describe it clearly, we first briefly recall the traditional approach of setting up initial conditions with a single transfer function.

\subsection{Baryonic initial conditions with a single transfer function}
Illustris, IllustrisTNG, MTNG, and \thesan used the total-matter (baryons + dark matter) transfer function computed with \camb, and MTNG and \thesan additionally employed the fixed-and-paired variance-suppression technique \citep{Angulo2016} to mitigate cosmic variance. First, a single-species gravitational glass was constructed in these projects with the exception of \thesan, which used a uniform Cartesian grid instead. Small displacements were then applied to particle positions using first- (or second-) order Lagrangian perturbation theory to imprint the target power spectrum. From this distribution, each particle was split into a gas and a dark-matter particle, with masses set by the cosmic baryon and total-matter fractions. The two child particles were offset from their shared origin in opposite directions in each coordinate dimension, by a distance equal to half of the original mean particle spacing while preserving the pair's centre of mass.

All simulations mentioned above compute the gravitational forces at Voronoi cell centres of mass, which is slightly inconsistent with the initial-condition setup that uses the mesh-generating-point positions. This mismatch induces an immediate loss of small-scale power in the baryonic component, starting at wavenumbers 
\begin{equation}
k \sim 0.1\,k_{\mathrm{Ny}} = 0.1\,\frac{\pi N}{L_{\rm box}} \, , 
\end{equation}
where the loss of power is of order 1--2 per cent, but becoming quickly larger towards the Nyquist frequency. While using the geometric centre of Voronoi cells is conceptually intuitive from the point of gas dynamics, this particular coordinate depends on the coordinates of all neighbouring mesh-generating points. As a consequence, one cannot prescribe the motion of the geometric centers of Voronoi cells directly at an individual level---one can only prescribe this motion easily for the mesh-generating points. When imposing initial density perturbations and initial velocities, we thus need to rely on imparting them onto the mesh generating points, assuming that the mass of cells is concentrated in them. The Voronoi cell centres arise only after the fact from the perturbed mesh-generating points, and are in general very slightly spatially offset to them. The gravitational accelerations obtained at the cell centres when associating the mass of cells with these points are for this reason not identical to the intended ones based on the imposed perturbations. Rather, these baryonic gravitational accelerations are slightly smoothed on scales close to the mean particle spacing.  While this smoothing effect will eventually be completely buried during non-linear evolution by the power transfer from large to small scales, the effect can slightly damp the growth of fluctuations close to the Nyquist frequency in the linear regime, and thus affect the timing of the formation of the first structures. We therefore associate the computation of the gravitational forces with the mesh generating point to eliminate this effect on early structure formation, at the price of being less well matched to the cell-centred approach for computing the hydrodynamical fluxes.

\begin{figure}
    \centering
    \includegraphics[width=1\linewidth]{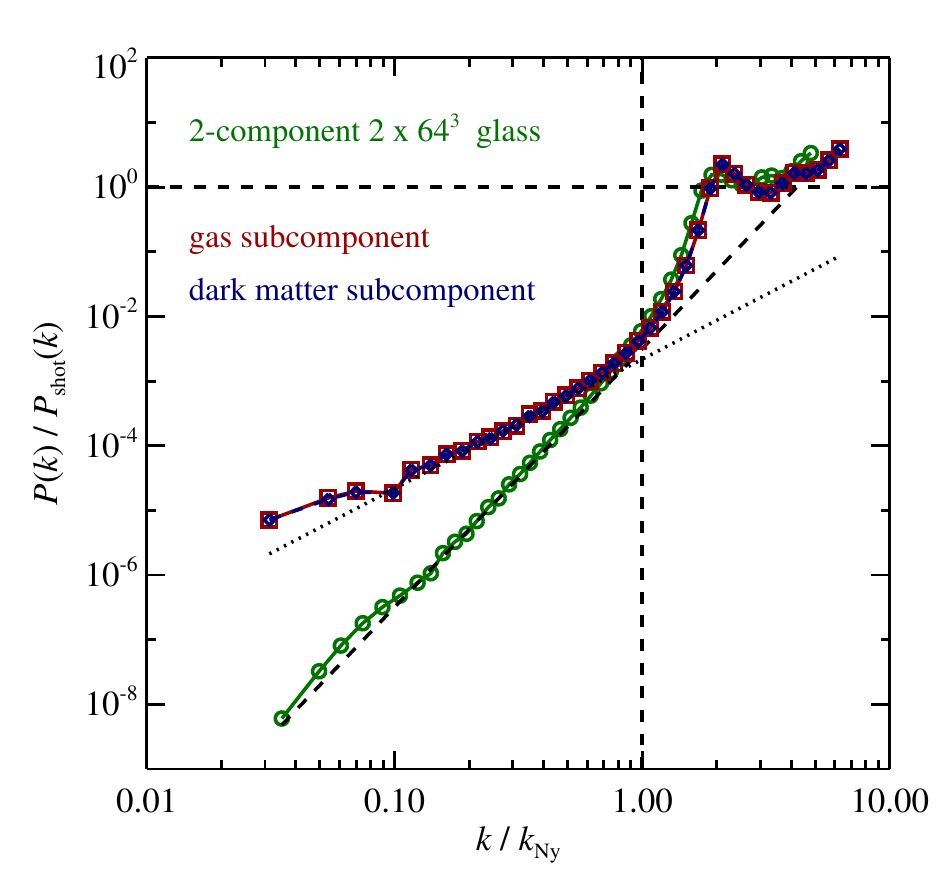}
 \caption{Power spectrum of a two-component glass with two times $64^3$ particles, showing the characteristic $P(k)\propto k^4$ spectrum (dashed inclined line) below the Nyquist frequency for the total particle distribution, while the two subcomponents drop off with $P(k)\propto k^2$ (dotted line). At the Nyquist frequency itself, the power is suppressed by about a factor $1/0.0032 \simeq 310$ below the shot-noise, and the suppression becomes very large towards larger scales.  \label{fig:FigGlassMultiple}}
\end{figure}

\subsection{Implementation of different transfer functions for baryons and dark matter}
\label{sec:self_consistent_TFs}

Imposing slightly different perturbation amplitudes for cold dark matter and baryons may appear straightforward in principle, but a number of subtle numerical issues make it surprisingly non-trivial in practice. Numerous studies have encountered significant discreteness effects when attempting this \citep[e.g.][]{Yoshida2003b,Bird2020,Liu2023}, often defeating the goal of following the linear growth of each component accurately. The typical outcome is a rapid erasure of the relative perturbations, with baryons and dark matter quickly settling into their joint mass-weighted perturbations. The subsequent growth of the joint mode then follows the mass-weighted linear-theory prediction, so the result is effectively equivalent to imposing a single mass-weighted transfer function in the first place.

The main cause of this behaviour is the appearance of spurious differential forces between alternating species of resolution elements along the direction of a perturbing mode once the two species carry different amplitudes. Consider, for example, two interleaved Cartesian grids, one for dark matter and one for baryons. If one imposes a perturbation on the baryons but not on the dark matter, the baryons should feel no force from the dark matter, since by construction the dark-matter distribution still represents a perfectly uniform density field. From the vantage point of a perturbed baryonic particle, however, the dark-matter grid does exert a non-zero force, because the baryons no longer sit at the centroids of the dark-matter grid: a single dark-matter particle invariably becomes the nearest neighbour of a given gas particle, producing a spurious force towards it. This happens because the dark matter is discretised into point particles rather than a continuum, and the resulting discreteness effects break the correct differential evolution between the two components.

An approach to alleviate this issue was proposed by \citet{Bird2020}. The idea is to use a two-component glass (or a Cartesian grid combined with a glass), where the glass is an irregular arrangement of particles that emerges when a Poisson distribution is evolved with the sign of gravity reversed. The spurious forces are then suppressed, because the lack of regularity of the glass causes them to average out statistically over the particles making up a given mode: the local geometry of the particle distribution around each particle differs slightly from that around any other.

Constructing a good two-component glass is, however, not without its own difficulties. One would ideally like to require that the two components be in equilibrium not only with respect to the total force, but also separately for each species. Violating this second criterion means that the unperturbed distribution already contains fluctuations in the local baryon fraction, which is undesirable.

A more accurate solution was put forward in an insightful study by \citet{Hahn2021}. Starting from an initially unperturbed, constant-density state, one can realise an initial density perturbation in two ways: by displacing particles from their equilibrium positions, or by leaving the particles in place and perturbing their masses. \citet{Hahn2021} pointed out that combining the two methods makes it possible to imprint distinct perturbations in baryons and dark matter while avoiding the spurious discreteness effects discussed above. The procedure is to first perturb the positions of dark-matter and baryonic particles with the total-matter perturbations---i.e.\ coherently, using a single displacement and velocity field, just as in the conventional single-transfer-function approach---and then, in a second step, to impose the relative density perturbations between the two species by varying their masses (and not their positions). This combination is largely free of the discreteness problem above and also allows the relative streaming motions between dark matter and baryons (the decaying mode) to be imposed cleanly.

We adopt this approach in \lumina, and therefore do not impose the dark-matter perturbations $\delta_{\rm dm}$ and the baryon perturbations $\delta_{\rm b}$ directly. Instead, we first impose the mass-weighted total perturbation $\delta_{\rm tot} = f_{\rm b}\,\delta_{\rm b} + (1-f_{\rm b})\,\delta_{\rm dm}$, with $f_{\rm b} \equiv \Omega_{\rm b}/\Omega_{\rm tot}$, on both species of an initially unperturbed density field. We do this in second-order Lagrangian perturbation theory (LPT) in the standard way \citep{Scoccimarro1998}, by perturbing the Lagrangian coordinates of the dark-matter and baryonic resolution elements and assigning corresponding velocities along the perturbations. This initialises the growing mode and does not excite the corresponding decaying mode (at least not deliberately---numerical transients can still arise, but they decay quickly, and using higher-order perturbation theory rather than only the Zel'dovich approximation helps here).

We then impose the relative density perturbation $\delta_{\rm b}-\delta_{\rm dm}$ between dark matter and baryons through mass variations of the dark-matter and baryonic resolution elements. These mass perturbations are likewise computed in second-order LPT and correspond to the constant mode in the evolution of $\delta_{\rm b}-\delta_{\rm dm}$, which is approximately frozen in \citep{Schmidt2016}. For typical cosmological parameters, the maximum initial distortion of the baryonic masses is $\approx 5\%$, which sets the level of the relative variations in the local baryon fraction. The mass distortion of the dark-matter particles is at most $\sim 1\%$ and is typically smaller. These relative mass perturbations are bounded: they do not grow significantly with higher resolution or larger box size. Any numerical disadvantages that might arise from using particles of variable mass---such as enhanced numerical heating of lighter dark-matter particles by heavier ones---are therefore negligible.

Finally, we impose the decaying mode of $\delta_{\rm b}-\delta_{\rm dm}$, which corresponds to the streaming velocities between the two components. As \citet{Hahn2021} pointed out, there is no straightforward way to compute these velocities in higher-order perturbation theory as a difference relative to the total-matter velocity. We therefore compute the total velocity of each species separately in second-order LPT and recover the streaming velocity of each component by subtracting the second-order-accurate mass-weighted velocity.

This still leaves the question of how the initial unperturbed distribution of tracers should be realised. The two most common options are a Cartesian grid and a glass. A two-component Cartesian setup is realised most conveniently by nesting two Cartesian grids; this is clean in the sense that it provides an exact unperturbed state, but the initial state contains preferred directions, and projections of the particle distribution can produce spurious aliasing artefacts because particles are regularly arranged along specific axes.

A glass is free of these artefacts, but its power spectrum does not vanish exactly on large scales, and an interpolation is required to map the perturbations from the Fourier grid used to compute them onto the particle coordinates of the glass. Furthermore, it is less obvious how a two-component glass should be constructed. In the IllustrisTNG and Auriga projects this was addressed approximately by duplicating a single-component glass and translating the second copy by half the mean grid spacing in each dimension, as described above. With this simple procedure the large-scale perturbations remain acceptably small, but the resulting two-component glass contains some spurious small-scale perturbations. \citet{Bird2020} reduced this problem by performing a few relaxation steps (i.e.\ with the sign of gravity reversed) after combining a glass with a Cartesian grid, but at the price of introducing spurious local density fluctuations in the individual components.

Here we address this with a more elaborate glass-construction procedure. First, we create a glass for component A (say, dark matter) and a separate glass for component B (say, gas particles). To accelerate the glass-making process we use a trick introduced by \citet{Springel2005}: when a perfect glass is slightly perturbed by small displacements $\vec{d}_i$, the Zel'dovich approximation predicts that the resulting forces are aligned with, and proportional to, the displacements. By computing the forces and pushing the particles back to their Lagrangian positions accordingly, the glass can in principle be recovered in a single step. Iterating this procedure with a high-accuracy force solver leads to a rapidly converging glass.

We then combine the two glasses for A and B in a single volume and apply the same glass-making procedure to the joint distribution, but with a small twist when computing the particle displacements at each step: at every iteration the displacement is averaged between two cases, one in which A and B feel only their own species (so that A sees only A and B sees only B), and one in which all particles see all particles (assuming equal masses). The combined particle set thus settles into a glass-like distribution in which components A and B are each individually close to force equilibrium, and in which the total particle set is also in equilibrium with respect to the total forces, independently of the precise mass ratio between A and B. This also minimises residual relative density variations between A and B in the final two-component glass.

\begin{table}
\centering
\addtolength{\tabcolsep}{15.0pt}
\begin{tabular}{ll}
\hline
Parameter & Value \\
\hline
$\Omega_{\rm m}$   & 0.3096 \\ 
$100\times\Omega_{\rm b}$  & 4.897 \\ 
$\Omega_\Lambda$   & 0.6903 \\
$H_0\,[\kms \Mpc^{-1}]$ & $67.66 $ \\
$\sigma_8$         & $0.8102$  \\
$n_{\rm s}$        & $0.9665$\\
$z_{\rm init}$     & 49 \\
$T_{\rm init}\,[\mathrm{K}]$ & $54.17$ \\
$T_{\rm CMB,0}\,[\mathrm{K}]$    & $2.7255$ \\
$T_{\nu,0}\,[\mathrm{K}]$        & $1.9524$ \\
$N_{\rm neutrino}$ & 3 \\
\hline
\end{tabular}
\caption{Key cosmological parameters from the TT,TE,EE+lowE + lensing + BAO data set of \protect\cite{planck2020}. We list the matter density $\Omega_{\rm m}$, baryon density $\Omega_{\rm b}$, dark-energy density $\Omega_\Lambda$, the Hubble constant $H_0$, the fluctuation amplitude $\sigma_8$ and spectral index $n_s$ of the initial power spectrum, the starting redshift of our simulations $z_{\rm init}$, the initial baryon temperature $T_{\rm init}$, the present-day CMB temperature $T_{\rm CMB,0}$, the present-day temperature of the cosmic neutrino background $T_{\nu,0}$, and the number of massless neutrino species used $N_{\rm neutrino}$. We absorb the slightly larger effective number of relativistic species, $N_{\rm eff}=3.046$, into the effective neutrino background temperature.}
\label{tab:cosmo}
\end{table}

 \begin{table*}
\centering
\caption{Numerical parameters of the \lumina simulation suite and of reference simulations used for comparison throughout this work.
In addition to the flagship radiation-hydrodynamics simulation, we carried out two smaller radiation-hydrodynamics test runs at the same resolution as \lumina, performed with (\luminafifty{}) and without (\luminafiftynox{}) the additional X-ray sources, which we use in \cref{app:impactPreheating} to quantify the impact of X-ray pre-heating, as well as two lower-resolution dark-matter-only reruns (DM-3000 and DM-1500), evolved to $z=0$ from the same initial phases.
We list the comoving box size $L$, the number of gas and dark-matter resolution elements ($N_{\rm gas}$, $N_{\rm dm}$), the mean baryonic and dark-matter mass resolution ($m_{\rm b}$, $m_{\rm dm}$), and the comoving Plummer-equivalent gravitational softening of dark-matter particles ($\epsilon$).
Gas softenings are adaptive (tied to the cell size) and bounded by the minimum value $\epsilon_{\rm gas,min}$.
For \thesantwo, the softening was previously reported incorrectly as $4.1\,\mathrm{ckpc}$ \citep{garaldi2026correction}.
We additionally state the volume of \lumina relative to each box as well as the baryon-mass ratio.}
\label{tab:simulation_overview}
\setlength{\tabcolsep}{6pt}
\begin{tabular}{@{}llccccc c c c cc@{}}
\toprule
Run & \multicolumn{2}{c}{Boxsize} & $N_{\rm gas}$ & $N_{\rm dm}$ & $m_{\rm b}$ & $m_{\rm dm}$ & $\epsilon$ & $\epsilon_{\rm gas, min}$ &$z_{\rm final}$ & $ V_{\rm Lumina} / V$ & $m_{\rm b} / m_{\rm b, Lumina}$ \\
\cmidrule(lr){2-3}
 & $(h^{-1}\,\mathrm{cMpc})$ & $(\mathrm{cMpc})$ & &  &
$(\Msun)$ & $(\Msun)$ & $[\,\mathrm{ckpc}]$ & $[\,\mathrm{ckpc}]$ \\
\midrule
\lumina & 338.3 & 500 & $6000^{3}$ & $6000^{3}$  &
$3.6\times10^{6}$ & $1.9\times10^{7}$ & 1.77 & 0.44 & 3 & 1 & 1\\
\luminafifty{} & 33.83 & 50 & $600^{3}$ & $600^{3}$  &
$3.6\times10^{6}$ & $1.9\times10^{7}$ & 1.77 & 0.44 & 5 & 1000 & 1\\
\luminafiftynox{} & 33.83 & 50 & $600^{3}$ & $600^{3}$  &
$3.6\times10^{6}$ & $1.9\times10^{7}$ & 1.77 & 0.44 & 5 & 1000 & 1\\
DM-3000 & 338.3 & 500 & 0 & $3000^{3}$  &
-- & $1.8\times10^{8}$ & 1.77 & -- & 0 & 1 &-\\
DM-1500 & 338.3 & 500 & 0 & $1500^{3}$  &
-- & $1.4\times10^{9}$ & 1.77 & -- & 0 & 1 & -\\
\addlinespace
\multicolumn{10}{l}{\textit{Reference simulations}}\\
\thesanone & 64.7& 95.5 & $2100^3$ & $2100^3$ & $5.8\times 10^5$ & $3.1\times 10^6$ & 2.2 & 2.2 & 5.5 &143.5 & 0.16\\
\thesantwo &  64.7& 95.5 & $1050^3$ & $1050^3$ & $4.7\times 10^6$  & $2.5\times10^7$ & 4.4 & 4.4 & 5.5 &143.5 & 1.3\\
TNG100 &75&110.7 &$1820^3$ & $1820^3$ &$1.4\times10^6$ & $7.5\times10^6$ & 0.74 & 0.19 & 0 &92.1 & 0.39\\
MTNG & 500 & 740 &$4320^{3}$ &$4320^{3}$ &$3.1\times 10^7$ & $1.7\times10^8$ & 3.7 & 0.37 & 0&0.308 & 8.6\\
\bottomrule
\end{tabular}
\end{table*}

In \cref{fig:FigGlassMultiple} we show the power spectra of both components for an example of a two-component glass created in this way. We also include the total power spectrum for the combined components. The latter shows the theoretically expected $P(k) \propto k^4$ power spectrum for a glass of discrete tracers \citep{Baugh1995, Liao2018} approximating the minimal possible power spectrum \citep[][chapter 28]{Peebles1980} in this case.  The individual subcomponents in our two-component glass do not drop off quite as fast towards larger scales. Importantly, however, both the total and the subcomponents show a substantial suppression at the Nyquist frequency compared to the shot-noise, only slightly worse than for a single component glass. Note that this shape of the power spectrum for a glass is universal when $k$ is expressed in units of the Nyquist frequency, and the power spectrum is measured in units of the shot-noise power. This highlights  that the residual noise in the glass is highest at the Nyquist frequency, where it is suppressed by about 2 orders of magnitude compared to the shot noise. The critical condition for using a glass as starting point for a cosmological simulation is hence that the imposed power spectrum at the starting redshift should everywhere be higher than the intrinsic glass power spectrum. If this is the case, the residual noise in the glass plays no role. Fortunately, this condition is typically met for most cosmological setups, unless one starts at very high redshift.

For \lumina, we have opted to use such a two-component glass for the initial unperturbed state. We computed individual transfer functions for dark matter and gas with \camb at the initial starting redshift, and imposed the corresponding perturbations with $2^{\rm nd}$-order accurate LPT via a combination of spatial and mass perturbations, as described above. The actual creation of the initial conditions was carried out by a novel version of \textsc{N-GenIC} integrated in our simulation code \arepo. To reduce the computational expense in making the glass distributions, we constructed a two--component glass at a resolution of $2\times 75^3$, which was then scaled and tiled to cover the full box size and particle number of \lumina.

\subsection{Radiation (photons) and massless neutrinos}
The relative impact of dark matter, baryons, and dark energy on the expansion history is described by the present-day density parameters $\Omega_{\rm cdm}$, $\Omega_{\rm b}$, and $\Omega_\Lambda$, respectively. In \lumina we additionally include the contribution from the CMB photon field,
\begin{equation}
  \Omega_\gamma
  = \frac{8\pi^3 G k_\text{B}^4}{45\,\hbar^{3} c^{5} H_0^{2}}\,T_{\rm CMB,0}^{4} \, ,
  \label{eq:Omega_gamma}
\end{equation}
where $T_{\rm CMB,0}$ is the present-day CMB temperature.

We further include (effectively) massless neutrinos, whose energy density scales in the same way as that of photons. Their present-day contribution can be written in terms of $\Omega_\gamma$ as
\begin{equation}
  \Omega_{\nu}
  = \Omega_\gamma\left[\frac{7}{8}\left(\frac{4}{11}\right)^{4/3} N_{\rm eff}\right] \, ,
  \label{eq:Omega_nu}
\end{equation}
so that the total radiation density is $\Omega_{\rm rad}=\Omega_\gamma+\Omega_\nu$.
In the Standard Model, the effective number of relativistic species is slightly larger than three, $N_{\rm eff}=3.046$, because the three neutrino flavours are not perfectly decoupled during $e^\pm$ annihilation \citep{gnedin1998cosmological,mangano2005relic,de2016relic}. Deviations in $N_{\rm eff}$ can equivalently be absorbed into an effective neutrino temperature via
\begin{equation}
  T_{\nu,0} = \left(\frac{4}{11}\right)^{1/3}
  \left(\frac{N_{\rm eff}}{3}\right)^{1/4} T_{\rm CMB,0}.
  \label{eq:Tnu}
\end{equation}

Both components scale as $\propto a^{-4}$ with the scale factor $a$ and therefore have only a minor impact over the redshift range covered by our simulations. Including them, however, ensures that the linear growth predicted by our code matches the results of Boltzmann solvers. For $z\gtrsim 200$ one would additionally need to account for Compton drag of the CMB on baryons; this effect is neglected here.

\subsection{Choice of cosmological parameters}
\label{sec:cosmo-params}
We adopt a flat $\Lambda$CDM cosmology consistent with the Planck 2018 TT,TE,EE+lowE+lensing+BAO constraints \citep[see \cref{tab:cosmo};][]{planck2020}. This updates the parameters relative to those used in IllustrisTNG and \thesan \citep{Planck2015}. The present-day CMB temperature is $T_{\rm CMB,0}=2.7255\,\mathrm{K}$ \citep{fixsen2009}. For an effective number of neutrino species $N_{\rm eff}=3.046$, the corresponding present-day neutrino background temperature is
\begin{equation}
        T_{\nu,0}
    = \left(\frac{4}{11}\right)^{1/3}
      \left(\frac{N_{\rm eff}}{3}\right)^{1/4}
      T_{\rm CMB,0}
    \simeq 1.952\,\mathrm{K} \, .
\end{equation}

We initialise the simulation at $z_{\rm init}=49$, which balances discreteness errors (which grow with $z_{\rm init}$) against the inaccuracy of second-order perturbation theory. Assuming thermal decoupling of the baryons from the CMB at $z_{\rm dec}=124.8$ \citep[e.g.][]{scott2009matter,venumadhav2018heating,Hergt2025}, followed by adiabatic cooling, the initial gas temperature is
\begin{equation}
       T_{\rm init}
    = T_{\rm CMB,0}\,
      \frac{(1+z_{\rm init})^{2}}{1+z_{\rm dec}}
    \simeq 54.17\,\mathrm{K} \, .
\end{equation}

\begin{figure}
    \centering

    \includegraphics[width=1\linewidth]{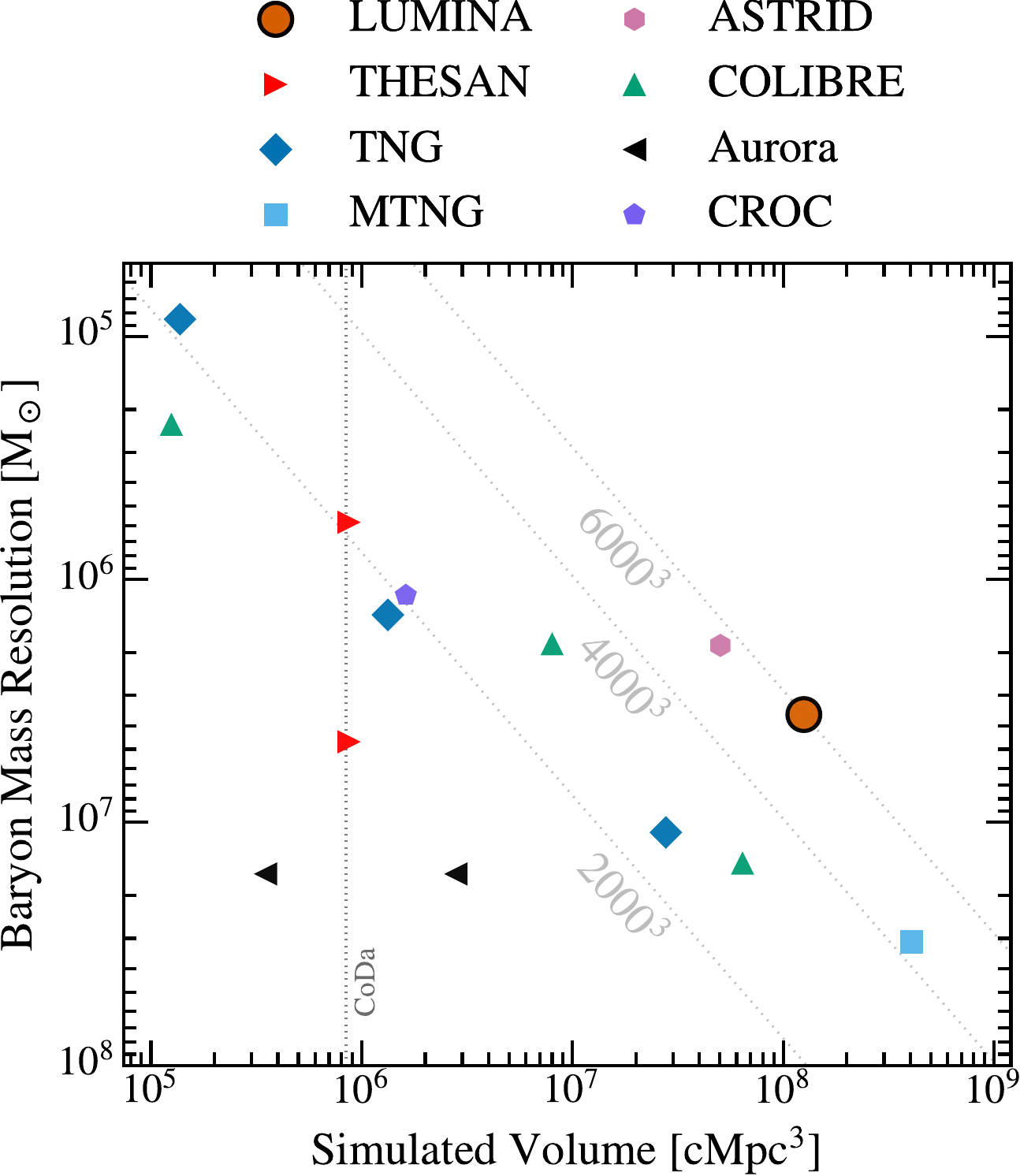}
    \caption{Comparison of the simulated volume and baryonic mass resolution of \lumina with other state-of-the-art galaxy-formation simulations. We include simulations evolved to $z=0$ without on-the-fly radiative transfer: TNG50, TNG100, and TNG300 from the \illustrisTNG\ suite; MTNG \citep{pakmor2023millenniumtng}; ASTRID \citep{ni2022astrid}; and the three highest-resolution COLIBRE runs \citep{schaye2025colibre}, namely COLIBRE:L050m5, COLIBRE:L200m6, and COLIBRE:L400m7. We also show RHD simulations using Lagrangian or quasi-Lagrangian methods and targeting the epoch of hydrogen reionization, each with its own end redshift $z_{\rm end}$: \thesanone\ and \thesantwo\ \citep[$z_{\rm end}=5.5$;][]{Thesan1}, and the two largest Aurora simulations \citep{Aurora2017}, L050N0512 ($z_{\rm end}=6$) and L100N1024 ($z_{\rm end}=8.4$). Finally, we indicate the box size of the largest CoDa run \citep[$z_{\rm end}=4.6$;][]{lewis2022short} as a vertical line, since its fixed Eulerian mesh has no single baryonic mass resolution, and show the largest CROC run \citep[$z_{\rm end}\approx 5$;][]{croc120cMpc} at the mean baryonic mass of its uniform base grid. The quasi-Lagrangian refinement of CROC keeps the mass per resolution element approximately constant up to a maximum refinement level; the underdense IGM is therefore sampled with a finer mass resolution, while in the densest gas the mass per cell grows with density. The dotted diagonal lines correspond to a constant number of resolution elements.}
    \label{fig:volume_vs_force_resolution}
\end{figure}

\section{Technical details and global properties}
\label{sec:TechnicalDetails}

\lumina was run with the model described above on Frontier \citep{atchley2023frontier}, an exascale-class system. We used 1{,}884 compute nodes (105{,}504 CPU cores and 7{,}536 AMD Instinct MI250X accelerators), corresponding to just over $20\%$ of the machine. The peak memory footprint reached $\simeq 600\,\mathrm{TB}$, and each full checkpoint occupied $\simeq 250\,\mathrm{TB}$. By leveraging \arepo's efficient parallel I/O together with the high-performance parallel file system, we sustained peak checkpoint write rates of $\simeq 800\,\mathrm{GB\,s^{-1}}$, which is essential at this scale. The key numerical parameters of \lumina are summarized in \cref{tab:simulation_overview}.

For the dark matter we adopt a comoving Plummer-equivalent gravitational softening of $1.2\,h^{-1}\,\mathrm{kpc}$, i.e.\ $\approx 1/47$ of the mean inter-particle spacing and close to the commonly used $\sim 1/50$ choice \citep[e.g.][]{millenium2005,Iannuzzi2011,Zhang2019}. Gas elements use the standard adaptive gravitational softening of \arepo, with an effective minimum of $0.3\,h^{-1}\,\mathrm{ckpc}$.

Because of the quasi-Lagrangian nature of the code, the spatial resolution of the hydrodynamics is a direct function of the local gas density,
\begin{equation}
\Delta x = \left(\frac{m_{\rm b}}{\rho}\right)^{1/3}
\approx 0.48\left(\frac{n_{\rm H}}{\mathrm{cm^{-3}}}\right)^{-1/3}\mathrm{kpc}
\approx 83\,\Delta_{\rm b}^{-1/3}\,\mathrm{ckpc}\,,
\label{eq:spatialResolution}
\end{equation}
where $n_{\rm H}$ is the physical hydrogen number density and $\Delta_{\rm b}$ the baryonic overdensity. It varies from ${\approx}\,80\,\mathrm{ckpc}$ at the cosmic mean density to roughly one physical kpc at the star-formation threshold.

In terms of particle count, \lumina is the largest simulation ever performed with \arepo. \Cref{fig:volume_vs_force_resolution} compares its volume and mass resolution with those of other large-box cosmological simulations evolved to $z=0$ without on-the-fly radiative transfer---IllustrisTNG \citep{pillepich2018first,springel2018first}, MTNG \citep{pakmor2023millenniumtng}, ASTRID \citep{ni2022astrid}, and COLIBRE \citep{schaye2025colibre}---and with radiative-transfer simulations focused on the EoR, such as \thesan\ \citep{Thesan1}, Aurora \citep{Aurora2017}, CoDa \citep{lewis2022short}, and CROC \citep{Gnedin2014design}. Relative to \textsc{TNG}300, \lumina achieves ${\sim}3\times$ better mass resolution in a ${\sim}4.5\times$ larger volume; compared with \textsc{TNG}100 (the original calibration target of the IllustrisTNG model) it is only ${\sim}2.6\times$ coarser in mass resolution in a ${\sim}92\times$ larger box. \textsc{MTNG} has a ${\sim}3.2\times$ larger volume than \lumina but at ${\sim}9\times$ poorer mass resolution. The only comparable simulation in particle count is \textsc{ASTRID} \citep{bird2022,ni2022astrid}, which has recently been extended to $z=0$ \citep{Chen2025}; it evolves $2\times 5500^{3}$ resolution elements in a $(250\,h^{-1}\,\mathrm{cMpc})^{3}$ volume, yielding ${\sim}2\times$ finer particle mass resolution in a ${\sim}2.5\times$ smaller volume than \lumina. The combination of volume and resolution makes \lumina particularly well suited for studying galaxy formation and reionization on cosmological scales over the redshift range considered here.

Compared with the \thesan\ project, \lumina has an $\approx 144\times$ larger volume, while its mass resolution is $\approx 6\times$ coarser than that of \thesanone\ and $\approx 1.3\times$ finer than that of \thesantwo. A direct comparison with CoDa is less straightforward because its uniform grid with $8192^{3}$ cells has no constant baryonic mass resolution: it provides high spatial resolution in the diffuse IGM but substantially worse effective resolution within and around galaxies; its volume is $\approx 147\times$ smaller than that of \lumina \citep{lewis2022short}. CROC, in contrast, combines a uniform base grid with quasi-Lagrangian, mass-based refinement up to a maximum refinement level, resulting in an approximately constant mass per resolution element, similar to \thesanone in a slightly larger box; its largest volume is ${\sim}76\times$ smaller than that of \lumina. Relative to a fully quasi-Lagrangian code, the uniform base grid provides a finer sampling of the underdense IGM, whereas at the maximum refinement level the mass per cell grows with the density in the densest gas.

All existing simulations that include on-the-fly radiative transfer end at $z\geq 4.6$ and therefore do not reach into \HeII reionization. \lumina is thus unique among RHD simulations in combining a large volume with coverage of \HeII reionization.

Throughout the remainder of this paper we compare with \textsc{TNG100} (the original IllustrisTNG calibration target), \textsc{MTNG} (the largest IllustrisTNG volume), and \thesanone\ and \thesantwo\ during the EoR.

\begin{figure*}
    \centering
    \includegraphics[width=1\linewidth]{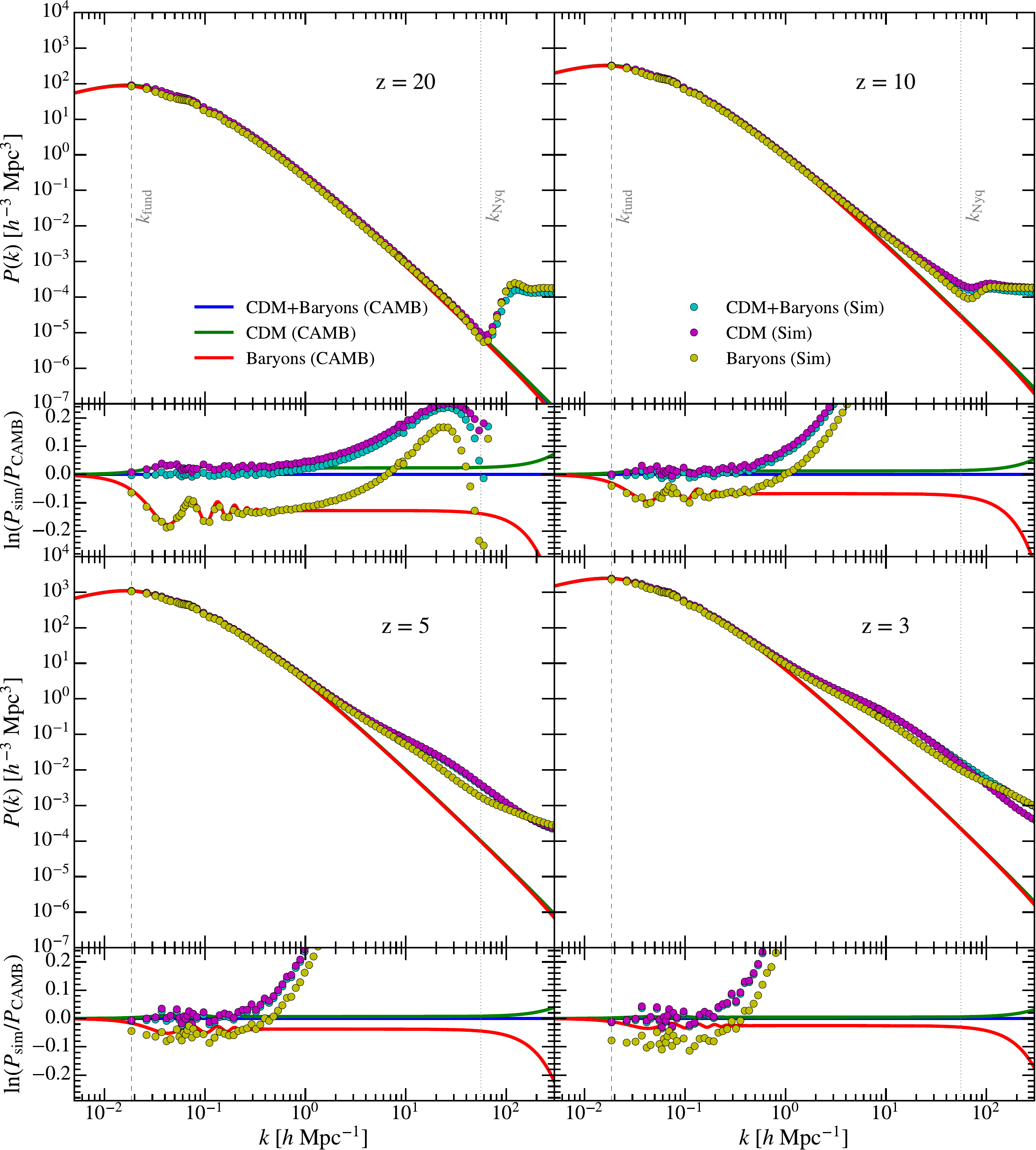}
    \caption{Matter power spectra in \lumina at $z=20$, 10, 5, and 3.
Top: absolute power spectra $P(k)$ for gas, cold dark matter (CDM), and total matter, compared to linear-theory predictions from \camb (solid curves).
Bottom: ratios of the measured spectra to the linear-theory total-matter prediction.
\lumina reproduces the expected large-scale suppression of baryonic power relative to CDM.
Vertical dashed and dotted lines mark the fundamental mode of the box and the Nyquist frequency, respectively.
}
    \label{fig:powerspec_xl}
\end{figure*}

\subsection{Matter power spectra}
We begin our analysis by examining the redshift evolution of the matter power spectrum, both for the individual species (gas and cold dark matter) and for the total matter field. The power spectra are measured on the fly using the approach of \citet{springel2018first}, summarized in \cref{subsec:powerSpectra}. \Cref{fig:powerspec_xl} shows the results at $z=20$, $10$, $5$, and $3$, together with linear-theory predictions computed with \camb.

At $z=20$, modes up to the Nyquist frequency are still in the linear or weakly non-linear regime. On large scales the baryonic power is suppressed relative to the dark-matter power by $\gtrsim 0.1\,\mathrm{dex}$, with the characteristic scale dependence imprinted by baryon acoustic oscillations, which \arepo\ tracks accurately. By $z=10$, power on small scales has become non-linear, coincident with the emergence of the first more massive haloes; the large-scale baryon suppression has decreased but is still at the ${\sim}0.1\,\mathrm{dex}$ level. At $z=5$ and $z=3$ a progressively wider range of modes becomes non-linear, and mode coupling reduces the interpretability of the residual scale dependence in the baryon-to-dark-matter power ratio.

Nevertheless, on sufficiently large scales the measured total-matter power spectrum remains in excellent agreement with linear theory computed without back-scaling, showing that relativistic species are correctly included in the background evolution.
This confirms that \lumina accurately reproduces the expected large-scale growth of structure over the redshift range considered here.

\begin{figure}
    \centering
    \includegraphics[width=1\linewidth]{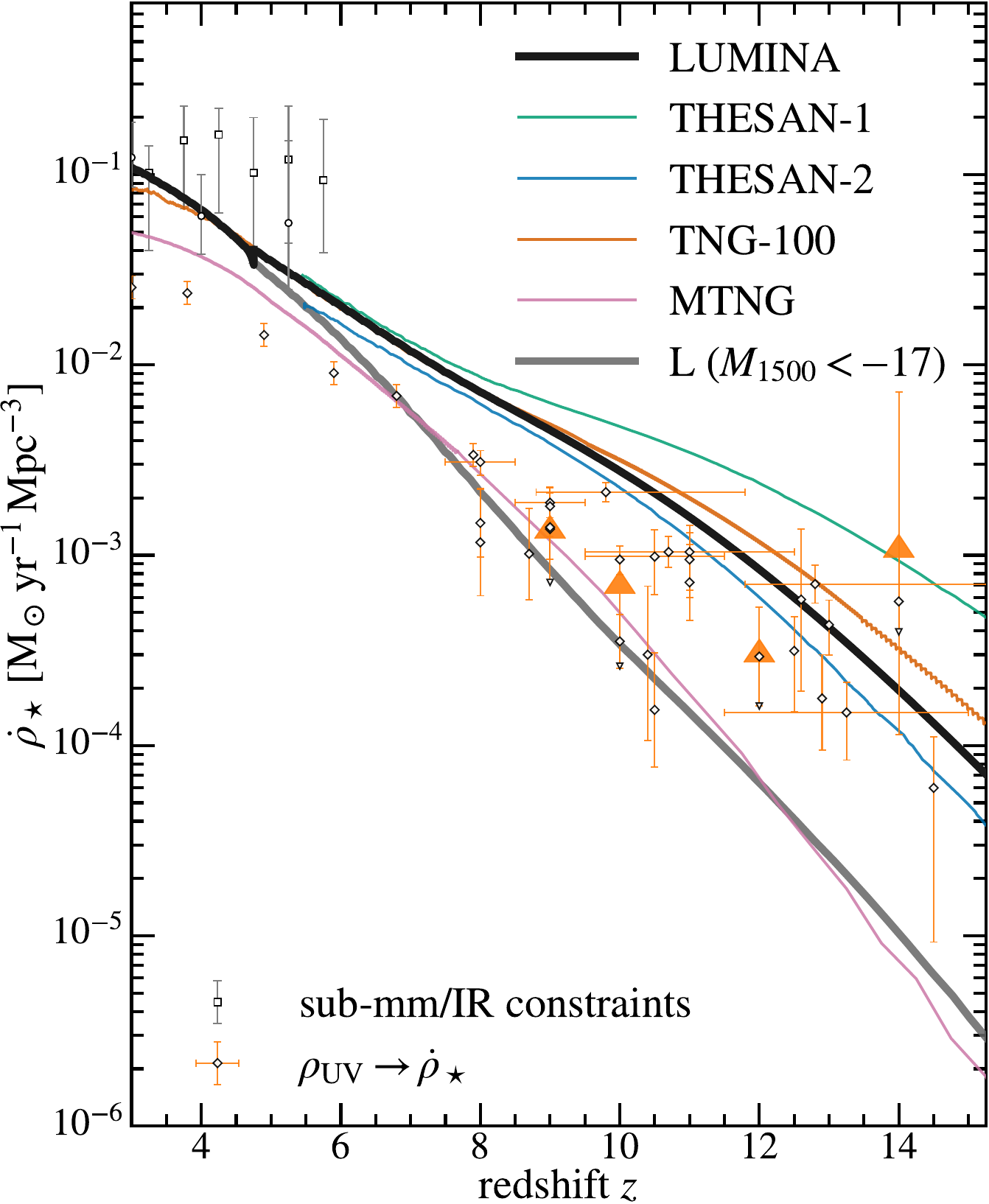}
    \caption{Evolution of the star-formation rate density as a function of redshift. We compare our results with \thesanone\ and \thesantwo\ \citep{Thesan1}, with TNG100 (the original calibration target of the IllustrisTNG project; \citealt{pillepich2018first}), and with the largest run of the MillenniumTNG project \citep{pakmor2023millenniumtng}. Higher resolution allows star formation to be resolved in smaller haloes and therefore produces a larger SFRD, especially at high redshift. \lumina agrees well with TNG100 and shows a small dip at $z=4.75$ caused by the change in our model. MTNG lies below the other simulations across the full redshift range shown owing to its coarser resolution. For a more direct comparison with observations, we additionally show the SFRD that includes only subhaloes brighter than $M_{1500}=-17$ (curve labelled ``L''); no dust extinction is applied, since at high redshift the SFRD is dominated by lower-mass galaxies. We also show observational constraints based on infrared and sub-millimetre data \citep{rowan2016star,gruppioni2020alpine}, as well as estimates based on the UV luminosity function \protect\citep{bouwens2020alma,harikane2023comprehensive,bouwens2023uv,bouwens2023evolution,mcleod2024galaxy,Donnan2023,perez2023life,donnan2024jwst,finkelstein2024complete,harikane2025jwst,whitler2025z}; the latter do not include dust extinction.}
    \label{fig:SFRD}
\end{figure}

\subsection{Star-formation rate density}
\label{sec:sfrd}

We show in \cref{fig:SFRD} the evolution of the cosmic star formation rate density (SFRD), $\dot{\rho}_\star(z)$. We compute $\dot{\rho}_\star$ directly from the instantaneous star-formation rates of all gas cells within the volume.
The large volume of \lumina strongly suppresses stochastic fluctuations from rare objects, yielding a smooth SFRD.

At $z\gtrsim 10$, the SFRD exhibits a monotonic resolution dependence: higher-resolution runs form stars earlier
and at a higher rate because they resolve star formation in lower-mass haloes and better capture the build-up of
dense, cold gas.
\lumina and TNG100 agree closely between $z \approx 9$ and $z \approx 4$, while by $z=3$ \lumina shows a slightly higher SFRD. One possible explanation is the absence of magnetic fields in \lumina, which can modestly suppress black-hole growth and thereby delay quenching. We also see at $z=4.75$ the switch to the longer depletion time in high-density gas, which leads to a brief reduction in the SFRD that persists until the gas settles to a new equilibrium.

The high-resolution \thesanone run converges with \lumina by $z \approx 7$. Interestingly, \thesantwo does not reach the SFRD of \lumina despite their similar mass resolution. This difference may be driven by the modified AGN feedback model in the \thesan suite, which quenches galaxies earlier \citep{Chittenden2025}. Finally, MTNG is the lowest-resolution simulation shown and remains systematically below the others.
For a more direct comparison to observed rest-frame UV luminosity functions, we compute on-the-fly for each FoF halo and subhalo the intrinsic non-ionizing UV continuum luminosity density at rest-frame $\SI{1500}{\angstrom}
$, denoted $L_{1500}\equiv L_\lambda(\SI{1500}{\angstrom}
)$ in units of $\si{\erg\per\second\per\angstrom}$. We obtain $L_{1500}$ by summing the contributions from all star particles, using the same BPASS v2.2.1 spectral energy distributions adopted for our stellar photon injection model.
We then convert $L_{1500}$ to an absolute AB magnitude at $\SI{1500}{\angstrom}$ via
\begin{equation}
    M_{1500} = -2.5\,\log_{10}\!\left(\frac{\lambda^2\,L_{1500}}{4\pi\,c\,(10\,\mathrm{pc})^2}\right) - 48.6 \, ,
\end{equation}
where $\lambda = \SI{1500}{\angstrom}$ and c is expressed in $\si{\angstrom\per\second}$.
To estimate a UV-selected star formation rate density, we include only subhaloes brighter than a fiducial detection limit, $M_{1500}\le -17$ \citep[e.g.][]{bouwens2015uv}, and sum their star formation rates. 
We emphasise that these magnitudes are intrinsic and do not include dust attenuation.
The SFRD gets significantly reduced at high redshift when star formation is dominated by faint galaxies but converges at $z\approx 4$ to  the full SFRD.

For comparison, we also show observationally inferred measurements of the cosmic star formation rate density, including reconstructions from sub-mm surveys \citep{gruppioni2020alpine}, infrared constraints at $z\sim1$--6 \citep{rowan2016star}, and recent UV-based determinations \citep{bouwens2020alma, harikane2023comprehensive, bouwens2023uv, bouwens2023evolution, mcleod2024galaxy, Donnan2023, perez2023life, donnan2024jwst, finkelstein2024complete, harikane2025jwst, whitler2025z}.
For the UV-based measurements, we convert the rest-frame UV luminosity density at $\SI{1500}{\angstrom}$, $\rho_{\rm UV}$, into an (unobscured) SFRD via
\begin{equation}
   \dot{\rho}_\star = K_{\rm FUV}\,\rho_{\rm UV} \, ,
\end{equation}
with
\begin{equation}
   K_{\rm FUV} = 7.2\times10^{-29}\;
   \frac{\Msun\,\mathrm{yr}^{-1}}{\mathrm{erg}\,\mathrm{s}^{-1}\,\mathrm{Hz}^{-1}} \, ,
\end{equation}
where we adopt the far-UV conversion factor from \citet{madau2014cosmic}, rescaled to a Chabrier IMF.

We stress that UV-based SFRDs derived in this way trace only the unobscured component and therefore constitute a lower limit once dust attenuation becomes non-negligible. In particular, dust corrections are already important for luminous and/or massive systems by $z\sim4$--6 \citep[e.g.][]{Fudamoto2020}, and the effective UV attenuation increases rapidly towards lower redshift \citep[e.g.][]{Bouwens2012, madau2014cosmic}.
With this caveat in mind, the unobscured SFRD predicted by \lumina is consistent with the UV-based constraints.

\begin{figure*}
    \centering
    \includegraphics[width=0.95\linewidth]{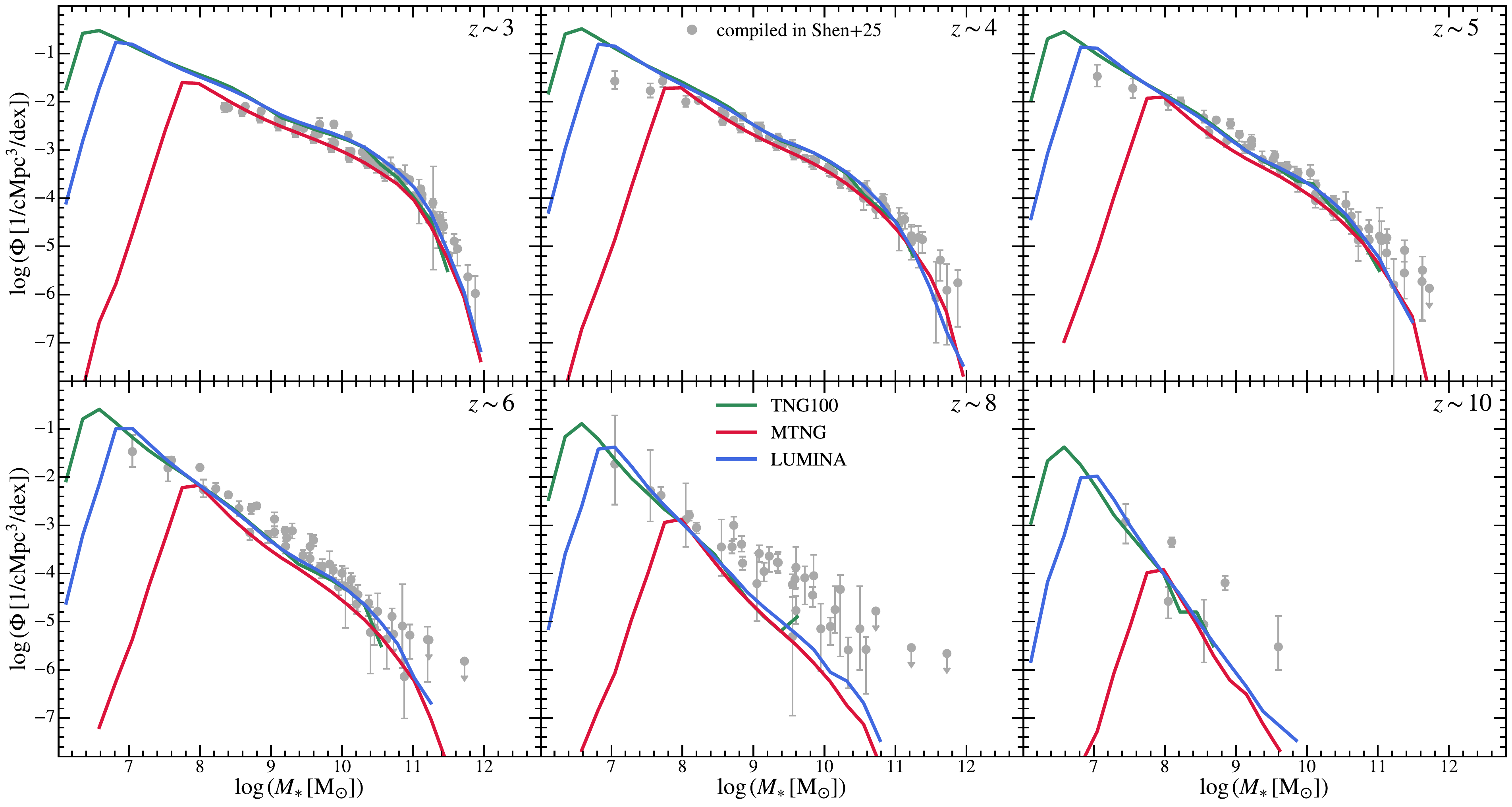}
 \caption{Galaxy stellar mass functions in \lumina at $z\simeq 3$--$10$. Results from TNG100 and MTNG are shown for reference, illustrating the impact of simulation volume and resolution across the full stellar-mass range. The three simulations show reasonable convergence within their shared dynamic range. The sharp cut-offs at the low-mass end reflect the resolution limits. We also compare the \lumina predictions with the observational compilation of \citet{Shen2025_EDE} (see the main text for the full list of references). \lumina starts to underpredict the abundance of very massive galaxies ($M_\ast\gtrsim 10^{11}\,\Msun$) at $z\gtrsim 5$, and the discrepancy becomes more pronounced at $z\gtrsim 8$. This further demonstrates that the massive galaxies found by \textit{JWST} at high redshift challenge canonical models of galaxy formation.}
 \label{fig:smf}
\end{figure*}

\begin{figure*}
    \centering
    \includegraphics[width=0.95\linewidth]{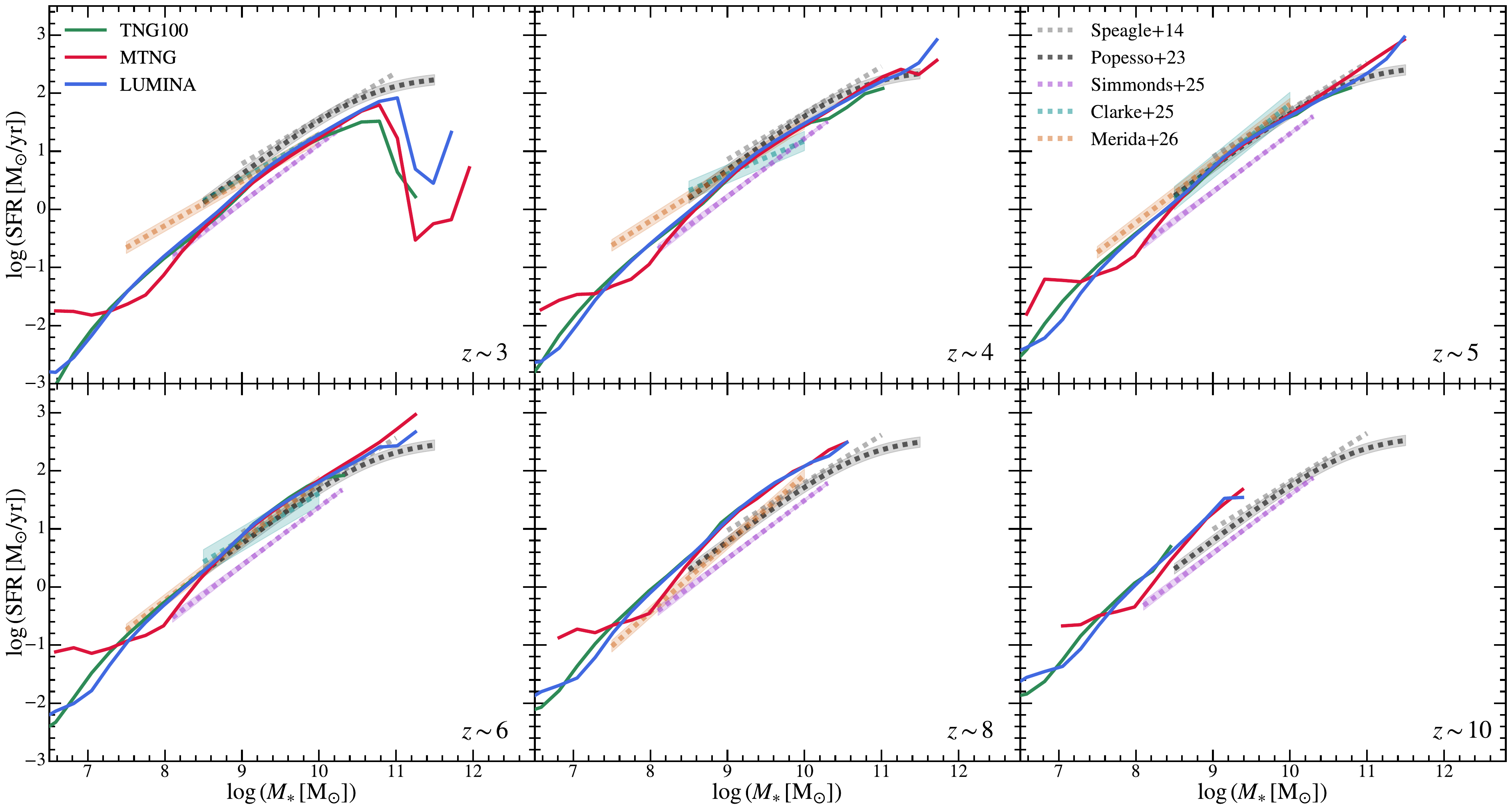}
    \caption{Galaxy star-formation main sequence in \lumina at $z\simeq 3$--$10$, with results from TNG100 and MTNG shown for reference. We compare the \lumina predictions with the latest \textit{JWST} measurements from \citet{Simmonds2025,Clarke2025,Merida2026}, together with the pre-\textit{JWST} constraints from \citet{Speagle2014,Popesso2023}; shaded bands indicate the uncertainties reported in the respective works. The comparison highlights the level of agreement between the simulations and current observational estimates over a wide range of stellar masses and redshifts. At $z\simeq 3$, galaxies above $M_\ast\sim 10^{11}\,\Msun$ begin to be quenched by AGN feedback and to deviate from the main sequence.}
    \label{fig:ms}
\end{figure*}

\section{Galaxy properties}
\label{sec:galaxies}

\begin{figure*}
    \centering
    \includegraphics[width=\linewidth]{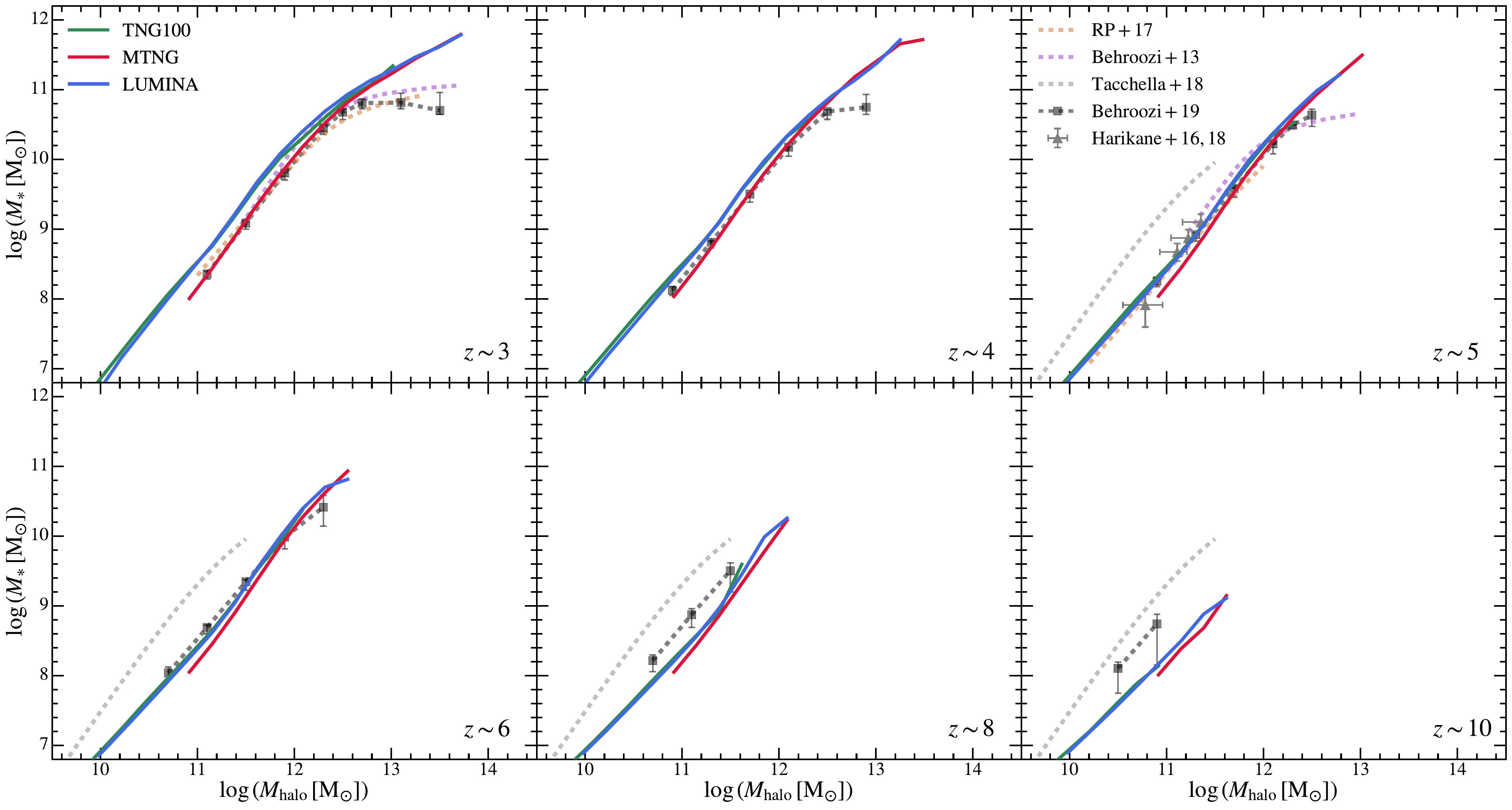}
\caption{Galaxy stellar-mass--halo-mass relations in \lumina at $z\simeq 3$--$10$, which can also be interpreted as the integrated star-formation efficiency. Results from TNG100 and MTNG are shown for reference and demonstrate that the galaxy stellar mass is converged with respect to numerical resolution. We compare the simulations with the observation-based constraints of \citet{Behroozi2013,Behroozi2019,RP2017,Harikane2016,Harikane2018,Tacchella2018}, finding reasonable agreement across redshifts. Error bars indicate the uncertainties reported by \citet{Behroozi2019} and \citet{Harikane2016,Harikane2018}. At $z\simeq 3$--$5$, however, all simulations overpredict the stellar masses in haloes more massive than ${\sim}10^{12.5}\,\Msun$, indicating that the IllustrisTNG model does not quench massive galaxies early enough. Supporting evidence comes from \textit{JWST}, which finds that IllustrisTNG does not well reproduce the abundance of quiescent galaxies at $z\gtrsim 4$ \citep[e.g.][]{deGraaff2025,Weibel2025}.}
    \label{fig:mstar_mhalo}
\end{figure*}

\begin{figure*}
    \centering
    \includegraphics[width=\linewidth]{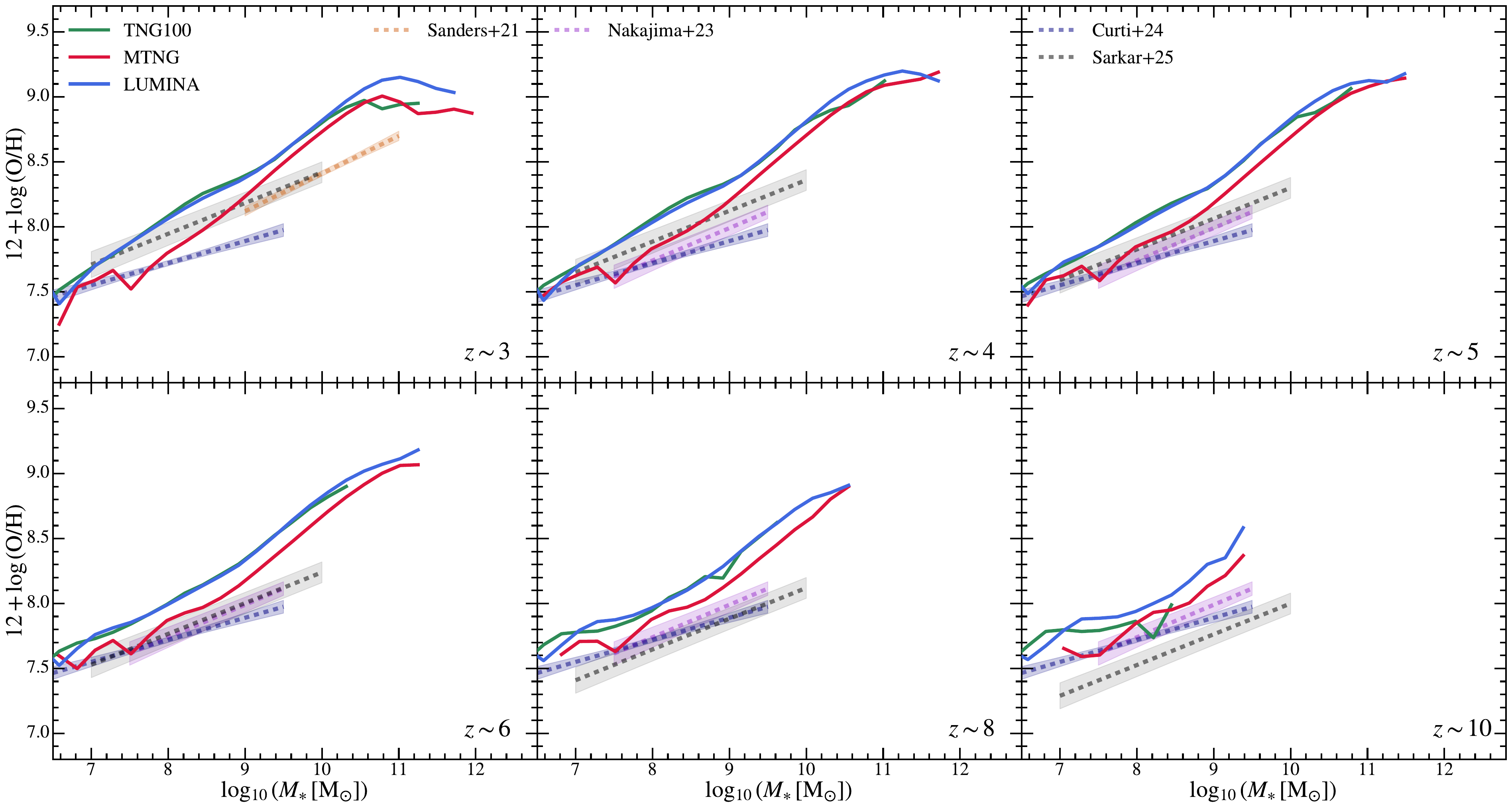}
\caption{Galaxy gas-phase mass--metallicity relations in \lumina at $z\simeq 3$--$10$, compared with TNG100 and MTNG. \lumina and TNG100 predict consistent metallicities, both of which are higher than MTNG owing to its lower resolution. The change in the metal-injection scheme at $z=4.75$ discussed in \cref{sec:methods} produces no noticeable signature in the mass--metallicity relation. We also compare with the observational constraints from \citet{Sanders2021,Nakajima2023,Curti2024,Sarkar2025}, with shaded bands indicating their reported uncertainties.}
    \label{fig:mzr}
\end{figure*}

\begin{figure*}
    \centering
    \includegraphics[width=\linewidth]{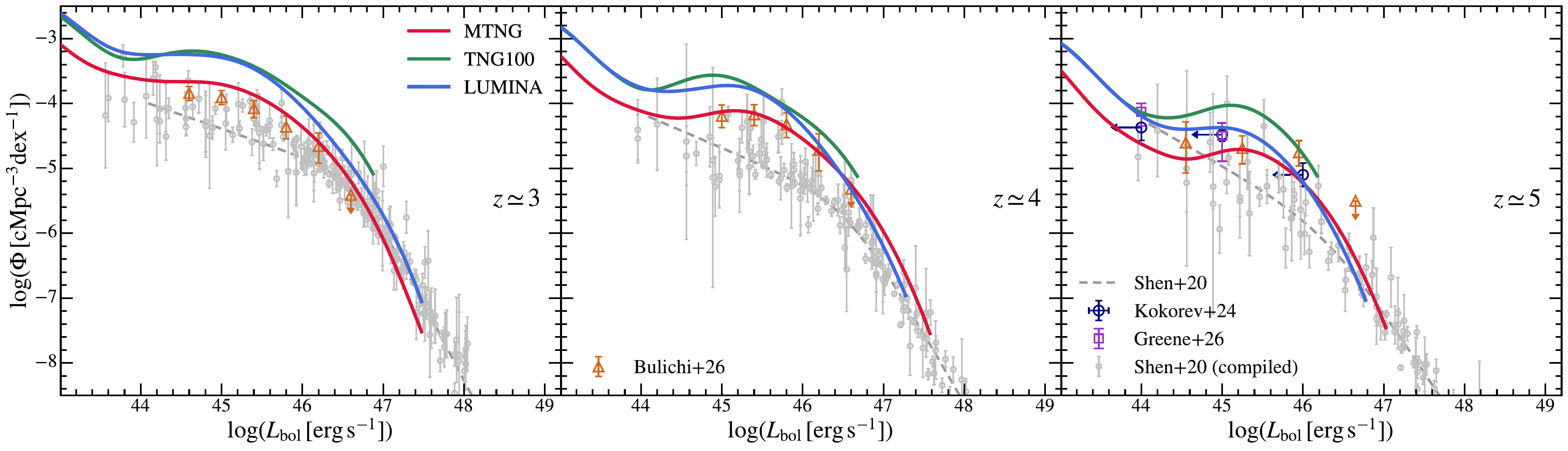}
\caption{Bolometric AGN luminosity functions in \lumina at $z\simeq 3$--$5$, compared with TNG100 and MTNG. \lumina agrees with MTNG at the bright end, while at the faint end MTNG predicts lower AGN number densities owing to its lower resolution. \lumina is more consistent with TNG100, but slightly underpredicts the abundance of faint AGN at $z\sim 5$, where BH masses are still close to the seeding scale. We compare \lumina with the multi-band observational quasar compilation of \citet{Shen2020} (transposed to the bolometric plane using their model), the \textit{JWST}/MIRI constraints from \citet{Bulichi2026}, and the LRD constraints at the faint end from \citet{Kokorev2024,Greene2026}.}
    \label{fig:bolqlf}
\end{figure*}

In this section, we present key diagnostics of the galaxy population in \lumina at $z\simeq 3$--$10$, comparing with TNG100 and MTNG and with the available observational constraints. \Cref{fig:smf} shows the galaxy stellar mass functions. Stellar mass here refers to the total mass of stellar particles within twice the stellar half-mass radius. \lumina is in nearly perfect agreement with TNG100 at the low-mass end, where the TNG100 volume still provides adequate statistics, while MTNG underpredicts the abundance of low-mass galaxies because of its coarser resolution. At the massive end, \lumina agrees broadly with MTNG, and \lumina captures the Schechter break well at $z\lesssim 6$. We compare the simulation predictions with the observational constraints compiled by \citet{Shen2025_EDE}, including pre-\textit{JWST} results from \citet{Santini2012,Ilbert2013,Song2016,Davidzon2017,Stefanon2021,Weaver2023} and more recent \textit{JWST}-based measurements from \citet{Weibel2024,WangT2024,Harvey2025,Shuntov2025}. All measurements have been converted to a \citet{Chabrier2003} IMF using the correction factors of \citet{madau2014cosmic}. Overall, the simulations reproduce the stellar mass functions well at $z\lesssim 4$. From $z\sim 5$ onwards, however, they underpredict the abundance of the most massive galaxies, $M_\ast\gtrsim 10^{11}\,\Msun$, and the discrepancy becomes more pronounced at $z\gtrsim 8$. This confirms that canonical galaxy-formation models calibrated at low redshift in $\Lambda$CDM underestimate the abundance of massive galaxies revealed by \textit{JWST} \citep[e.g.][]{Casey2024,Weibel2024,Shuntov2025}, potentially pointing to enhanced star-formation efficiency in massive systems \citep[e.g.][]{Dekel2023,Somerville2025,Shen2025}, a top-heavy stellar IMF \citep[e.g.][]{Wang2024a}, and/or extensions to the standard cosmological framework \citep[e.g.][]{Shen2024b,Shen2025_EDE}. This tension is distinct from---though potentially connected to---the overabundance of bright galaxies at $z\gtrsim 10$ \citep[e.g.][]{finkelstein2024complete}, for which bursty star formation \citep[e.g.][]{Mason2023,Shen2023,Sun2023} and the absence of dust attenuation \citep{Ferrara2023} may also play a role. However, the uncertainties in observationally inferred stellar masses may be underestimated, in which case the Eddington bias would be stronger than assumed, potentially explaining the discrepancy between model predictions and observations~\citep{Chaikin2026}.

\Cref{fig:ms} shows the galaxy star-formation rate (SFR) as a function of stellar mass in \lumina. The SFR is computed from the sum of the instantaneous SFRs of all gas cells within twice the stellar half-mass radius. We show the median SFR in each stellar mass bin, including both star-forming and quiescent galaxies, although the latter appear only at $z<4$ in \lumina. The \lumina results agree well with TNG100 and MTNG over their shared dynamic range. In MTNG, the SFR flattens at $M_\ast\lesssim 10^{8}\,\Msun$ owing to its limited mass resolution. At $z\simeq 3$, galaxies more massive than $10^{11}\,\Msun$ are significantly quenched. The characteristic stellar mass at which quenching sets in is slightly higher in \lumina than in TNG100, most likely reflecting a modestly suppressed early BH growth caused by the combination of our steeper star-formation efficiency at $z>4.75$ and the absence of magnetic fields, which would otherwise contribute to the pressure correction in the BH accretion model. We compare these results with the observational star-formation main sequence from \citet{Speagle2014,Popesso2023,Clarke2025,Simmonds2025,Merida2026}. Because the effective equation of state and our star-formation recipe may suppress short-timescale variability of the SFR, we adopt UV-based SFR measurements for the comparison. Overall, we find good agreement between the simulated main sequence in \lumina and the observations.

\begin{figure*}
    \centering
    \includegraphics[width=1\linewidth]{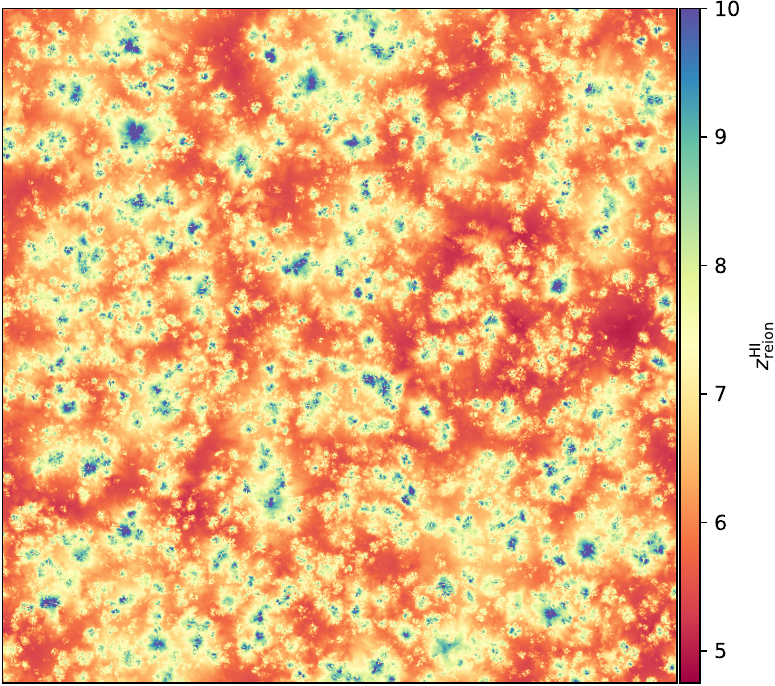}
    \caption{Hydrogen reionization redshift in a single-cell slice through the midplane of the simulation box.
    We define the reionization redshift of each grid cell as the last time at which the local neutral-hydrogen fraction exceeded $1\%$, using a Cartesian grid with resolution $1280^3$.}
    \label{fig:z_reion_slice}
\end{figure*}

In \cref{fig:mstar_mhalo}, we present the stellar-mass--halo-mass relation, which is equivalently a measure of the integrated star-formation efficiency of galaxies. The halo mass used here is the total mass of particles that are gravitationally bound to a subhalo (for central galaxies, they are roughly the same as the $M_{200{\rm c}}$ used in other parts of the paper). We compare the predictions of \lumina, TNG100, and MTNG with observation-based constraints from \citet{Behroozi2013,Harikane2016,Harikane2018,RP2017,Tacchella2018,Behroozi2019}. At $z\gtrsim 8$, \lumina slightly underpredicts the stellar masses of galaxies relative to the constraints from \citet{Behroozi2019}, while the results from \citet{Tacchella2018} appear somewhat higher than the rest of the comparison set at these redshifts. One should be cautious that these empirical constraints, based on pre-\jwst observations, are mostly extrapolations of best-fit models at lower redshifts. Nevertheless, the relatively low star-formation efficiency in low-mass haloes in \lumina likely contributes to the underprediction of cosmic SFRD shown in \cref{fig:SFRD}. At $z\simeq 3$, \lumina overpredicts the stellar masses of galaxies hosted by haloes with $M_{\rm halo}\gtrsim 10^{13}\,\Msun$. This indicates that the IllustrisTNG model does not quench massive galaxies early enough~\citep[e.g.][]{Nanayakkara2025,Lagos2025}. Supporting evidence has been revealed by \jwst that IllustrisTNG does not well reproduce the abundance of quiescent galaxies at $z\gtrsim 4$~\citep[e.g.][]{deGraaff2025,Weibel2025}.

In \cref{fig:mzr} we present the gas-phase mass--metallicity relation of galaxies. We convert the observed oxygen abundances to total gas metallicities using $\log(Z_{\rm gas}/Z_\odot) = 12 + \log({\rm O/H}) - 8.69$, with $Z_\odot = 0.0139$ from \citet{Asplund2021}. \lumina agrees well with TNG100, while MTNG produces lower metallicities across the full stellar-mass range owing to its lower resolution. Compared with the observational constraints of \citet{Sanders2021,Nakajima2023,Curti2024,Sarkar2025}, the simulations tend to overpredict the gas-phase metallicities at $M_\ast \gtrsim 10^{10}\,\Msun$.

\begin{figure*}
    \centering
    \includegraphics[width=1\linewidth]{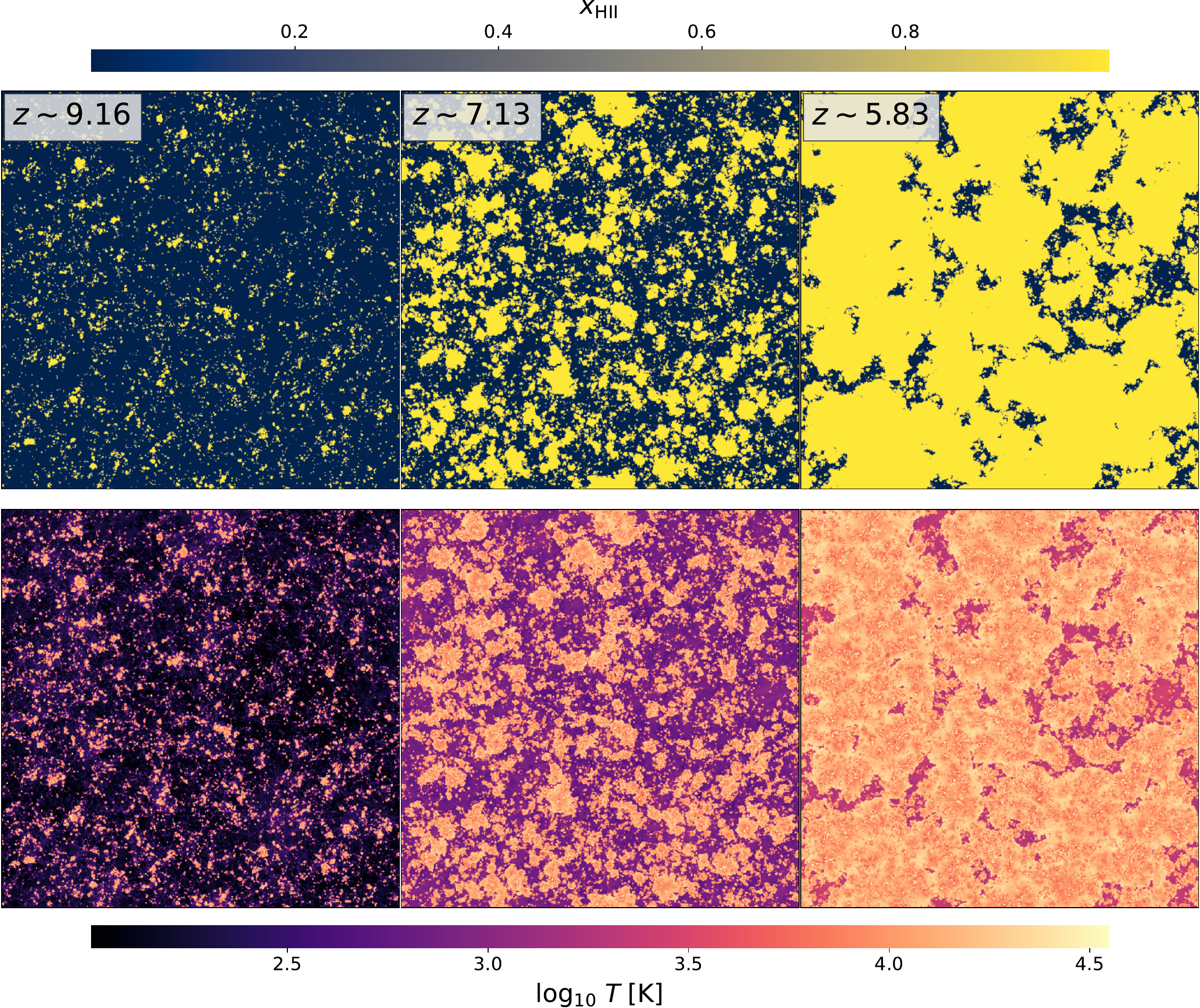}
  \caption{Volume-weighted \HII fraction (top) and gas temperature (bottom) in a single-cell slice through the midplane of the \lumina\ volume, computed on the $1280^{3}$ Cartesian grid. The columns correspond to the early ($\langle x_{\HII}\rangle_{V}=0.1$), intermediate ($\langle x_{\HII}\rangle_{V}=0.5$), and late ($\langle x_{\HII}\rangle_{V}=0.9$) stages of reionization. At early times, reionization proceeds through small ionized bubbles around individual galaxies that subsequently grow and merge. Ionized regions are heated to $T\simeq 10^{4}\,\mathrm{K}$, while the neutral regions experience only modest pre-heating, primarily from X-ray photons.}
    \label{fig:hydrogen_reionization_hii_and_temp}
\end{figure*}

In \cref{fig:bolqlf} we show the bolometric AGN luminosity functions in the simulations at $z\simeq 3$--$5$; we refer the reader to \cite{Shen2026} for details of how the bolometric luminosity function is derived. \lumina matches TNG100 at the faint end and MTNG at the bright end, as expected from their relative resolution and volume. We compare the simulation predictions with the pre-\textit{JWST} multi-band constraints of \citet{Shen2020} and find good agreement at $L_{\rm bol}\gtrsim 10^{46}\,{\rm erg\,s^{-1}}$. The overprediction at the low-luminosity end is not entirely unexpected, since those constraints are derived primarily from X-ray surveys. The recent discovery of X-ray-faint LRDs and mid-IR \textit{JWST} surveys of AGN~\citep{Bulichi2026} suggest that such X-ray-based constraints are likely incomplete. We also compare with recent observations of LRDs at $z\simeq 5$ from \citet{Greene2026} and \citet{Kokorev2024}, for which the inferred $L_{\rm bol}$ values should be interpreted as upper limits because canonical optical bolometric corrections have been applied. Overall, we find encouraging agreement.

Across all of these diagnostics, \lumina is broadly consistent with the IllustrisTNG expectations and with the currently available observational constraints, demonstrating that the galaxy-formation model performs well at the resolution and volume of this simulation.

\section{Evolution of the intergalactic medium}
\label{sec:IGM}
The intergalactic medium is reshaped by the last two major phase transitions of the cosmic baryons: hydrogen and helium reionization. Beyond changing the ionization fractions, these transitions inject energy via photoheating and leave clear, long-lived signatures on the thermal state of the low-density gas. \lumina follows both processes self-consistently across its full $500\,\mathrm{cMpc}$ volume, capturing the topology, timing, and thermal impact of each transition. In this section we first analyse the topology and progression of \HI reionization (\cref{subsec:HIreionization}), then turn to \HeII reionization (\cref{subsec:heliumReionization}), examine the resulting thermal history of the low-density IGM (\cref{subsec:IGMthermal}), and close with the integrated Thomson optical depth to the CMB (\cref{subsec:tauCMB}) as a consistency check on the combined reionization history.

\begin{figure}
    \centering
    \includegraphics[width=1\linewidth]{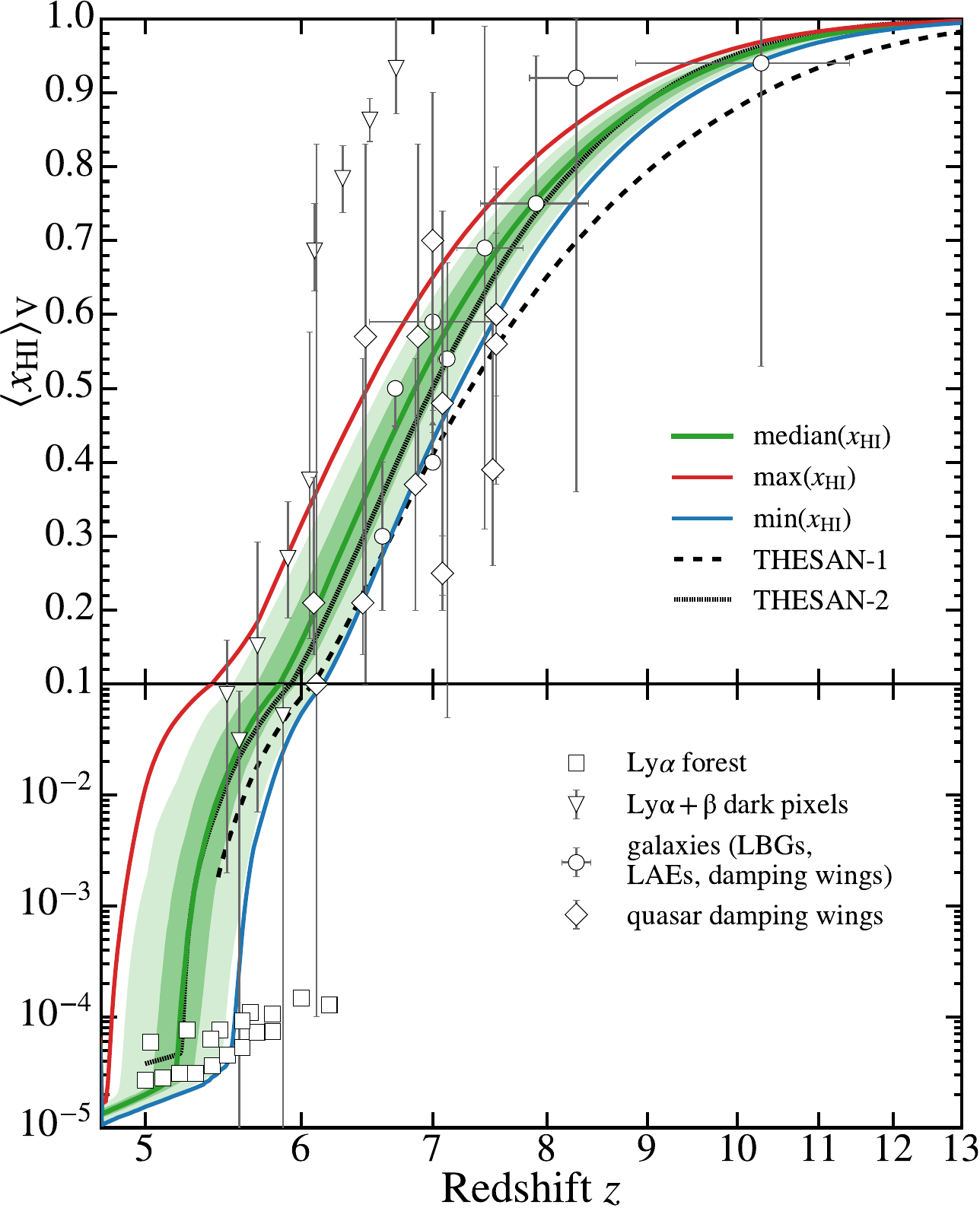}
 \caption{Evolution of the volume-weighted neutral-hydrogen (\HI) fraction. We split the full simulation box into 125 sub-boxes of side length $100\,\mathrm{cMpc}$ and compute the reionization history of each independently. The median \HI fraction across the sub-boxes is shown, together with the central 68.3\% and 95.4\% percentile ranges (shaded bands), as well as the maximum (red) and minimum (blue) values. We also show the box-averaged values from the \thesanone\ and \thesantwo\ simulations \protect\citep{Thesan1}, together with observational constraints from quasar damping wings \citep{Banados2018,Davies2018,Yang2020-dw,Durovcikova2020,Wang2021,Durovcikova2024}, galaxy damping wings \citep{Mason2018,Ouchi2010,SobacchiMesinger2015,Mason2019,Ning2022,Umeda2023}, the Ly$\alpha$ forest \citep{fan2006constraining,Yang2020,Bosman2021}, and Ly$\alpha+\beta$ dark pixels \citep{McGreer2015,Jin2023}.}
    \label{fig:x_HI_reionization}
\end{figure}

\begin{figure}
    \centering
    \includegraphics[width=1\linewidth]{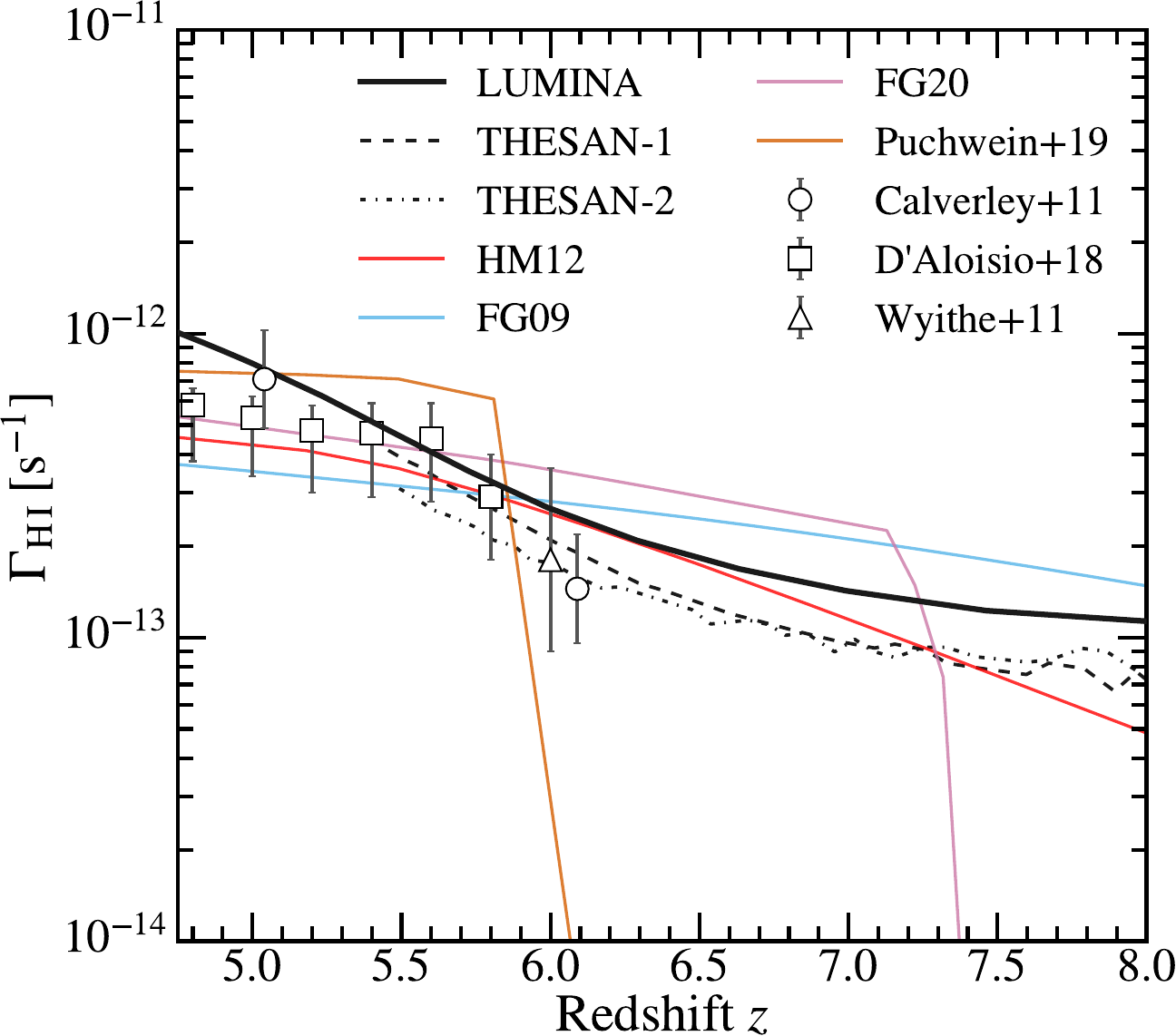}
 \caption{Evolution of the volume-weighted \HI photoionization rate $\Gamma_{\rm HI}$, measured in ionized regions ($x_{\rm HII}>0.5$) and restricted to gas with $n_{\rm H}<0.106\,\mathrm{cm^{-3}}$. We show \lumina\ together with \thesanone\ and \thesantwo\ \citep{ThesanEnrico}, the UV-background models of \citet[December 2011 update]{FG09}, \citet{faucher2020cosmic}, \citet{Haardt2012}, and \citet{puchwein2019consistent}, and the observational constraints of \citet{calverley2011measurements}, \citet{d2018large}, and \citet{wyithe2011near}.}
    \label{fig:GammaHI_vs_z}
\end{figure}

\begin{figure*}
    \centering
    \includegraphics[width=1\linewidth]{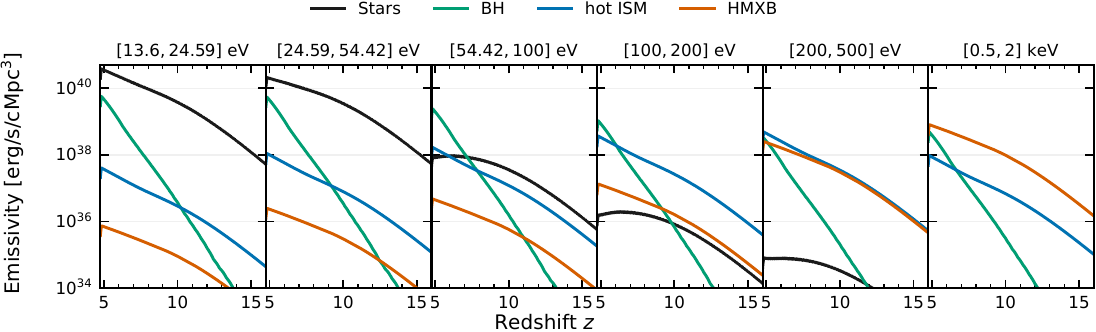}
     \includegraphics[width=1\linewidth]{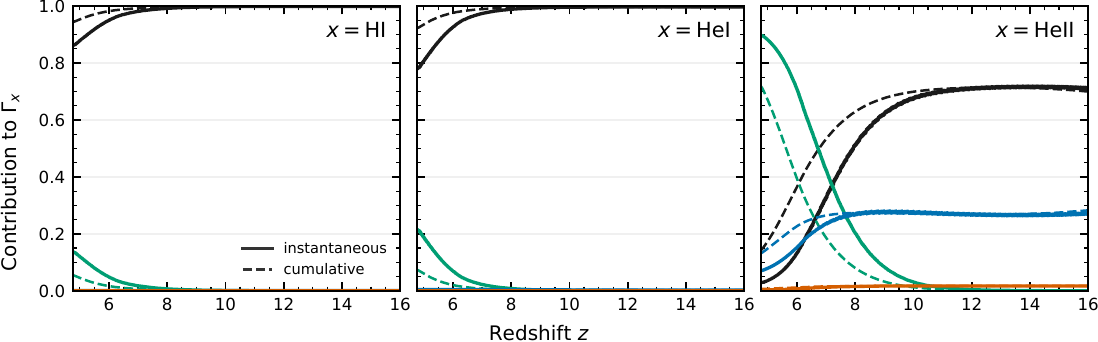}
  \caption{\textit{Top:} global emissivity for each radiation bin and source class as a function of redshift $z$. \textit{Bottom:} fractional contribution of each source class to the photoionization rate of \HI, \HeI, and \HeII. We show the instantaneous value (solid) as well as the cumulative contribution (dashed). Stars dominate the \HI- and \HeI-ionizing bins over the redshift range shown; AGN dominate the \HeII-ionizing bin below $z\simeq 8$, and their contribution to the other bins rises rapidly towards lower redshift. At high redshift the X-ray bins are dominated by the shock-heated ISM and HMXBs.}
 
    \label{fig:radiation_input_hydrogen}
\end{figure*}

\subsection{Hydrogen reionization}
\label{subsec:HIreionization}
The large box size of \lumina makes the spatial inhomogeneity of \HI reionization directly visible. \Cref{fig:z_reion_slice} shows the redshift at which each cell of a $1280^{3}$ Cartesian grid (constructed from the renders described in \cref{subsec:uniformGrid}) contains more than $1\%$ \HI for the last time. Reionization proceeds inside-out around the most massive sources: the first ionized regions appear already at $z\simeq 10$ (blue), while underdense regions---including a prominent void on the right-hand side of the slice---remain neutral much longer (red). The bulk of the volume reionizes between $z\simeq 8$ and $z\simeq 6$ (yellow), with strong cell-to-cell variations across the box.

The corresponding temporal sequence is shown in \cref{fig:hydrogen_reionization_hii_and_temp}, which displays the volume-weighted \HII fraction (top row) and the gas temperature (bottom row) in the same midplane slice at $\langle x_{\HII}\rangle_{V} = 0.1$, $0.5$, and $0.9$. Reionization begins as a population of small ionized bubbles surrounding individual galaxies that subsequently grow and merge, leaving only a handful of neutral voids at late times. Inside ionized regions the gas is photoheated to $T\gtrsim 10^{4}\,\mathrm{K}$, whereas the surrounding neutral IGM is initially cold and is only pre-heated by X-rays before being fully ionized later. A more detailed analysis of the reionization topology, including bubble sizes and shapes, will be presented in a forthcoming paper.

Because reionization is so spatially variable, a single volume-averaged history over the whole domain is of limited interpretive value: a single persistent neutral island can bias the global mean in ways that smaller boxes cannot reveal. To quantify the reionization history more robustly, we partition \lumina into $5^{3}=125$ sub-boxes of side length $100\,\mathrm{cMpc}$, comparable in size to the volumes used in CoDa \citep{ocvirk2016cosmic}, CROC \citep{croc120cMpc}, and \thesan\ \citep{Thesan1}. \Cref{fig:x_HI_reionization} shows the resulting volume-weighted neutral-hydrogen fraction in each sub-box, summarized by the median, the central $68.3\%$ and $95.4\%$ ranges, and the running minimum and maximum. \lumina exhibits a late end to reionization, as required by the high Ly$\alpha$-forest opacity at $z\approx 5.5$ \citep[e.g.][]{Kulkarni2019,keating2020long,ThesanEnrico}: the median sub-box is fully ionized by $z=5.2$, and the last neutral patches do not disappear until $z=4.75$. The median history closely follows \thesantwo, while \thesanone\ reaches full ionization slightly earlier and starts to ionize earlier, reflecting its higher resolution and its ability to capture smaller haloes. The agreement with the \thesan\ project is reassuring given our different treatment of radiation on the effective equation of state and the correspondingly different escape-fraction calibration (see \cref{subsec:diffThesan}). \lumina is also consistent with the bulk of the observational compilation shown in the figure---quasar damping wings \citep{Banados2018,Davies2018,Yang2020-dw,Durovcikova2020,Wang2021,Durovcikova2024}, galaxy damping wings \citep{Mason2018,Ouchi2010,SobacchiMesinger2015,Mason2019,Ning2022,Umeda2023}, the Ly$\alpha$ forest \citep{fan2006constraining,Yang2020,Bosman2021}, and Ly$\alpha{+}\beta$ dark pixels \citep{McGreer2015,Jin2023}---which is partly a direct consequence of the escape-fraction calibration. We caution that small neutral regions may persist below $z=5$ in part because the reduced speed of light delays the irradiation of remote voids; a more direct comparison with observations, including measurements of the mean free path of ionizing photons and mock Ly$\alpha$-forest analyses, will be presented in future work.

The strength of the ionizing UV background within already-ionized regions is captured by the hydrogen photoionization rate,
\begin{equation}
\Gamma_{\mathrm{HI}}
= \tilde{c}\sum_{i=1}^6 N_i \sigma_{i, \mathrm{HI}} \, ,
\label{eq:GammaHI}
\end{equation}
where $\tilde{c}$ is the reduced speed of light, $N_i$ is the photon density in bin $i$, and $\sigma_{i,\mathrm{HI}}$ is the corresponding cross-section for neutral hydrogen. We compute $\Gamma_{\mathrm{HI}}$ in post-processing from the full snapshots as a volume-weighted average over Voronoi cells that are predominantly ionized ($x_{\mathrm{HI}}<0.5$) and have densities below the effective equation-of-state threshold (i.e.\ excluding star-forming gas). \Cref{fig:GammaHI_vs_z} compares the resulting $\Gamma_{\mathrm{HI}}(z)$ with the values from \thesanone\ and \thesantwo, with several uniform UV-background models \citep{FG09,faucher2020cosmic,Haardt2012,puchwein2019consistent}, and with Ly$\alpha$-forest constraints \citep{calverley2011measurements,d2018large,wyithe2011near}. A direct comparison with UVB models during reionization should be interpreted with caution: those models assume a spatially uniform background, while the simulated $\Gamma_{\mathrm{HI}}$ is strongly bimodal, with high values inside ionized bubbles and nearly zero in the neutral IGM, so the inferred mean depends on the chosen \HI threshold, particularly during the early stages. Within these caveats, our results are consistent with the observational constraints and lie slightly higher than the \thesan\ suite, possibly because of our different treatment of the effective equation of state.

The source budget responsible for reionization is shown in \cref{fig:radiation_input_hydrogen}, which decomposes the energy injection rate and the resulting photoionization rates into source classes (stellar, AGN, HMXB, and hot ISM) for each frequency bin. Globally, \HI and \HeI reionization are driven by stellar emission, with AGN contributing only at the per-cent level (around $14\%$ at $z=5$), in agreement with previous studies \citep[e.g.][]{madau1999radiative,kulkarni2019evolution,jiang2022definitive,matsuoka2023quasar}. AGN dominate the \HeII-ionizing band below $z\simeq 7$. In the $100$--$200\,\mathrm{eV}$ bin the shock-heated ISM is the leading emitter, while HMXBs dominate the highest-energy X-ray bin; together these channels supply most of the X-ray pre-heating of the neutral IGM seen in \cref{fig:hydrogen_reionization_hii_and_temp}. By $z=4.75$, AGN take over the combined emissivity above $100\,\mathrm{eV}$ as well, motivating our choice to retain only AGN as sources of photons with $h\nu>54.42\,\mathrm{eV}$ at $z\le 4.75$.

\begin{figure*}
    \centering
    \includegraphics[width=1\linewidth]{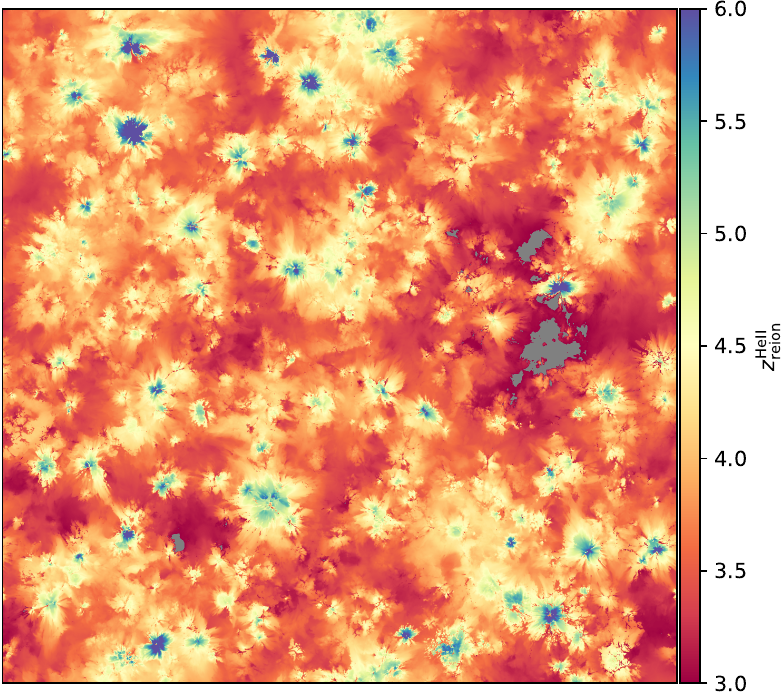}
\caption{\HeII reionization redshift in a single-cell slice through the midplane of the simulation box. We define the reionization redshift of each grid cell as the last time at which its volume-weighted \HeIII fraction fell below $90\%$, evaluated on a $1280^{3}$ Cartesian grid (equivalent to a top-hat smoothing on $390\,\mathrm{ckpc}$ cubes). The grey regions are not yet ionized at $z=3$.}
    \label{fig:z_reion_slice_he}
\end{figure*}

\begin{figure*}
    \centering
    \includegraphics[width=1\linewidth]{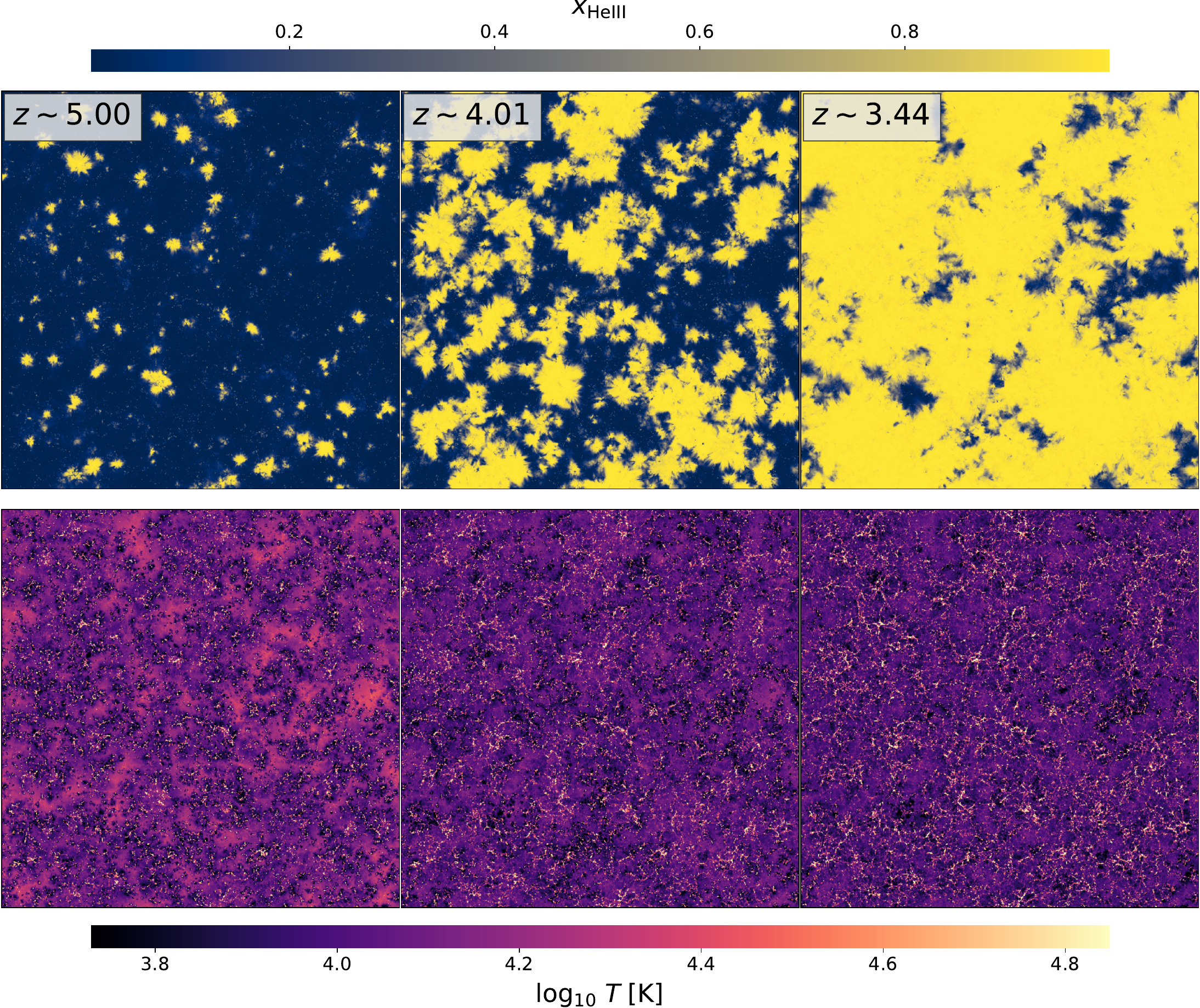}
\caption{Volume-weighted \HeIII fraction (top) and gas temperature (bottom) in a single-cell slice through the midplane of the \lumina\ volume, computed on the $1280^{3}$ Cartesian grid. The columns correspond to the early ($\langle x_{\HeIII}\rangle_{V}=0.1$), intermediate ($\langle x_{\HeIII}\rangle_{V}=0.5$), and late ($\langle x_{\HeIII}\rangle_{V}=0.9$) stages of reionization. Unlike \HI reionization, \HeII reionization is driven by a smaller number of large bubbles powered by luminous AGN.}
    \label{fig:helium_reionization_heiii_and_temp}
\end{figure*}

\begin{figure}
    \centering
    \includegraphics[width=1\linewidth]{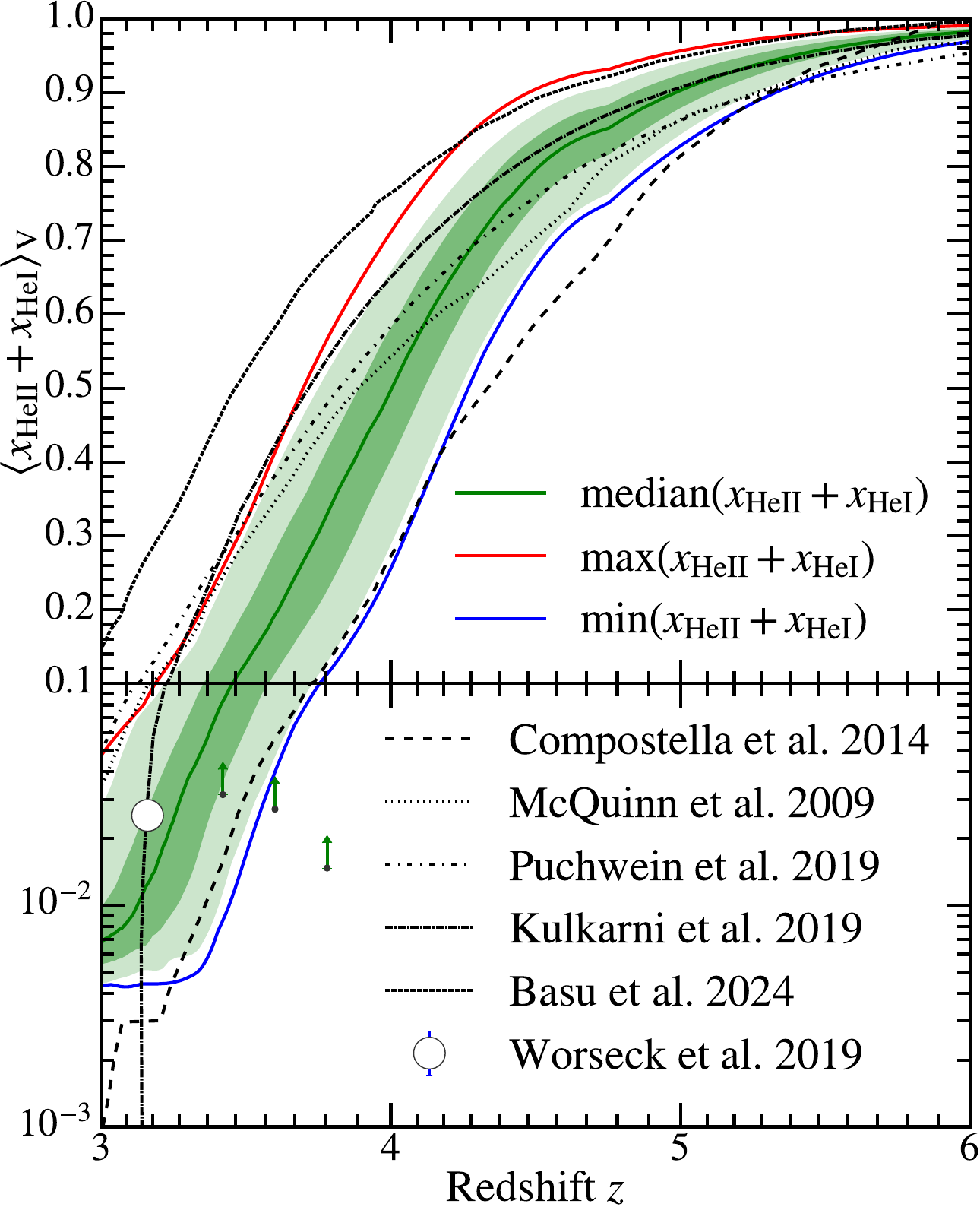}
\caption{Redshift evolution of the volume-averaged sum of the neutral and singly ionised helium fractions in \lumina\ (green). We partition the simulation volume into 125 sub-boxes of side length $100\,\mathrm{cMpc}$ and compute the reionization history of each independently. The solid green line shows the median across sub-boxes, and the shaded bands mark the central $68.3\%$ and $95.4\%$ intervals. The red and blue curves trace the maximum and minimum of $x_{\mathrm{He\,I}}+x_{\mathrm{He\,II}}$ across sub-boxes at each redshift. Black curves with different line styles show theoretical predictions from \protect\citet{mcquinn2009he,compostella2014agn,kulkarni2019evolution,puchwein2019consistent,basu2024helium}; since most of these studies focusing on \HeII reionization report only $x_{\mathrm{He\,II}}$, we assume $x_{\mathrm{He\,I}}=0$ for them. This choice allows us to separate the reionization of \HeI from that of \HeII in \lumina, which overlap at $z\approx 6$. Observational estimates from \protect\citet{worseck2019evolution} are also shown.}
    \label{fig:he_ii_fraction}
\end{figure}

\begin{figure}
    \centering
    \includegraphics[width=1\linewidth]{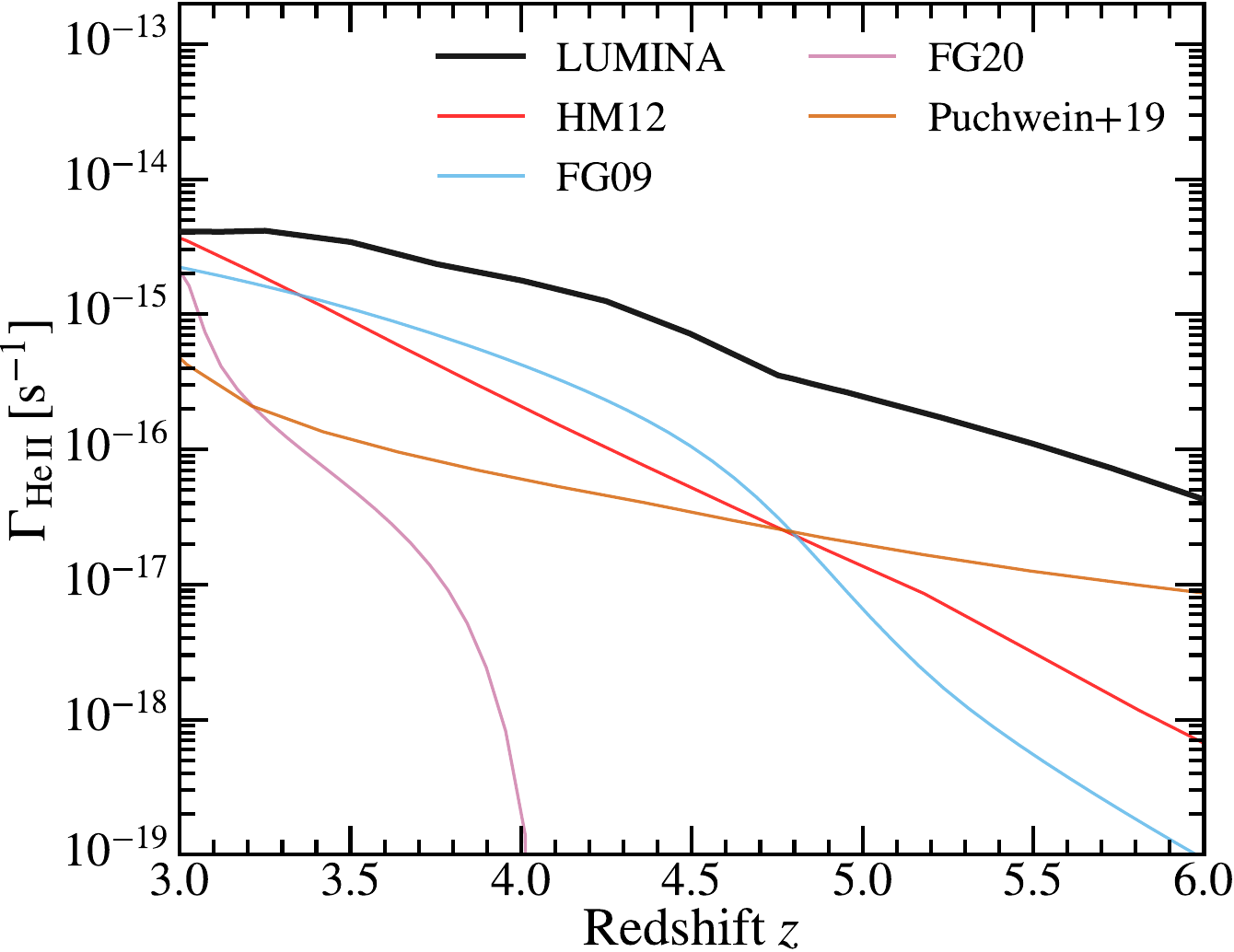}
    \caption{Evolution of the volume-weighted \HeII photoionization rate, $\Gamma_{\rm HeII}$, measured from gas cells with $n_{\rm H}<0.106\,\mathrm{cm^{-3}}$. For comparison we also show the uniform UV-background models of \protect\citet{Haardt2012}, \protect\citet{FG09}, \protect\citet{faucher2020cosmic}, and \protect\citet{puchwein2019consistent}. The assumption of a uniform UVB makes a direct quantitative comparison difficult before the end of \HeII reionization.}
    \label{fig:GammaHeII_vs_z}
\end{figure}

\subsection{Helium reionization}
\label{subsec:heliumReionization}
\HeII reionization proceeds under a substantially different source budget. \Cref{fig:z_reion_slice_he} shows, on the same $1280^{3}$ Cartesian grid, the redshift at which each cell is below $90\%$ \HeIII for the last time. This threshold is less stringent than the one used for \HI because the higher \HeIII recombination rate leaves a non-negligible residual \HeII fraction after reionization, so a stricter cut would be too sensitive to post-reionization fluctuations. The most luminous quasars ionize their surroundings as early as $z\simeq 6$ (blue); the global midpoint occurs near $z\simeq 4$ (yellow), and the last extended \HeII patches disappear by $z\approx 3$ (red), although one large region on the right of the slice remains not fully reionized by the end of the simulation (grey). Both the ionized bubbles and the surviving \HeII regions are systematically larger than during \HI reionization, reflecting the rarity and high luminosity of the AGN sources.

The corresponding evolution of the gas state is shown in \cref{fig:helium_reionization_heiii_and_temp}: the top row displays the volume-weighted \HeIII fraction and the bottom row the temperature in the same slice at $\langle x_{\mathrm{HeIII}}\rangle_{V} = 0.1$ ($z\approx 5$), at the midpoint ($z\approx 4.01$), and at $\langle x_{\mathrm{HeIII}}\rangle_{V} = 0.9$ ($z\approx 3.44$). The onset of \HeII reionization overlaps with the tail of \HI reionization, and regions that have only just been hydrogen-ionized at $z\approx 5$ are still hot. Adiabatic expansion subsequently cools the IGM, while the propagating \HeII fronts deposit additional thermal energy. The interplay between these two processes is most visible in the centre-right region of the slice: it is among the hottest at $z\approx 5$ but among the coldest by $z\approx 3.44$, because it was heated late during \HI reionization but had not yet been swept by a \HeII front.

To quantify the timing and variance of \HeII reionization, we again partition the volume into $5^{3}=125$ sub-boxes of side length $100\,\mathrm{cMpc}$ and compute the combined volume-weighted \HeI and \HeII fraction in each. \Cref{fig:he_ii_fraction} shows the median, the central $68.3\%$ and $95.4\%$ ranges, and the running minimum and maximum across sub-boxes. Consistent with the slice maps, \HeII reionization starts around $z\simeq 6$ and reaches its midpoint at $z\simeq 4$. The earliest sub-box drops below $x_{\mathrm{HeI}} + x_{\mathrm{HeII}} = 0.01$ by $z\simeq 3.4$, the median crosses the same threshold near $z\simeq 3$, and a small number of sub-boxes still retain up to ${\sim}4\%$ \HeII at the end of the simulation; as in the \HI case, these residual reservoirs are expected to be fully ionized at later times. We overlay results from earlier simulations and from semi-analytic or post-processing models \citep{mcquinn2009he,compostella2014agn,kulkarni2019evolution,puchwein2019consistent,basu2024helium}; for a consistent comparison we assume $x_{\mathrm{HeI}}\approx 0$ during \HeII reionization, since several of those works report only $x_{\mathrm{HeII}}$. Our reionization is slightly earlier---and in particular faster---than most of these models, with the exception of \citet{compostella2014agn}, but it is consistent with the current \HeII Ly$\alpha$-forest constraints from \citet{worseck2016early,worseck2019evolution}. Because inferring $x_{\mathrm{HeII}}$ from \HeII Ly$\alpha$ absorption requires forward-modelling mock spectra to handle noise, resolution, continuum uncertainties, and selection effects \citep[e.g.][]{worseck2016early,worseck2019evolution}, we defer a detailed like-for-like comparison to a future paper based on synthetic skewers through \lumina.

The build-up of the \HeII-ionizing background is summarized in \cref{fig:GammaHeII_vs_z}, where we plot the volume-averaged photoionization rate $\Gamma_{\mathrm{HeII}}$ computed from all Voronoi cells with $n_{\rm H}<0.106\,\mathrm{cm^{-3}}$. To facilitate the comparison with the spatially uniform UVB models of \citet{FG09,Haardt2012,faucher2020cosmic,puchwein2019consistent}, we do not additionally mask on ionization state. The simulated values are systematically higher than the uniform-UVB predictions, consistent with our somewhat earlier \HeII reionization. However, $\Gamma_{\mathrm{HeII}}$ is strongly bimodal---approximately constant inside ionized regions and nearly zero in \HeII-dominated gas---so the volume-averaged evolution largely tracks the progression of \HeII reionization itself; as in the \HI case, a quantitative comparison with uniform UVB models before overlap should be interpreted with caution.

\begin{figure}
    \centering
    \includegraphics[width=\linewidth]{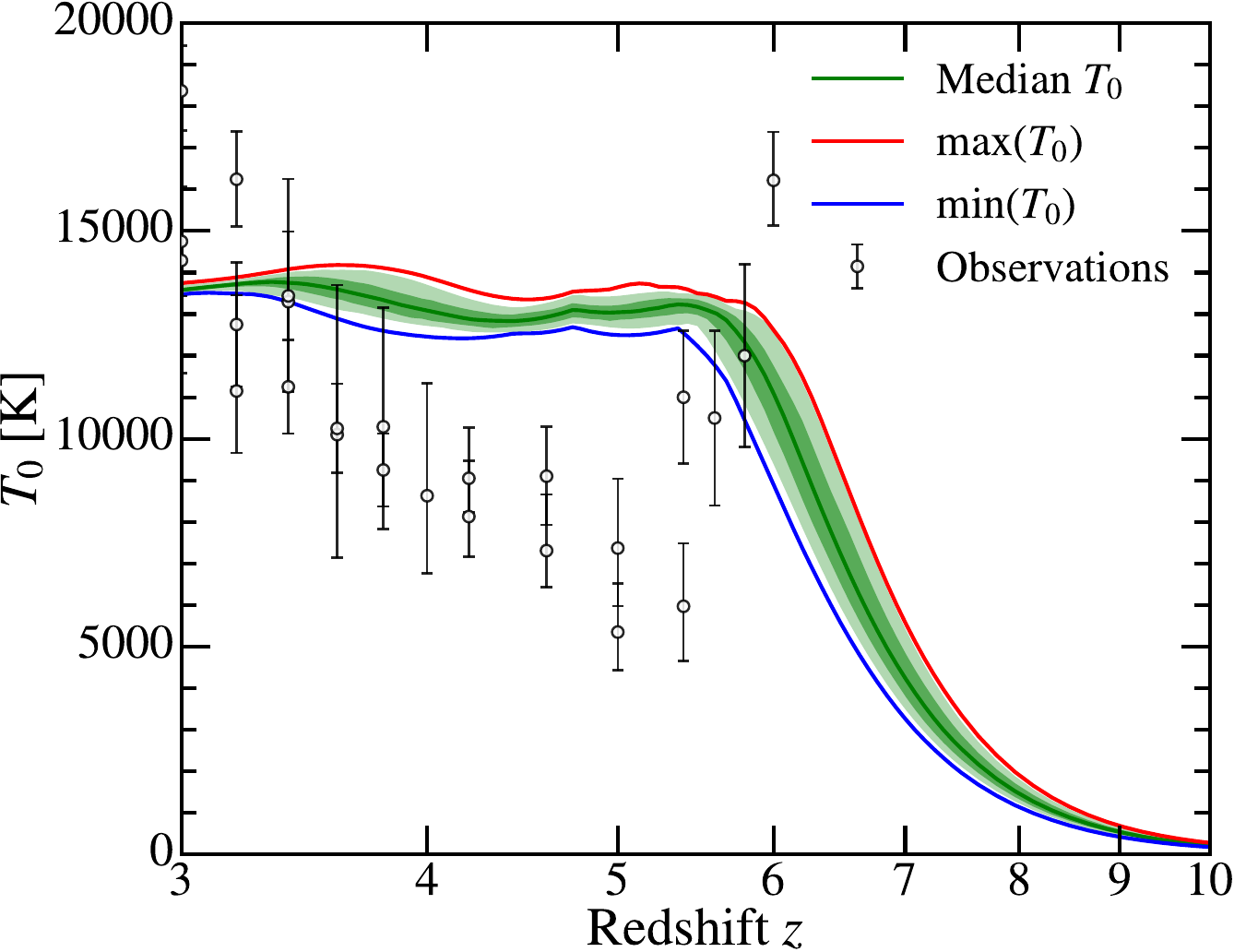}
    \includegraphics[width=\linewidth]{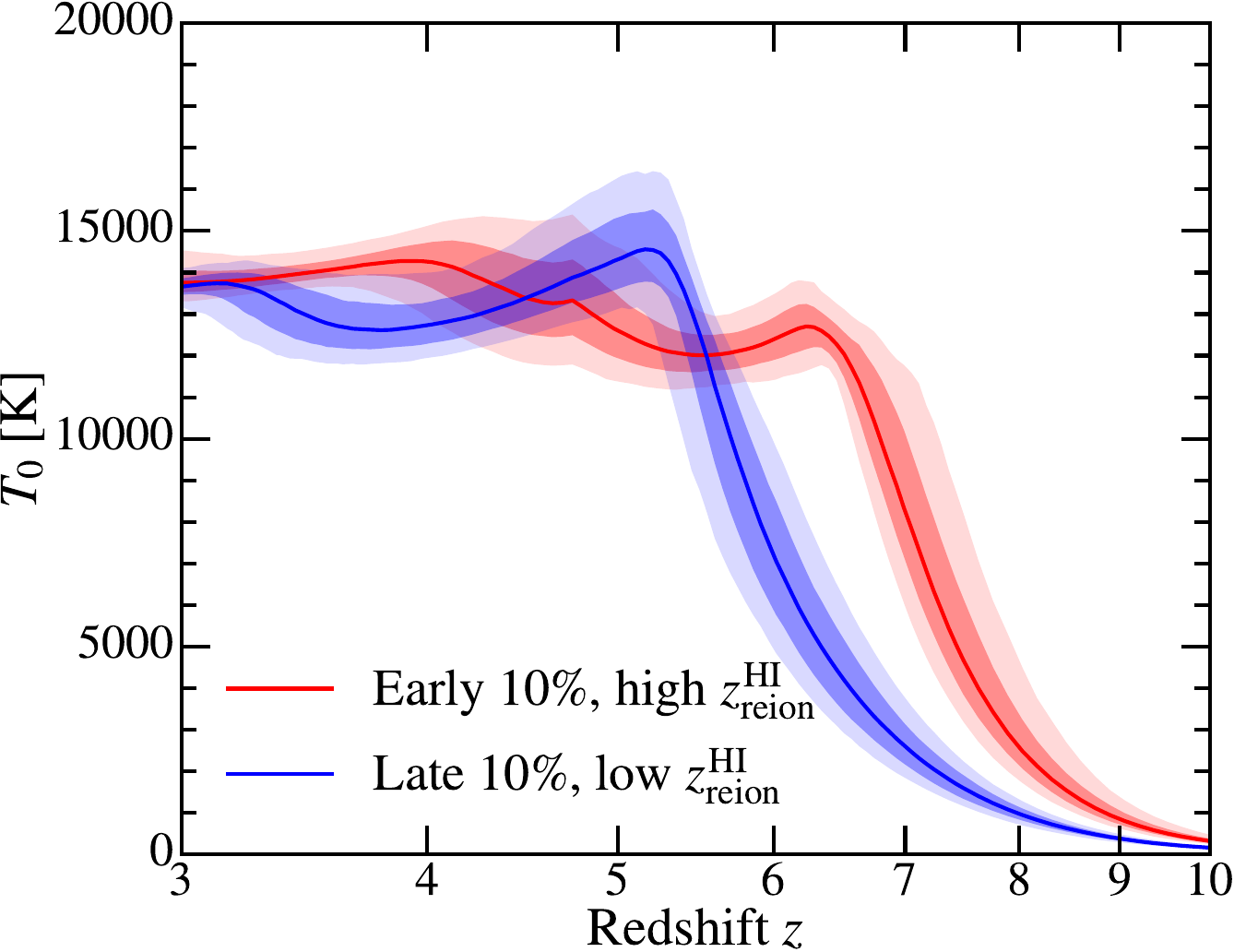}
  \caption{Temperature at mean density, $T_{0}(z)$. Top: the simulation volume is divided into 125 sub-boxes of side length $100\,\mathrm{cMpc}$, and $T_{0}$ is determined independently in each. The green curve shows the median over all sub-boxes, with shaded bands marking the $16^{\rm th}$--$84^{\rm th}$ and $2.5^{\rm th}$--$97.5^{\rm th}$ percentile ranges. The red and blue curves denote the maximum and minimum values across sub-boxes. Observational constraints from \protect\citet{bolton2012improved,hiss2018new,walther2019new,boera2019revealing,gaikwad2020probing,gaikwad2021consistent} are shown for comparison. Bottom: $T_{0}(z)$ in regions with early \HI reionization (red) and late \HI reionization (blue). Both samples exhibit a temperature peak during \HI reionization, followed by adiabatic cooling and a second heating phase during \HeII reionization. Regions that reionize early in \HI also tend to reionize early in \HeII. For this analysis the simulation volume is partitioned into 8000 sub-volumes of side length $25\,\mathrm{cMpc}$, and $T_{0}$ is computed in each as described in the main text. We define $z_{\rm reion}^{\rm HI}$ as the last redshift at which the volume-weighted \HI fraction in a sub-volume exceeds $1\%$. Solid lines show the median, and shaded bands the $16^{\rm th}$--$84^{\rm th}$ and $2.5^{\rm th}$--$97.5^{\rm th}$ percentile ranges.}
    \label{fig:temperature_evolution_igm}
\end{figure}

\begin{figure*}
    \centering
    \includegraphics[width=1\linewidth]{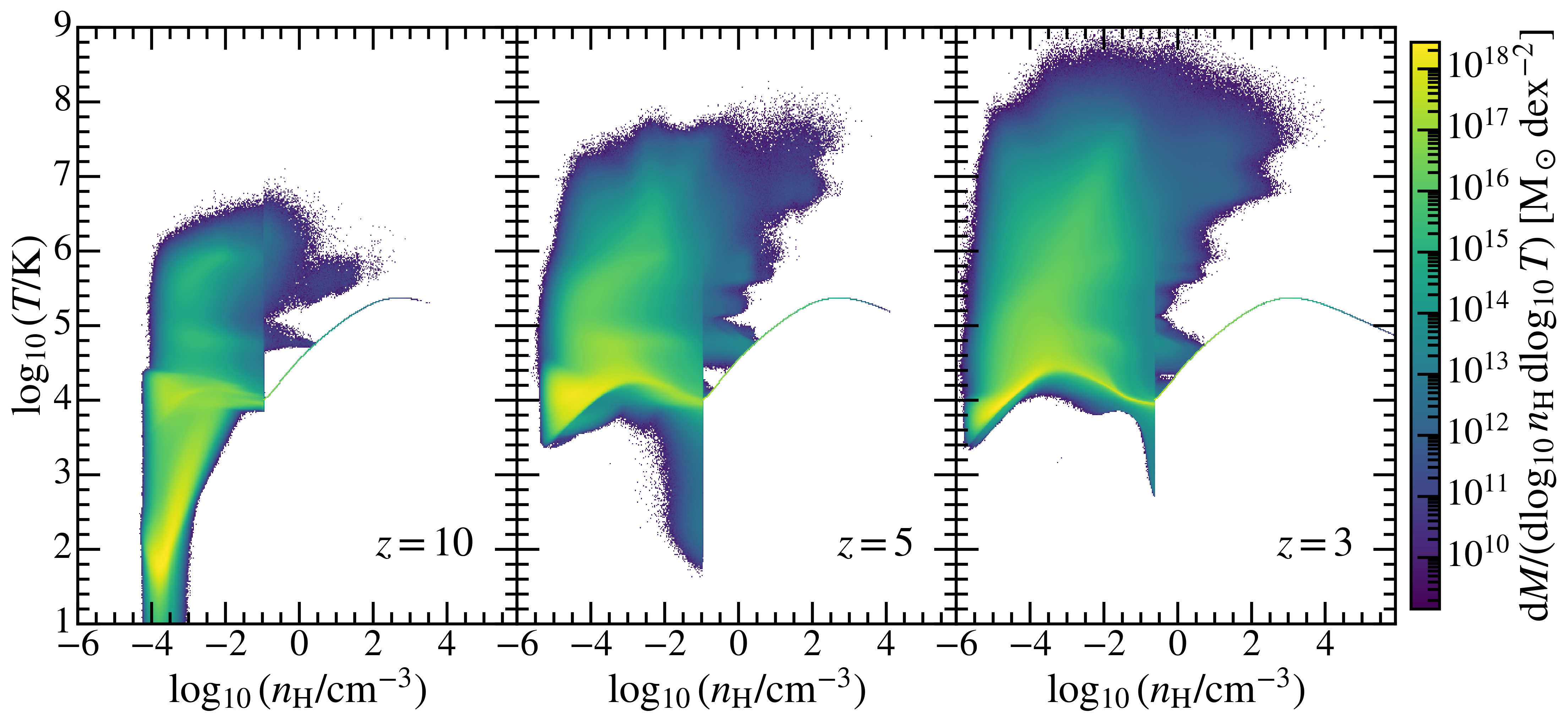}
    \caption{Temperature--number-density phase diagram of all gas in \lumina\ at three redshifts: early hydrogen reionization, near the completion of hydrogen reionization, and near the completion of \HeII reionization. The pixel colour encodes the gas mass per logarithmic bin in $n_{\rm H}$ and $T$, $\mathrm{d}M/(\mathrm{d}\log_{10}n_{\rm H}\,\mathrm{d}\log_{10}T)$, in units of $\mathrm{M}_\odot\,\mathrm{dex}^{-2}$. In all panels the star-forming equation of state is visible above the density threshold $n_{\rm H}\approx 0.106\,\mathrm{cm^{-3}}$, with additional shock- and AGN-heated gas extending to higher temperatures. Before \HI reionization, the IGM is predominantly cold; it is then pre-heated by X-rays and subsequently heated further during reionization.}
    \label{fig:density_temperature_3panel}
\end{figure*}

\subsection{Thermal history of the IGM}
\label{subsec:IGMthermal}
The two reionization histories described above leave a coherent imprint on the thermal state of the low-density IGM. We characterise this state through the temperature at mean density, $T_0$, and the polytropic index, $\gamma$, defined through
\begin{equation}
    T(\Delta) \;=\; T_0\,\Delta^{\gamma-1} \, ,
    \label{eq:temperatureIGM}
\end{equation}
which yields $\gamma = 5/3$ for purely adiabatic cooling but is modified by the timing and inhomogeneity of photoheating. To trace $T_0(z)$ we use the high-cadence three-dimensional renders described in \cref{subsec:uniformGrid} on a uniform $1280^{3}$ Cartesian grid, retain only cells with baryon overdensity $0.3<\Delta<10$ to exclude gas in haloes, partition the volume into $5^{3}=125$ sub-boxes of side length $100\,\mathrm{cMpc}$ (slightly larger than the \thesan\ box), and fit \cref{eq:temperatureIGM} in each sub-box at each snapshot. At every redshift we report the median across sub-boxes together with the $16^{\rm th}$--$84^{\rm th}$ and $2.5^{\rm th}$--$97.5^{\rm th}$ percentile ranges.

\Cref{fig:temperature_evolution_igm} (top) shows the resulting thermal history together with observational constraints. For $z<10$ the IGM heats up during \HI reionization, with the median peaking around $z\approx 5.5$; it cools by adiabatic expansion thereafter, undergoes a second heating phase during \HeII reionization, and finally returns to adiabatic cooling. The ensemble median does not exhibit a sharply defined \HI peak because it contains sub-boxes that are simultaneously undergoing \HI reionization, \HeII reionization, or adiabatic cooling, which washes out the distinct features of each phase. To bring out the underlying signal, in the bottom panel we further subdivide the box into $20^{3}=8000$ sub-regions of side length $25\,\mathrm{cMpc}$, define $z_{\rm reion}^{\rm HI}$ as the last time the volume-weighted \HI fraction exceeded $1\%$, and contrast the upper decile (``early'', the $800$ regions with the largest $z_{\rm reion}^{\rm HI}$) with the lower decile (``late''). Both subsets show a \HI peak, an intermediate adiabatic-cooling phase, and a \HeII peak, and the full sequence---early adiabatic cooling, X-ray pre-heating, \HI/\HeI heating, post-\HI cooling, renewed \HeII heating, and late-time adiabatic cooling---only emerges once this regional split is applied. Crucially, regions that reionize early in \HI also reionize early in \HeII, in line with the spatial correlation between the stellar-mass density and the black-hole mass density that drives the source budgets identified in \cref{subsec:HIreionization,subsec:heliumReionization}.

The same picture emerges from the density--temperature plane. \Cref{fig:density_temperature_3panel} shows temperature--number-density phase diagrams of all gas in \lumina at three representative redshifts: $z=10$ (early \HI reionization), $z=5$ (end of \HI reionization), and $z=3$ (end of \HeII reionization). At all epochs, gas above $n_{\rm H}=0.106\,\mathrm{cm^{-3}}$ follows the effective equation of state, which sets the lower temperature envelope. At $z=10$ the IGM is multiphase, with cold gas near the temperature floor ($T\simeq 5\,\mathrm{K}$) cooled predominantly by adiabatic expansion, photoionised gas at $T\simeq (1$--$2)\times 10^{4}\,\mathrm{K}$, and a shock- and feedback-heated component reaching $T\gtrsim 10^{6}\,\mathrm{K}$. The hot component shows two prominent gaps, which we associate with temperature ranges in which the net cooling rate changes rapidly owing to atomic line-cooling features in the primordial cooling function---specifically, the collisional excitation of \HI near $T\simeq 2\times 10^{4}\,\mathrm{K}$ and of \HeII near $T\simeq 10^{5}\,\mathrm{K}$ \citep{weinberg1997photoionization}. By $z=5$, most of the IGM is ionized, and even the still-neutral regions have been heated to several thousand kelvin by X-rays---in contrast to \thesan, which retained cold gas ($T<100\,\mathrm{K}$) until the end of \HI reionization \citep{ThesanEnrico}. A small fraction of dense, self-shielded gas persists just below the star-forming density threshold; the maximum temperatures rise as accretion shocks become stronger and AGN feedback turns on in the most massive haloes; and the additional structure in the hot component above the effective equation of state most likely reflects features in the metal-line cooling function \citep[e.g.][]{sutherland1993cooling,Wiersma2009}. By $z=3$, the originally self-shielded gas has been ionized and heated---which we attribute to the adoption of a uniform UV background for photons with $h\nu<54.4\,\mathrm{eV}$---and a very hot ($T\sim 10^{8}\,\mathrm{K}$), high-density component appears above the effective equation of state, consistent with AGN heating. The maximum gas densities at $z=3$ are higher than at earlier times, reflecting both the longer consumption time adopted at $z\le 4.75$ and the increasingly extreme protocluster environments.

\begin{figure}
    \centering
    \includegraphics[width=1\linewidth]{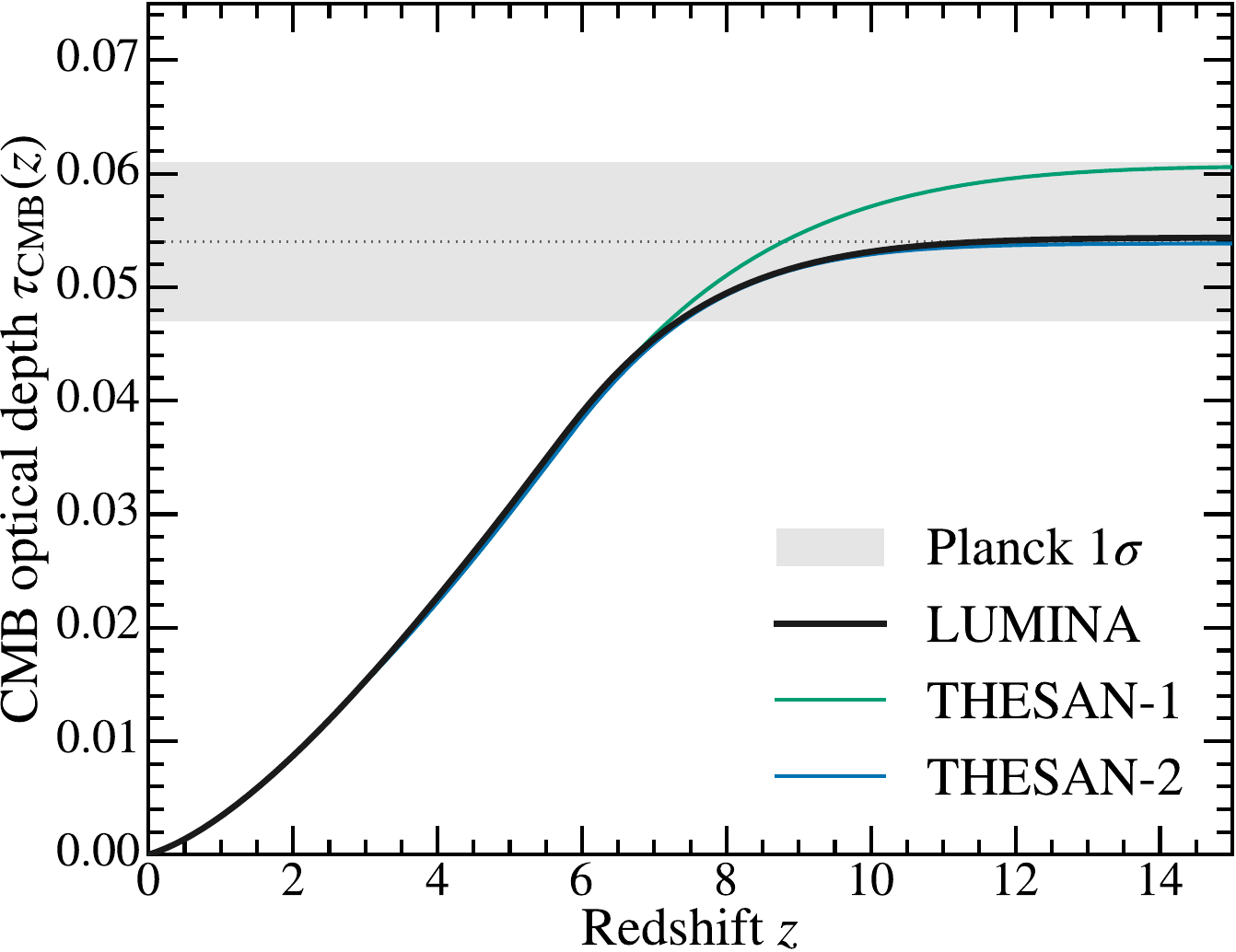}
    \caption{CMB optical depth $\tau_{\rm CMB}(z)$, compared with the constraint from \protect\citet[][TT,TE,EE+lowE]{planck2020} (grey band, $1\sigma$). We use \cref{eq:opticalDepthCMB} together with the simulated electron density for $z>3$, and assume full ionization of H and He for $z<3$. We compare \lumina\ with \thesanone\ and \thesantwo\ from the original \thesan\ suite. \lumina\ shows nearly perfect agreement with \thesantwo, while \thesanone\ yields a slightly larger $\tau_{\rm CMB}$.}

    \label{fig:cmb_depth}
\end{figure}

\subsection{Thomson optical depth to the CMB}
\label{subsec:tauCMB}
As an integrated cross-check on the combined reionization history, we compute the Thomson optical depth of the CMB,
\begin{equation}
    \tau_{\rm CMB}(z)
    = c\,\sigma_{\rm T}\int_{0}^{z} \frac{n_e(z')}{(1+z')\,H(z')}\,\mathrm{d}z' \,,
    \label{eq:opticalDepthCMB}
\end{equation}
where $c$ is the speed of light, $\sigma_{\rm T}$ is the Thomson cross-section, $n_e$ is the proper free-electron number density, and $H(z)$ is the Hubble parameter. We obtain $n_e(z)$ self-consistently from the high-cadence mass-averaged ion fractions for $z\ge z_{\rm end}=3$; as shown in a companion paper \citep{Smith2026}, the mass-weighted average is better suited than the volume-weighted one for reproducing light-cone-based $\tau_{\rm CMB}$ measurements, because reionization redshift and overdensity are correlated. For $z<z_{\rm end}$ we assume that hydrogen and helium are fully ionized (i.e.\ \HII and \HeIII), implying $n_e/n_{\rm H} = 1 + 2\,n_{\rm He}/n_{\rm H} \simeq 1.158$ for a primordial helium mass fraction $Y_{\rm p}=0.24$.

The resulting $\tau_{\rm CMB}$ is shown in \cref{fig:cmb_depth} alongside the Planck constraint and the predictions of \thesan. Our value is consistent with $\tau=0.054\pm 0.007$ \citep[TT,TE,EE+lowE;][]{planck2020}. Recent re-analyses of the Planck HFI data find slightly larger optical depths, $\tau=0.059\pm 0.006$ \citep{pagano2020reionization} and $\tau=0.0627^{+0.0050}_{-0.0058}$ \citep{de2021inference}, which are modestly higher than ours and would favour an earlier onset and/or a more extended high-redshift tail of reionization. Neither \thesanone\ nor \thesantwo\ followed the \HeII$\rightarrow$\HeIII transition self-consistently, so for a consistent comparison we assume (\textit{i}) full ionization of \HI and \HeI below the final redshift of each \thesan\ run and (\textit{ii}) full ionization of \HeII below $z=3$. Under these assumptions \lumina closely matches \thesantwo, reflecting their similar globally averaged reionization histories, while \thesanone\ yields a slightly larger $\tau_{\rm CMB}$ owing to its earlier onset of reionization. We note that \citet{Thesan1} used a volume-weighted average and therefore reported slightly lower $\tau_{\rm CMB}$ values for the same simulations. Finally, \citet{sailer2025} argued that, when low-$\ell$ CMB polarization information is omitted, the combination of BAO with high-$\ell$ CMB data and CMB lensing can prefer a substantially larger optical depth, $\tau\simeq 0.090\pm 0.012$. Such a high value would be in tension with the low-$\ell$ polarization-based Planck constraints and would require significantly more ionization at very high redshift, for example from time-varying escape fractions, additional early sources, or other non-standard ionization channels. The sightline-to-sightline fluctuations of $\tau_{\rm CMB}$ in a patchy reionization scenario \citep[e.g.][]{dvorkin2009reconstructing,Mesut2024} will be quantified in a companion paper \citep{Smith2026}.

\section{Data products}
\label{sec:dataProducts}
The large number of resolution elements implies substantial storage requirements. For illustration, storing one single-precision value (\SI{4}{B}) per particle already amounts to \SI{1.7}{TB}. Beyond the direct storage cost, data volumes of this size make it impractical to load a significant fraction of the simulation into the memory of a single node and increase I/O overheads. We therefore carry out many analyses on the fly and generate auxiliary data products that enable post-processing without repeatedly reading the full raw snapshots. Below we summarize the data products produced.

\subsection{Traditional snapshots}
Starting at $z=23$, we output 32 full snapshots containing the primary quantities for all resolution elements. By default, each snapshot produced by \arepo\ is distributed across several hundred HDF5 files, with every file storing all fields for a subset of particles or cells; this layout requires opening many files in order to access a single field across the full volume. We therefore repackage each snapshot into a smaller set of files in a columnar layout, where each file contains a single field for all resolution elements.

In addition, for each snapshot we provide an HDF5 virtual dataset that exposes the repackaged files as a single logical snapshot. We use HDF5 chunking to accelerate partial reads. To reduce storage further, we store particle coordinates as 32-bit integers, yielding a uniform absolute error in position; physical coordinates are recovered by multiplying by the box size and dividing by $2^{32}$.

\subsection{Halo catalogues}
\label{subsec:haloCatalogue}
To obtain a finer time sampling for studies of galaxy evolution, we compute halo catalogues using FoF and \textsc{SUBFIND-HBT} (see \cref{subsec:structureFinding}). In addition to the standard fields provided for IllustrisTNG \citep{nelson2019illustristng}, we compute the angular momentum and spin of each halo and subhalo following \citet{zjupa2017angular}, as well as their luminosities. Merger trees are constructed on the fly as described in \citet{Gadget4}.

\textsc{SUBFIND-HBT} allows only the central subhalo to grow, which can be problematic during mergers involving satellite galaxies. This occurs especially at high redshift and can cause parts of a physically distinct satellite to be incorrectly assigned to the central subhalo. We therefore additionally post-process the full snapshots with the traditional \textsc{SUBFIND} algorithm and adopt those results throughout this paper.

\subsection{High-cadence Cartesian output}
\label{subsec:uniformGrid}
To obtain a finer output cadence for IGM studies, we also deposit selected particle and cell properties on to uniform three-dimensional Cartesian grids. These gridded products, already used in \thesan, are described in detail in \citet{ThesanDR}. Primary fields (e.g.\ gas density and ionization fractions) are stored on a $2560^{3}$ grid, while additional quantities are stored at $1280^{3}$ resolution. Grids are written at global (synchronised) time steps, targeting a maximum cadence of $2\,\mathrm{Myr}$ in physical time; owing to the power-of-two time-step hierarchy, adjacent outputs are typically separated by $2$--$4\,\mathrm{Myr}$. In total we produced 708 Cartesian outputs between $z=29.4$ and $z=3$.

Whereas \thesan\ assigned each source entirely to the nearest Cartesian cell (nearest-grid-point assignment), which can leave cells empty in underdense regions, we use a modified adaptive cloud-in-cell (CIC) scheme. Collisionless particles are represented as cubes whose side length equals the grid spacing, and gas cells are approximated by Cartesian cubes of the same volume as their Voronoi counterparts. We then compute geometric overlaps with the uniform grid and distribute extensive quantities in proportion to the intersected volume. This conservative, volume-weighted deposition yields a smooth mapping---particularly for Voronoi-defined fields---and reduces sampling noise in low-density regions.

\subsection{On-the-fly light-cone construction}
During the simulation we construct a pixelised light cone in a small-angle Cartesian-like approximation, with coordinates $(\theta_x,\theta_y)$ and $5120^{2}$ perspective rays oriented according to the cosmological distance and angular position relative to the $+z$ viewing direction. The transverse pixel size is chosen such that the light cone subtends the full simulation box at $z=5$, corresponding to a field of view of $3.6\,\mathrm{deg}$. At $z>5$ the light cone is filled using periodic replications of the simulation volume, while at $z<5$ a slightly smaller fraction of the box lies within the field of view.

Along each ray we discretise the line of sight into volumetric pixels (voxels) whose depth is matched to the transverse pixel size, yielding approximately cubic sampling elements. For each voxel we compute both volume-weighted and mass-weighted quantities by depositing contributions from the intersected Voronoi cells. To reduce small-scale sampling noise and produce smoother maps, we approximate each Voronoi cell by a $(\Delta\theta_x,\Delta\theta_y,\Delta z)$ cube of equal volume, analogous to our adaptive-CIC Cartesian output procedure.

At every global simulation time step we determine which voxels intersect the current light-cone shell and record the corresponding derived properties. Selected quantities are provided both in real space and in redshift space, including line-of-sight peculiar-velocity distortions where appropriate. Each ray contains $47{,}169 = 32{,}081\,(\text{high-}z) + 15{,}088\,(\text{low-}z)$ voxels from the end of the simulation at $z=3$ up to the maximum redshift $z=30$ covered by the cone.

Our approach differs from the light-cone outputs adopted in MTNG \citep{Gadget4,pakmor2023millenniumtng}, where the properties of individual particles are recorded at the instant they cross the light-cone surface. The MTNG approach enables an exact reconstruction of light-cone fields without additional smoothing, but at the price of substantially larger storage requirements and more complex post-processing.

\subsection{High-frequency black-hole output}
In sync with the light-cone outputs, we also record the full set of properties for every black-hole particle. This high-cadence output makes it possible to construct black-hole merger trees and to study the temporal variability of accretion, feedback, and clustering. The merger trees can also serve as input to post-processing codes that model black-hole binary evolution, such as Holodeck \citep{Agazie2023}; these can be used, for instance, to predict the high-redshift binary population and the stochastic gravitational-wave background. Such post-processing is needed because black-hole particles are instantaneously repositioned to the local potential minimum, resulting in merger timescales that are shorter than physically realistic. In total, we produce 909 black-hole outputs spanning the period from the first seeding at $z=14.3$ down to $z=3$.

\subsection{Power spectra on the fly}
\label{subsec:powerSpectra}
At each group-catalogue output we compute a suite of power spectra following \citet{springel2018first}. Particles are deposited on to a uniform $16384^{3}$ Cartesian mesh using non-adaptive CIC assignment, and we evaluate the Fourier transform of the resulting density field. To extend the dynamic range to smaller scales, we additionally apply the self-folding technique of \citet{jenkins1998evolution}, in which the full simulation volume is mapped onto a sub-volume of linear size $L/f_{\rm fold}$. We adopt $f_{\rm fold}=16$ and apply two successive folding iterations.

This approach yields: (\textit{i}) the total matter power spectrum; (\textit{ii}) matter power spectra for the individual particle species; (\textit{iii}) the power spectrum of the neutral-hydrogen density field during hydrogen reionization; and (\textit{iv}) the power spectrum of the ionized-fraction field. The latter two are directly relevant for predictions of the 21\,cm power spectrum.

\subsection{High-cadence reionization history and radiation output}
\label{subsec:highCadenceReionHistory}
At every global time step we compute volume- and mass-weighted ionization fractions, together with volume- and mass-weighted gas temperatures, averaged over the full simulation volume. In addition, we record the radiative energy injected in each frequency bin separately for every source class. This enables a global, time-resolved assessment of the relative contributions of the different radiation sources to hydrogen reionization.

\section{Summary and discussion}
\label{sec:summary}
We have introduced \lumina, a large-volume radiation-hydrodynamic simulation with comoving side length $L_{\rm box}=500\,\mathrm{cMpc}$ that self-consistently follows hydrogen and \HeII reionization, together with their sources, down to $z=3$. The calculation evolves $2\times 6000^{3}$ resolution elements, corresponding to a dark-matter particle mass of $m_{\rm DM}=1.9\times 10^{7}\,\Msun$ and a target baryonic mass resolution of $m_{\rm b}=3.6\times 10^{6}\,\Msun$. We combine the well-tested \textsc{IllustrisTNG} galaxy-formation model with a GPU-accelerated M1 radiation-transport solver.

For low-density gas with $n_{\rm H}<0.106\,\mathrm{cm^{-3}}$ we replace the equilibrium cooling treatment of \textsc{TNG} with a non-equilibrium primordial chemistry network coupled to the local radiation field. Above this threshold, where the sub-grid ISM model assumes a multi-phase unresolved medium, we treat the gas as optically thin. We include three classes of radiation sources: (\textit{i}) stellar populations, modelled with BPASS v2.2.1 spectra and an effective escape fraction $f_{\mathrm{esc},\star}=0.18$ calibrated to reproduce a realistic \HI reionization history; (\textit{ii}) high-mass X-ray binaries and shocked, hot interstellar gas, whose emissivities scale with the local star-formation rate and dominate the X-ray budget during \HI reionization; and (\textit{iii}) active galactic nuclei, which provide the hard photons that drive \HeII reionization.

For $z>4.75$ we transport radiation in six frequency groups covering \HI, \HeI, and \HeII ionization, together with three additional higher-energy groups. At $z<4.75$ we replace the two lowest-energy groups with a spatially uniform UV background and merge the remaining transported high-energy bins to reduce the computational cost. We also use a new initial-conditions generator that self-consistently includes distinct baryon and dark-matter transfer functions as well as the relative baryon--dark-matter streaming velocity.

We have presented initial results for the evolution of the matter power spectrum, the \HI and \HeII reionization histories, the thermal history of the IGM, and the galaxy population. Our main findings are:
\begin{enumerate}
    \item Adopting distinct baryon and dark-matter transfer functions accounts for the ${\sim}10$~per cent suppression of the baryonic power spectrum relative to the total matter at $z=10$, which persists across a broad range of scales. \lumina propagates this suppression forward in time in good agreement with linear theory, with implications for early galaxy formation and for IGM observables such as the 21\,cm power spectrum.

    \item The IGM exhibits a multi-stage thermal history: early adiabatic cooling, X-ray pre-heating above the CMB temperature, rapid photoheating to $T\simeq (1$--$2)\times 10^{4}\,\mathrm{K}$ during \HI and \HeI reionization, subsequent cooling, renewed heating during \HeII reionization, and late-time adiabatic cooling. As we show in a future paper, regions that reionize early in \HI also tend to reionize early in \HeII, producing correlated temperature peaks.

    \item \lumina predicts a comparatively late and spatially extended \HI reionization history. After subdividing the full volume into $5^{3}=125$ sub-boxes of side length $100\,\mathrm{cMpc}$, we find that the median sub-volume completes \HI reionization well before the latest-reionizing regions; the final neutral reservoir disappears only by $z=4.75$. The large simulation volume enables a more robust assessment of the impact of cosmic variance on the timing and spatial structure of reionization.

    \item \HI and \HeI reionization are driven primarily by stellar sources. AGN contribute a sub-dominant fraction of the ionizing budget towards the end of hydrogen reionization at $z=5$ (${\sim}15\%$). The X-ray pre-heating is dominated by high-mass X-ray binaries and by the shocked hot gas associated with star formation.

    \item \HeII reionization begins at $z\simeq 6$ and is driven by the hard radiation field from AGN. We find a midpoint at $z\simeq 4$ and near-completion by $z=3$, with a residual \HeII volume filling fraction below $1$~per cent, in agreement with current observational constraints. This demonstrates that \lumina can self-consistently follow the onset, topology, and completion of helium reionization alongside hydrogen reionization.
\end{enumerate}

These results establish \lumina as a framework for jointly modelling galaxy formation, hydrogen reionization, and \HeII reionization in a cosmological volume large enough to capture substantial cosmic variance. The present paper provides the context for a broader set of companion studies based on \lumina, including initial analyses of the spatial variance of the CMB optical depth $\tau_{\rm CMB}$ \citep{Smith2026}, of the AGN luminosity function across multiple bands \citep{Shen2026} and intergalactic clumping \citep{Sadain2026}; further investigations of the simulation and its observational implications will follow.

\section*{Acknowledgements}
OZ thanks Lan and Nori for their help in developing the project website and for their support throughout the simulation campaign. We thank Yueying Ni, Ewald Puchwein, Meredith Neyer, William McClymont, Laura Keating, Josh Borrow, Enrico Garaldi, Koki Kakiichi, Harley Katz, Nick Gnedin, and Christopher Cain for useful discussions.

An award of computer time was provided by the INCITE program. This research used resources of the Oak Ridge Leadership Computing Facility at the Oak Ridge National Laboratory, which is supported by the Advanced Scientific Computing Research programs in the Office of Science of the U.S. Department of Energy under Contract No.\ DE-AC05-00OR22725. We thank our OLCF liaison, Reuben D. Budiardja, for technical support during the simulation campaign.
The authors acknowledge the MIT Office of Research Computing and Data and the FAS Division of Science Research Computing Group at Harvard University for providing computational resources that contributed to the results reported in this paper.
The GPU implementation was optimized during the NASA GPU Hackathon 2024, part of the Open Hackathons program. The authors acknowledge OpenACC-Standard.org for their support and thank Christopher Brissette for his mentorship during the hackathon.
Support for OZ was provided by Harvard University through the Institute for Theory and Computation Fellowship.
Support for programs JWST-AR-08709 (AS) and JWST-AR-04814 (XS, MV) were provided by NASA through a grant from the Space Telescope Science Institute, which is operated by the Association of Universities for Research in Astronomy, Inc., under NASA contract NAS 5-03127.
RK acknowledges support from the Natural Sciences and Engineering Research Council of Canada (NSERC) through a Discovery Grant and a Discovery Launch Supplement, funding reference numbers RGPIN-2024-06222 and DGECR-2024-00144, and from York University's Global Research Excellence Initiative.
MV acknowledges support through NASA ATP Grant 23-ATP23-149 and NSF AAG Grant AST-2307699.
VS and LH acknowledge support from the Simons Foundation through the ``Learning the Universe" initiative.

The AI tools Claude and ChatGPT were used for language editing and proofreading of the manuscript.

\section*{Data Availability}
All \lumina data will be made publicly available in 2028.
Some of the data presented in the figures of this article are already available at \url{https://www.lumina-simulation.com}.
Additional data are available from the corresponding author upon reasonable request.



\bibliographystyle{mnras}
\bibliography{main} 



\appendix

\section{Dark-matter-only simulations}
\label{app:haloPropertiesAtZ0}

Because the computational cost increases steeply towards low redshift, we do not evolve the \lumina\ flagship simulation down to $z=0$. Instead, we perform a set of dark-matter-only (DM-only) simulations to $z=0$ with $3000^{3}$ and $1500^{3}$ particles, respectively. These runs are carried out with the \textsc{Gadget-4} code \citep{Gadget4} using exactly the same cosmological parameters and random phases as \lumina, which enables a straightforward, object-by-object matching of resolved haloes between \lumina\ and the DM-only runs. Their mass resolution is sufficient to resolve haloes of mass $10^{10}\,\Msun$ and $10^{11}\,\Msun$ with at least 50 dark-matter particles.

The DM-only simulations serve as ideal parent volumes for constructing zoom-in initial conditions of objects identified in \lumina, which can then be followed efficiently to lower redshift and/or at higher resolution. The initial conditions are generated with the \textsc{N-GenIC} package included in \textsc{Gadget-4}, supporting the multi-zoom-in technique described in \citet{burger2025applying}, in which several haloes can be simultaneously targeted within a single simulation volume.

\begin{figure}
    \centering
    \includegraphics[width=1\linewidth]{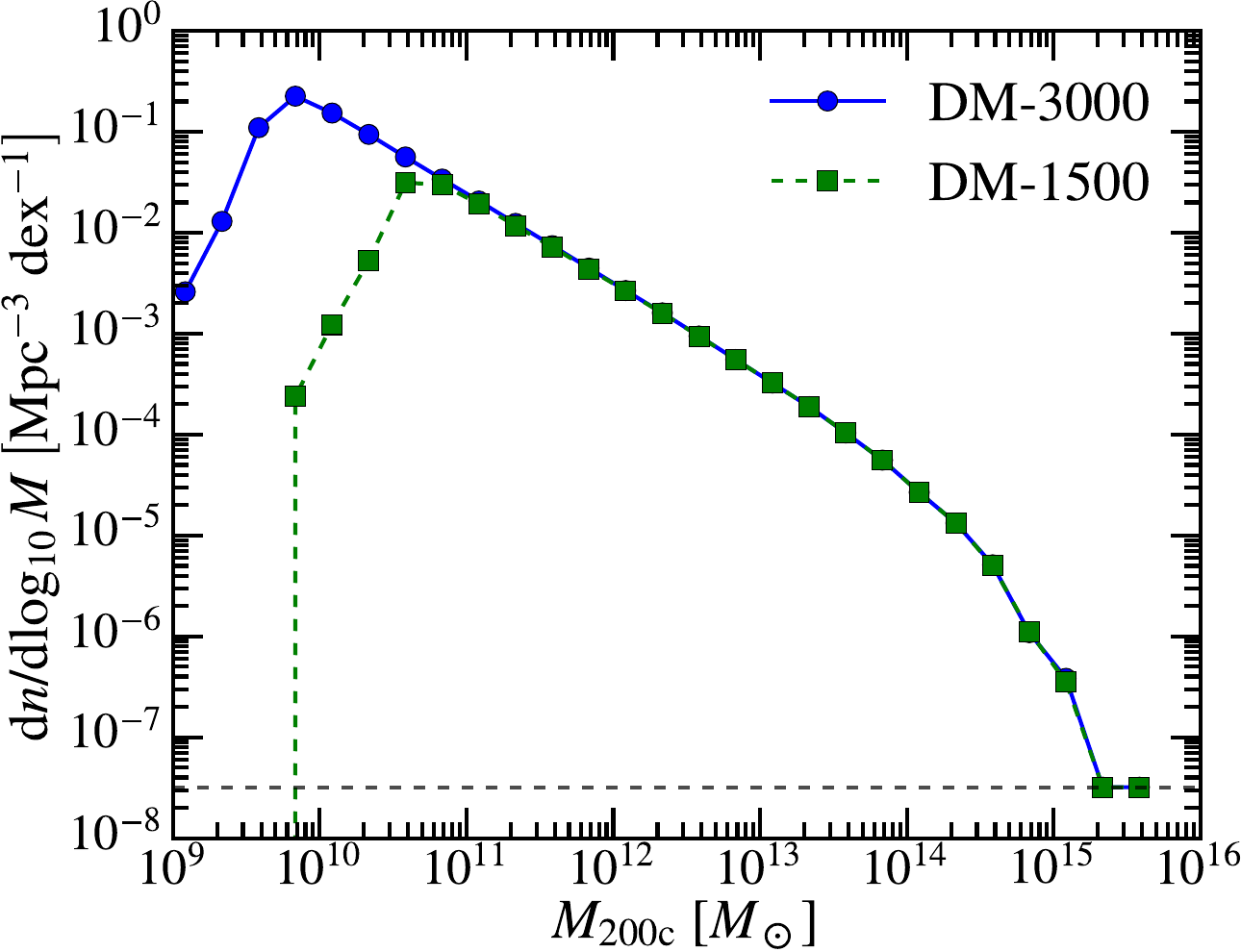}
    \caption{Halo mass function at $z=0$ for the dark-matter-only \lumina\ simulations with $3000^{3}$ (blue, solid) and $1500^{3}$ (green, dashed) dark-matter particles. The two runs agree well in the regime where haloes are resolved with at least 50 particles ($M_{200{\rm c}} \gtrsim 10^{11}\,\Msun$). The horizontal dashed line indicates the number density that corresponds to a single halo within the simulation volume.}
    \label{fig:halo_mass_function}
\end{figure}

To illustrate the range of potential zoom-in targets, \cref{fig:halo_mass_function} shows the halo mass function at $z=0$ for both DM-only runs, and \cref{tab:DMonly} lists the number of haloes in different mass ranges as a function of redshift. The most massive halo reaches $M_{200{\rm c}} = 3.3\times 10^{15}\,\Msun$ at $z=0$.

\begin{table}
  \centering
  \begin{tabular}{cccc}
    \hline
    $\log_{10}\!\left(M_{200{\rm c}}/\Msun\right)$ 
      & $N(z=0)$ & $N(z=3)$ & $N(z=5)$ \\
    \hline
    $9.5$--$10.0$  & 11217263  & 13143136 & 8774818  \\
    $10.0$--$10.5$ &  7114802  & 7874229  & 4373250  \\
    $10.5$--$11.0$ &  2615381  & 2537192  & 1097052  \\
    $11.0$--$11.5$ &   947595  & 740153   & 222012   \\
    $11.5$--$12.0$ &   342382  & 191355   & 32788    \\
    $12.0$--$12.5$ &   123017  & 40257    & 3073     \\
    $12.5$--$13.0$ &    43020  & 6109     & 111      \\
    $13.0$--$13.5$ &    14622  & 483      & 2        \\
    $13.5$--$14.0$ &     4493  & 16       & 0        \\
    $14.0$--$14.5$ &     1118  & 0        & 0        \\
    $14.5$--$15.0$ &      157  & 0        & 0        \\
    $15.0$--$15.5$ &        9  & 0        & 0        \\
    $15.5$--$16.0$ &        1  & 0        & 0        \\
    \hline
  \end{tabular}
  \caption{Halo counts in mass bins of width 0.5\,dex at different redshifts in the DM-only simulation with $3000^{3}$ particles.}
  \label{tab:DMonly}
\end{table}

\section{Impact of X-ray pre-heating}
\label{app:impactPreheating}

\begin{figure}
    \centering
    \includegraphics[width=1\linewidth]{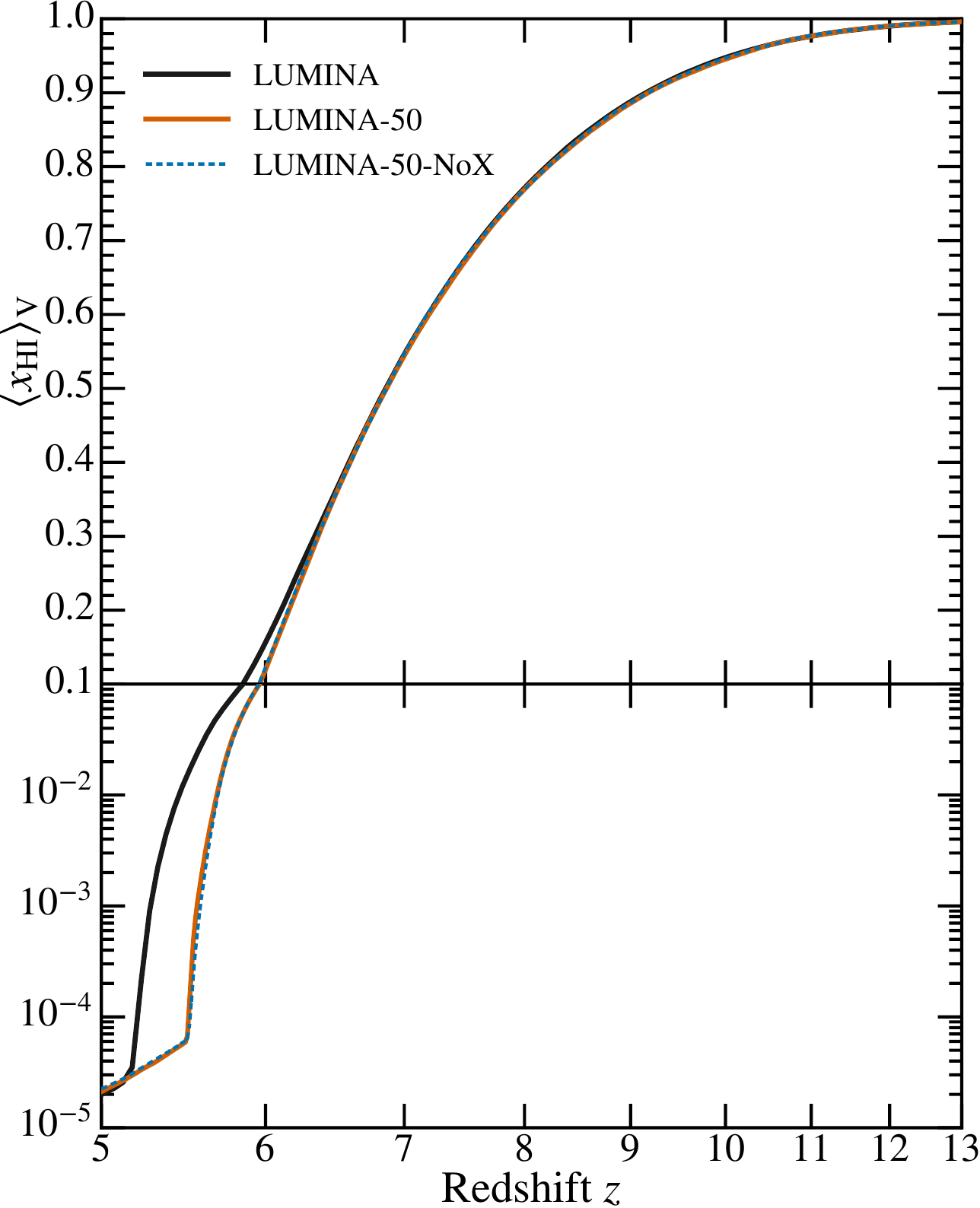}
    \caption{Volume-weighted neutral-hydrogen fraction as a function of redshift in the \luminafifty{} (solid orange) and \luminafiftynox{} (dashed blue) test simulations. The black curve shows the median reionization history of the \lumina flagship run, computed from 125 sub-boxes of side length $100\,\mathrm{cMpc}$ (see ~\cref{fig:x_HI_reionization}). The two test runs are nearly indistinguishable, demonstrating that X-ray pre-heating does not affect the progression of reionization; their slightly earlier end of reionization relative to the flagship median reflects their smaller box size.}
    \label{fig:app_reion_history_xray}
\end{figure}

\begin{figure}
    \centering
    \includegraphics[width=1\linewidth]{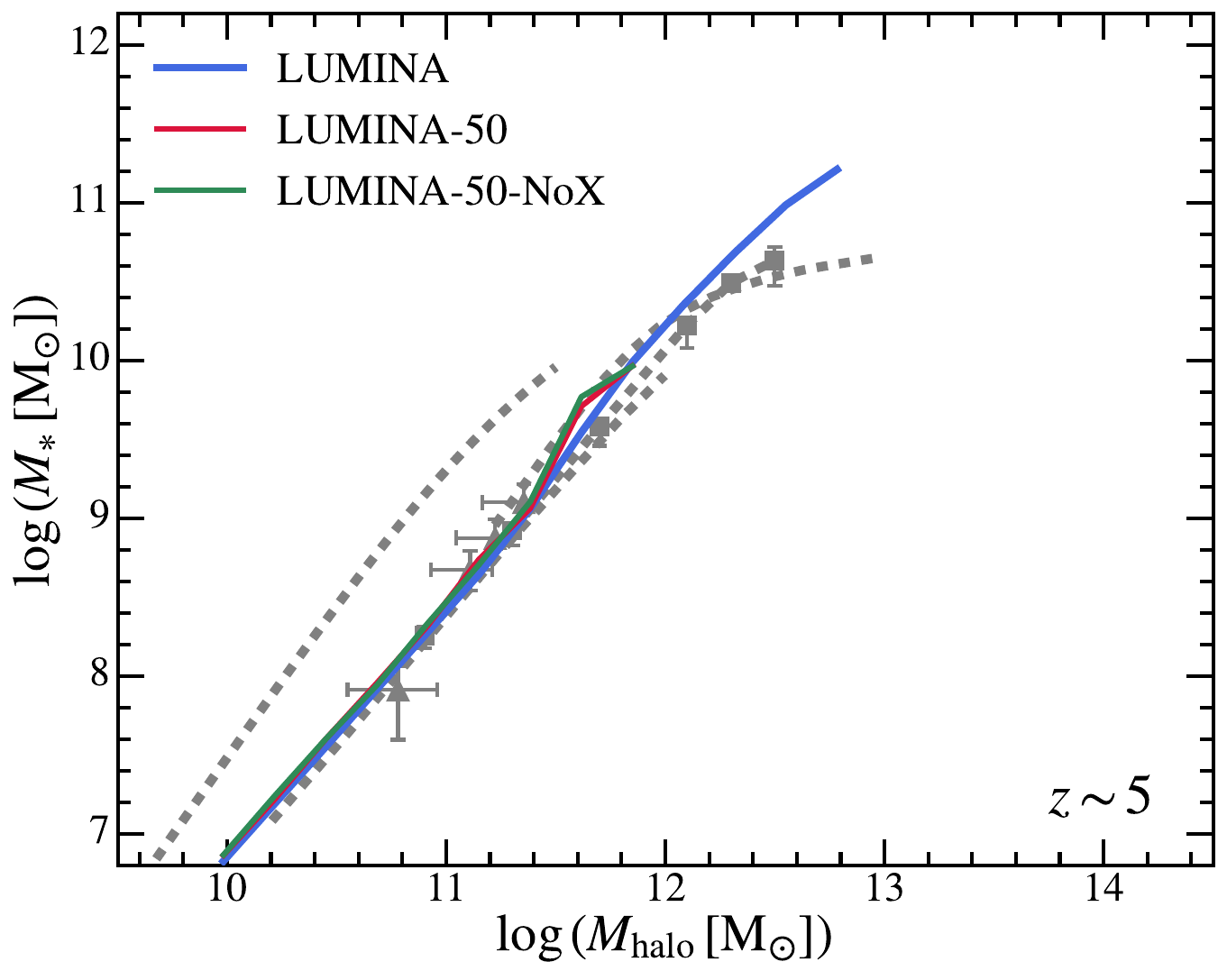}
    \caption{Stellar-mass--halo-mass relation at $z=5$ in the \luminafifty{} (red) and \luminafiftynox{} (green) runs, compared with the flagship \lumina simulation (blue). Lines show the median stellar mass as a function of halo mass. The two test runs are indistinguishable, demonstrating that X-ray pre-heating does not measurably alter the star-formation efficiency of resolved haloes; the flagship run extends to more massive haloes owing to its much larger volume. Grey lines and symbols show the observation-based constraints from \citet{Behroozi2013,Behroozi2019,RP2017,Harikane2016,Harikane2018,Tacchella2018}, as in \cref{fig:mstar_mhalo}.}
    \label{fig:app_shmr_xray}
\end{figure}

\begin{figure}
    \centering
    \includegraphics[width=1\linewidth]{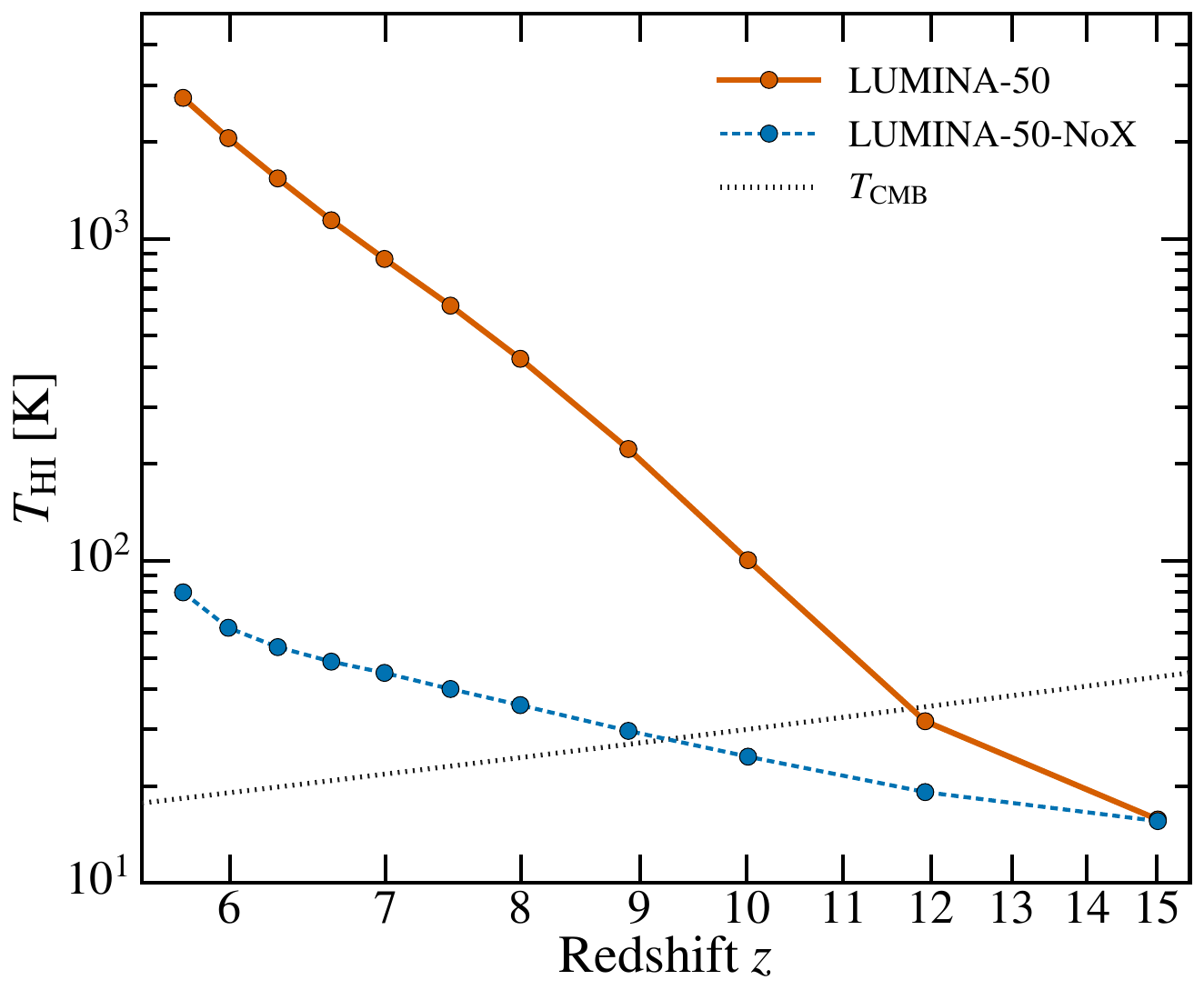}
    \caption{Median temperature of the neutral, mean-density IGM in the \luminafifty{} (solid orange) and \luminafiftynox{} (dashed blue) runs as a function of redshift. In each snapshot we select gas cells with baryon overdensity $0.5<\Delta<2$ and ionized fraction $x_{\HII}<10^{-3}$ and compute the median of their temperature distribution. The dotted line shows the CMB temperature $T_{\rm CMB}(z)=2.7255\,(1+z)\,\mathrm{K}$. Snapshots in which fewer than $1000$ cells satisfy the selection criteria ($z\leq 5.5$, once reionization is essentially complete) are omitted.}
    \label{fig:app_mean_IGM_temp_xray}
\end{figure}

As discussed in \cref{subsubsec:xraySource}, \lumina includes high-mass X-ray binaries and the shock-heated ISM as additional X-ray sources. Owing to their long mean free paths, the photons emitted by these sources penetrate deep into the neutral IGM and pre-heat it well before the arrival of the ionization fronts. In this appendix we quantify the impact of this pre-heating on the main science results presented in this work. To this end, we performed two additional radiation-hydrodynamics simulations in a periodic box of side length $50\,\mathrm{cMpc}$, using the same mass resolution and initial-conditions methodology as the flagship run (see \cref{tab:simulation_overview}). The run \luminafifty{} employs the full \lumina model, with all six frequency bins and all four source classes, while \luminafiftynox{} includes only stellar and AGN sources together with the three frequency bins below $100\,\mathrm{eV}$, and therefore lacks the X-ray pre-heating channel. Both simulations are evolved to $z=5$, covering hydrogen reionization.

\Cref{fig:app_reion_history_xray} compares the reionization histories of the two runs, together with the median history of the flagship run. The volume-weighted mean \HI fractions of \luminafifty{} and \luminafiftynox{} are nearly indistinguishable over the entire redshift range, as expected given that hydrogen reionization is driven by stellar EUV emission; both also track the flagship median closely, ending reionization slightly earlier owing to their smaller box size. 

The galaxy population is similarly unaffected. \Cref{fig:app_shmr_xray} shows the stellar-mass--halo-mass relation in both runs at $z=5$, after the galaxy population has been exposed to the X-ray background for the longest time: the median relations agree very well, and both are consistent with the flagship run, which extends to more massive haloes owing to its much larger volume. We find equally good agreement for the cosmic star-formation history of the two runs.

The one quantity that changes substantially is the temperature of the still-neutral IGM. \Cref{fig:app_mean_IGM_temp_xray} shows the median temperature of neutral ($x_{\HII}<10^{-3}$) gas at mean density ($0.5<\Delta<2$) in both runs, measured from the Voronoi cells in 14 snapshots between $z=15$ and $z=5$. Starting from ${\sim}16\,\mathrm{K}$ at $z=15$, this gas warms only slowly in \luminafiftynox{}, exceeding the CMB temperature at $z\approx 9$ and reaching just ${\sim}80\,\mathrm{K}$ by $z=5.7$, shortly before the last neutral regions are overrun by ionization fronts. In \luminafifty{}, by contrast, the neutral IGM is heated to ${\sim}2700\,\mathrm{K}$ by $z=5.7$ -- a difference of more than an order of magnitude. This difference is invisible in most of the diagnostics presented in this paper, which are dominated by ionized or dense gas, but it is crucial for predictions of the 21\,cm signal, which depend on the contrast between the spin temperature of the neutral IGM and the CMB \citep{Pritchard2007}.


\bsp 
\label{lastpage}
\end{document}